\documentclass[aip,cha,reprint,nofootinbib,a4paper,sort&compress]{revtex4-1}

\setcitestyle{numbers,square}
\usepackage{color}
\usepackage[]{graphicx}
\usepackage{chngcntr}
\usepackage{latexsym}
\usepackage{amsmath}
\usepackage[caption=false]{subfig}
\usepackage{multirow}
\usepackage{mathtools}
\usepackage{textcomp}
\usepackage{xr}
\usepackage{multibib}
\usepackage{booktabs}
\usepackage{caption}
\usepackage{comment}
\usepackage{minitoc}

\usepackage[pdftex,hyperfigures,breaklinks,colorlinks,citecolor=black,linkcolor=black]{hyperref} 

\newcommand{\be}{\begin{equation}}
\newcommand{\ee}{\end{equation}}
\newcommand{\bc}{\begin{cases}}
\newcommand{\ec}{\end{cases}}

\begin{document}

\title{Integrating behavioral experimental findings into dynamical models to inform social change interventions}

\author{Radu Tanase}
\affiliation{URPP Social Networks, University of Zurich, CH-8050 Zurich, Switzerland}

\author{René Algesheimer}
\affiliation{URPP Social Networks, University of Zurich, CH-8050 Zurich, Switzerland}

\author{Manuel S. Mariani}
\affiliation{URPP Social Networks, University of Zurich, CH-8050 Zurich, Switzerland}

\begin{abstract}
Addressing global challenges often involves stimulating the large-scale adoption of new products or behaviors. Research traditions that focus on individual decision making suggest that achieving this objective requires identifying the drivers of individual discrete adoption choices. On the other hand, computational approaches rooted in complexity science focus on maximizing the propagation of a given product or behavior throughout social networks of interconnected adopters. Here, by integrating discrete choice modeling into the complex contagion theory, we propose a method to estimate individual-level thresholds to adoption. We validate the predictive power of this approach in two choice experiments. By integrating the estimated thresholds into computational simulations, we show that state-of-the-art seeding policies for initiating large-scale behavioral change might be suboptimal if they neglect individual-level behavioral drivers, which can be corrected through the proposed experimental method.
\end{abstract}

\maketitle

\section*{Introduction}

Preventing, mitigating, and solving some of the most pressing challenges faced by mankind often requires the design of effective technological and behavioral solutions. But even the most effective solutions would fail to make an impact without their large-scale adoption by the world population.
For some global challenges, such as the COVID-19 pandemic, policymakers around the world have often enforced the required behaviors to counteract the pandemic’s most severe effects in a top-down fashion, e.g., by restricting travel, requiring mandatory mask-wearing in public spaces, and isolating positive cases. 
For other global challenges, even when technological and behavioral solutions to mitigate the challenge are available, policymakers are unwilling to or cannot fully oblige the citizens to adopt them~\cite{constantino2022scaling}. This is the case, for example, for shifting individuals’ transportation, nutrition, and building design choices toward sustainable alternatives, which might substantially reduce greenhouse gas emissions~\cite{creutzig2022demand}. Many sustainable technologies and behaviors are well-known, but triggering their large-scale adoption remains elusive~\cite{constantino2022scaling}.

To generate scientific insights that could inform societal change policies, a tension between two approaches exists. A well-established research tradition -- e.g., in microeconomics, consumer behavior, and psychology -- focuses on modeling the individual-level decision-making process, seeking to capture the drivers of individual-level choices and potentially nudge new behaviors~\cite{train2009discrete,kahneman2011thinking,peterson2021using,elster2015explaining}. These approaches typically capture the complexity of individual decision-making processes, which leads in the best case to a precise image of individual processes. However, because of the feedback mechanisms and consequent nonlinearities that characterize collective behaviors, their conclusions may not be directly extrapolated to the collective level~\cite{granovetter1978threshold,watts2002simple,miller2009complex,helbing2015saving,bouchaud2013crises,siegenfeld2020models,smith2018simulating}.

On the other hand, computational approaches to understanding collective behavior in complex systems typically formulate parsimonious models on individual-level behaviors under simplifying assumptions. 
These approaches -- often rooted in statistical physics and agent-based modeling -- have found applications ranging from materials composed of simple particles to systems composed of more sophisticated units such as flocks~\cite{cavagna2010scale}, swarms~\cite{brambilla2013swarm}, crowds~\cite{helbing2015saving}, and societies~\cite{bouchaud2013crises}.
In the context of social change, computational approaches leverage agent-based simulations to determine the system-level implications of interconnected individual-level choices~\cite{granovetter1978threshold,watts2002simple,miller2009complex,helbing2015saving,bouchaud2013crises,flache2017models,siegenfeld2020models,smith2018simulating,efferson2020promise}.
These models enable a comprehensive understanding of the possible macro-level behaviors of a system, but when not calibrated with empirical individual processes, they may fail to identify the best intervention for a given scenario~\cite{aral2018social}.
Individual-level and computational approaches are mainly developed in parallel. Scholars from various disciplines have argued that the disconnect between our understanding of individual decision-making and social network dynamics limits policymakers' ability to stimulate effective collective behavioral responses to increasingly complex global challenges ~\cite{bak2021stewardship}, and called for data-driven efforts to integrate the two perspectives~\cite{peres2010innovation,flache2017models,smith2018simulating,bak2021stewardship,galesic2021human,galesic2021integrating}.

Here, we demonstrate how the two approaches can be integrated in the context of new product or behavior adoption, and how such an integration can inform network interventions for social change.
The leading framework to model social change is provided by the complex contagion theory~\cite{granovetter1978threshold,watts2002simple,centola2007complex,guilbeault2018complex,centola2018behavior,centola2021change,mariani2024collective}, which posits that when an adoption choice involves a substantial level of personal investment (e.g., effort, monetary cost, personal or reputational risk), it requires a sufficient level of social reinforcement from the decision-maker's social contacts who already adopted the new product or behavior~\cite{centola2007complex,centola2018behavior,guilbeault2018complex,constantino2022scaling}. 
This idea is incorporated in the threshold model of new product diffusion~\cite{watts2002simple,centola2007complex}, where individual $n$ adopts a given product (or behavior) $i$ if and only if at least a threshold fraction $\tau_{ni}$ of her social contacts already adopted the product. According to the theory, threshold-based diffusion models describe accurately how new ideas and behaviors spread in many policy-relevant contexts, including the adoption of healthy behaviors~\cite{christakis2008collective,centola2010spread,myneni2015content,aral2017exercise}, online social platforms~\cite{ugander2012structural,toole2012modeling,karsai2014complex}, new technologies~\cite{oster2012determinants,beaman2021can}, views on political and controversial topics~\cite{romero2011differences,barash2012critical,fink2016complex,notarmuzi2022universality}, social movements and political protests~\cite{granovetter1978threshold,gonzalez2011dynamics,steinert2017spontaneous}, the abandonment of harmful cultural traditions~\cite{efferson2020promise}  -- see ref.~\cite{guilbeault2018complex} for a comprehensive review.
The relevance of the threshold distribution for diffusion outcomes~\cite{granovetter1978threshold,valente1996social,watts2002simple,valente2020diffusion} seems to directly call for measuring individual-level thresholds from empirical data to calibrate threshold-based simulations.
Yet surprisingly, despite early empirical works on historical diffusion data~\cite{valente1996social,valente1998mass}, there is currently
limited evidence on the empirical individual-level threshold levels, and a lack of empirically validated methods to estimate individual thresholds and integrate them into social network interventions~\cite{peres2010innovation,guilbeault2018complex,guilbeault2021topological}.

Here we develop a model-based approach to estimate individual-level thresholds without requiring historical adoption nor social network data, we validate it across two distinct empirical domains, and we demonstrate its usefulness to inform network interventions for social change. 
Pioneering empirical efforts~\cite{valente1995network,valente1996social,valente1998mass} calculated individual thresholds from historical adoption data by relying on simple heuristics such as the percentage of adopters in one's personal network at adoption time.
While intuitive and practical, these heuristics are difficult to refute in practice, as observed in ref.~\cite{valente2010social}.
Significant challenges occur when one seeks to measure a given individual’s threshold for a given product from historical adoption data:
If a given individual never adopted the product, we cannot determine at which higher level of social signal, if any, she would have adopted it; if she did adopt, we cannot infer for which lower level of social signal she would have still adopted, as there is typically a time lag between when an individual's threshold is reached and their actual adoption~\cite{valente1995network,valente1996social}. 
Besides, if the product is new, no historical adoption data exist.

To address these challenges, we draw on utility-based theories of behavioral change~\cite{gavrilets2024modelling}, historically rooted in economics~\cite{durlauf1999can} and consumer behavior~\cite{kohli1991reservation,miller2011should}. These theories 
assume that by adopting a certain new behavior,
individuals can gain or lose utility depending on whether the behavior conforms with others’ behavior or expectations~\cite{gavrilets2024modelling}. Importantly, they provide analytical expressions linking individual thresholds to underlying utility parameters in various scenarios~\cite{kuran1989sparks,goldenberg2010chilling,yang2022sociocultural}. Motivated by these works, we adopt here a simple utility-based formulation that enables convenient estimation from choice data~\cite{goldenberg2010chilling}. This perspective connects the complex contagion literature with the individual-level perspective (see Fig.~\ref{fig:individual}A).

We leverage a simple formula that links individuals’ thresholds to the drivers of individual-level adoption choices, which enables the measurement of individual-level thresholds from discrete-choice experiments without requiring detailed social network data. 
Results from two choice experiments
demonstrate that individual-level thresholds can be estimated from choice data and used to make accurate out-of-sample choice predictions at the individual level. We show how the estimated thresholds can be integrated into policies to promote large-scale behavioral change. To this end, we compare the performance of policies aware of the estimated thresholds against traditional ones based on social network centrality, and we determine under which conditions the threshold-aware policies outperform existing ones.
For example, our findings reveal that the recently-introduced complex centrality~\cite{guilbeault2021topological} consistently outperform state-of-the-art policies only when its calculation takes into account the estimated thresholds.

\addtocontents{toc}{\protect\setcounter{tocdepth}{-1}}
\section*{Results}

\subsection*{From individual-level preferences to thresholds}

\begin{figure*}
    \centering
    \includegraphics[width=\textwidth]{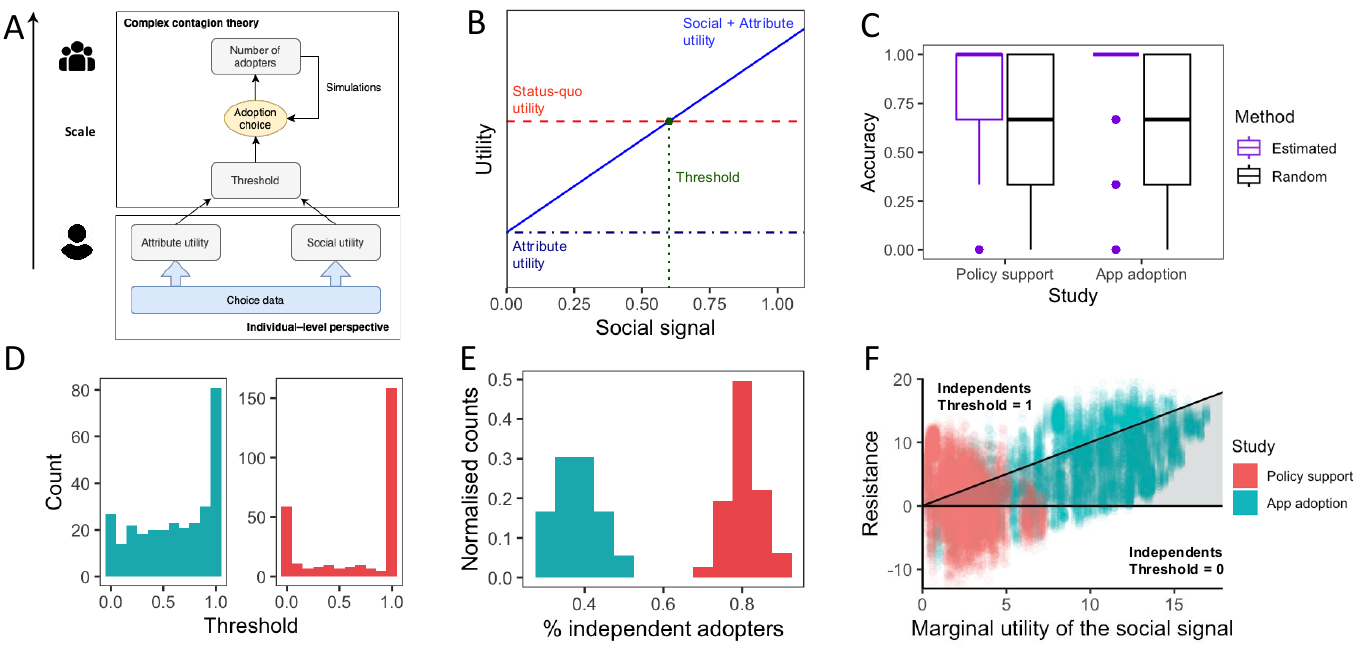}
    \caption{\textbf{Estimating individual–level thresholds.} \textbf{(A)} Threshold-based diffusion models assume that individual–level adoption choices are determined by the decision-makers' threshold. On the other hand, individual-level perspectives focus on estimating individuals' utilities of adopting from choice data. Utility-based approaches to behavioral change reconcile the two perspectives by reinterpreting the individual-level thresholds in terms of individuals' attribute and social utilities~\cite{goldenberg2010chilling,yang2022sociocultural}. \textbf{(B)} For a susceptible adopter, the status-quo utility is initially larger than the utility from adopting. As the number of adopters increases, so does her social utility. The threshold is defined as the minimal level of social signal at which the utility from adopting exceeds the utility from not adopting. \textbf{(C)} For both experiments, the individual-level thresholds estimated from experimental data hold out-of-sample predictive power, as illustrated by their superior accuracy compared to a random-threshold baseline. In these box plots, the central line indicates the median; box limits represent the first and third quartiles (interquartile range, IQR); whiskers go from the edges of the box to the farthest data points that are within $1.5\times\text{IQR}$. Any points beyond that limit are shown as individual outliers. The number of observations used to calculate the accuracy of each method is 4,363 and 4,183 for the PS and AA Study, respectively. \textbf{(D)} In general, different products exhibit different threshold distributions, as illustrated by the two examples provided here (instant messaging app in blue; energy policy in orange). \textbf{(E)} The distribution of the proportion of independent adopters (as opposed to susceptible adopters) is significantly lower for the app adoption experiment (AA, in blue) than for the policy support experiment (PS, in orange), which highlights the importance of context for the distribution of individual thresholds. \textbf{(F)} An individual is susceptible for adopting a given product when her resistance is positive and lower than the marginal utility of  social signal, which corresponds to the gray stripe in the $\gamma-R$ diagram. There are significantly more observations that fall within the gray stripe in the AA experiment than in the PS experiment, which explains the higher percentage of susceptible adopters in the AA experiment. Data in this panel is based on a sample of products, as described in Supplementary Note~\ref{secSI:product_sampling}.
    }
    \label{fig:individual}
\end{figure*}

We assume that when evaluating a certain product (or behavior) $i$, an individual is influenced by both the product's attributes and the received social signal about the product. We assume that individual $n$'s utility from adopting product (or behavior) $i$ consists of two terms: the utility of the product attributes for $n$ $(U^{(A)}_{ni})$ and the ``social utility"~\cite{durlauf1999can} derived from the social signal received by $n$ about $i$ ($U^{(S)}_{ni}$): $U_{ni} = U^{(A)}_{ni} + U^{(S)}_{ni}$. 
Under this formulation, we define the adoption threshold as the minimal level of the social signal for which the utility of $n$ from adopting $i$  exceeds $n$'s status-quo utility, i.e., the utility from not adopting ($U^{(0)}_{n}$, see Fig.~\ref{fig:individual}B). 
In the following, it is convenient to define $n$'s resistance to adopting $i$ as the gap between her status-quo utility and the attribute utility from adopting $i$, i.e., $R_{ni}=U_{n}^{(0)}-U^{(A)}_{ni}$
Assuming that $R_{ni}>0$ and the social utility is linear in the number of adopters ($U^{(S)}_{ni} = \gamma_{n}s_{ni}$, where $s_{ni}$ denotes the percentage of adopters of $i$ within $n$'s social neighborhood, and $\gamma_{n}>0$ represents $n$'s marginal utility of the social signal), the adoption threshold can be expressed as~\cite{goldenberg2010chilling}:
\begin{equation}
\label{eq:th}
    \tau_{ni} = \frac{U_{n}^{(0)}-U^{(A)}_{ni}}{\gamma_{n}} = \frac{R_{ni}}{\gamma_{n}}.
\end{equation}
This result is intuitive: The higher the individual's resistance, or the lower the weight of the social signals on her choices, the higher their threshold.
We assume -- as typical in discrete choice modeling~\cite{train2009discrete} -- that the attribute utility is linear and separable in the attributes: $U^{(A)}_{ni} = \sum_{k=1}^{K} \beta_{nk}\,x_{ki}$, where $x_{ki}$ encodes the extent to which alternative $i$ exhibits attribute $k$ and $\beta_{nk}$ is referred to as the partworth utility of $n$ for attribute $k$. 

An implication of Eq.~\eqref{eq:th} is that, if one is able to estimate individuals' $R$ and $\gamma$ from choice data, one can directly calculate their threshold as well.
Before providing the results of this procedure, we anticipate that, given estimates $\hat{R}_{ni}$ and $\hat{\gamma}_n$ of individual resistance and marginal utility from social signal, Eq.~\eqref{eq:th} can lead to multiple scenarios.
We refer to individuals with $0<\hat{R}_{ni}<\hat{\gamma}_n$ as \textit{susceptible} adopters: For these individuals, Eq.~\eqref{eq:th} delivers $\hat{\tau}_{ni}\in(0,1)$, which implies that their adoption choices for product $i$ are influenced by different levels of social signals $s_{ni}\in(0,1)$.
For individuals with $\hat{R}_{ni}<0$, Eq.~\ref{eq:th} delivers $\hat{\tau}_{ni}<0$; 
Using the terminology of the innovation diffusion theory~\cite{ryan1950acceptance,coleman1957diffusion,lionberger1960adoption,rogers2010diffusion},
these individuals 
act as innovators with respect to their personal network~\cite{valente1996social}, as they adopt as soon as they become aware of product $i$ regardless of the social signal's strength; for practical purposes, their estimated threshold can be set to zero. Note that they may not necessarily be innovators with respect to the whole social system, as they may still adopt late if they are not exposed early on to the product~\cite{valente1996social}.
For individuals with $\hat{R}_{ni}>\hat{\gamma}_n$, Eq.~\ref{eq:th} delivers $\hat{\tau}_{ni}>1$; these individuals will never adopt product $i$ regardless of the social signal's strength; for practical purposes, their estimated threshold can be set to one. We refer to individuals with $\hat{R}_{ni}<0$ or $\hat{R}_{ni}>\hat{\gamma}_n$ as \textit{independent} adopters, as their adoption choices are independent of the level of social signal. 

This interpretation of the threshold has two important implications. First, we can unlock the connection between individual-level utilities and threshold-based diffusion processes, thereby connecting macro-level patterns of adoption with their micro-level drivers (see Figs.~\ref{fig:individual}A--B). 
Second, given empirical choice data, we can estimate the individual-level thresholds from choice data through well-established utility estimation algorithms~\cite{allenby2006hierarchical}, which we demonstrate next.

\subsection*{Experimental validation} 

To demonstrate the detectability of individual-level thresholds from choice data, we rely on choice-based conjoint experiments \cite{rao2014applied} (see Supplementary Note \ref{secSI:intro_conjoint} for an overview). 
In this class of experiments, widespread in consumer behavior research, participants are exposed to a set of choices among hypothetical products (represented by combinations of attributes) and a status quo option (selecting none of the presented choices).
In the designed experiments, detailed below, participants are exposed to product alternatives characterized additionally by varying levels of social signal.
By analyzing the resulting choice data through a well-established Hierarchical Bayes (HB) algorithm~\cite{allenby2006hierarchical}, we recover the relative weight of each attribute in determining each participant's choices (i.e., the parameters $\beta$ above), the marginal utility the social signal ($\gamma$), and the status quo utility ($U^{(0)}$), which enables the calculation of individual thresholds through Eq.~\eqref{eq:th} (see Methods for the details).

We run two choice-based conjoint experiments, which cover two choice contexts where the complex contagion theory is expected to apply~\cite{guilbeault2018complex}: political views (energy policy support experiment, hereafter PS experiment), and the adoption of a new technology under network externalities (instant messaging app adoption, hereafter AA experiment).
In the PS experiment, participants are presented with a set of energy policies regarding carbon-capturing technologies, and in each choice task, they are asked to select which policy they would support (if any). 
The policies differ in terms of 5 attributes (policy instrument; costs; beginning of policy implementation; required distance to residential areas; policy endorsers)~\cite{pianta2021carbon} and in their level of social signal (presented as the percentage of peers supporting the policy). The policies' attributes -- except for the social signal -- are identical to those in ref.~\cite{pianta2021carbon}. 
In the AA experiment, participants are presented with a set of messaging apps and are asked to select which app they would consider installing (if any). 
The apps differ in terms of 4 attributes (accessibility; authentication method; customization level; support for video calls) and their level of social signal (percentage of friends already using the app). 
For both experiments, we collect choice data from 
participants recruited through Prolific, an online platform commonly used for sourcing high-quality respondents for research studies~\cite{douglas2023data}. We recruit $N_R=296$ participants in the PS experiment and $N_R=300$ in the AA experiment, and we estimate the individual-level thresholds through the HB algorithm and Eq.~\eqref{eq:th}. We refer to Methods and Supplementary Note~\ref{secSI:conjoint_exp_description} for details about the two studies and the parameter estimation.

Before attempting to understand the properties of the estimated thresholds, it is essential to validate the estimation method. If the estimated thresholds are accurate, we expect that they hold out-of-sample predictive power for individual choices in hold-out tasks that have not been used for the estimation. 
In particular, for an accurate set of estimated thresholds, we expect that the alternatives selected by the participants in hold-out tasks are predicted by the gap between the alternative's social signal and the individual's threshold for that alternative (see Methods). 
By translating this remark into an accuracy function (see Eq.~\eqref{accuracy} in Methods), we 
can compare the accuracy of choice predictions based on the estimated thresholds against those based on randomly-extracted thresholds from a uniform distribution. We refer to Methods for the validation procedure details.

We find that for both the PS and AA experiments, the accuracy of the estimated thresholds on hold-out tasks is significantly larger than that of randomly-extracted thresholds (see Fig.~\ref{fig:individual}C; 
one-tailed Wilcoxon signed-rank tests,both p-values smaller than $p = 0.001$; effect sizes $r = 0.535$ and $r=0.494$ for the PS and AA experiments, respectively).
This result indicates that despite the parsimonious assumptions needed to obtain the simple expression in Eq.~\eqref{eq:th}, the estimated thresholds hold significant out-of-sample predictive power, which points to their ability to predict individual choices and motivates a detailed analysis of their properties.
Future studies could use the utility-based framework to develop and validate more sophisticated threshold estimation methods.

\subsection*{Detection of independent and susceptible individuals} 

The debate around the shape of the threshold distribution has long stagnated, partly because of the inaccessibility of individual thresholds. The estimated threshold can, therefore, shed light on both the distribution of individual thresholds and its implications for social influence maximization policies.
Visual inspections of the threshold distributions for different products reveal that the threshold distribution is strongly product-dependent (see Fig.~\ref{fig:individual}D), which is often neglected in simulation-based studies of social influence maximization. For example, in Fig.~\ref{fig:individual}D, the illustrative energy policy exhibits a strongly polarized, bimodal distribution, with most participants being either supportive of the policy (peak at $0$) or unsupportive (peak at $1$), regardless of the social signal. 
By contrast, the instant messaging app represented in the same panel shows a more uniform threshold distribution, with a sizeable number of individuals lying in $(0,1)$ and, therefore, susceptible to different levels of social signal. 

Moving from these case studies to the general properties of the threshold distributions across the two contexts, we quantify the proportion of independent adopters in the two experiments, i.e., the proportion of estimated thresholds equal to zero or one, which denotes a lack of susceptibility to social influence. 
We find that the proportion of independent adopters is significantly larger in the PS experiment than in the AA experiment (Fig.~\ref{fig:individual}E). 
This result can be explained in terms of the estimated parameters (Fig.~\ref{fig:individual}F). 
Assuming that $\hat{\gamma}_n>0$, an adopter is susceptible if and only if $0<\hat{R}_{ni}<\hat{\gamma}_n$. This condition corresponds to a well-defined ``susceptibility stripe" in the two-dimensional $\gamma-R$ diagram (gray area in Fig.~\ref{fig:individual}F) where each individual-product pair is represented by a dot in the plane spanned by the marginal utility of social signal and the resistance.
Outside of the susceptibility stripe, the adopter's choice for the product is independent of the social signal. 
Significantly more individual-product pairs lie within the susceptibility stripe in the AA experiment ($61\%$) than in the PS experiment ($20\%$, see gray area in Fig.~\ref{fig:individual}F; 
two-sample proportion Z-test, $p < 0.001$), which explains the lower proportion of independent adopters in the AA experiment (Fig.~\ref{fig:individual}E).

In sum, the estimated thresholds reveal that in different choice contexts, the number of individuals who choose independently of social signals may differ substantially, which is often neglected in extant social infuence maximization studies. 
The high proportion of independent adopters in the PS experiments qualitatively agrees with previous research~\cite{rinscheid2021shapes} that found that social norm messages communicating the policies’ support level may be ineffective to stimulate 
carbon--capturing policy support.
The two analyzed contexts suggest that the number of independent adopters is smaller in contexts where network externalities influence the adoption more strongly. As the main objective of the two experiments is to showcase the feasibility of the threshold estimation method and validate its out-of-sample predictive power, 
we leave to future research to determine whether and how 
 the proportion of susceptible adopters is affected by the framing of the social signal (e.g., using a different description of the reference group~\cite{bicchieri2016norms}) and other contextual factors.

\subsection*{Implications for seeding policies}

Do the estimated thresholds challenge what we know about social influence maximization? 
We consider here the seeding problem where a social change practitioner~\cite{constantino2022scaling} (e.g., a government or an organization) seeks to determine which nodes in a social network to incentivize first to promote the large-scale adoption of a new product or behavior (hereafter, product)~\cite{valente1999accelerating,banerjee2013diffusion,valente2020diffusion}. This problem represents a specific type of network intervention, which refers to purposeful actions taken by a change practitioner to leverage the structure of social networks in achieving a predefined objective~\cite{valente2012network,robins2023multilevel}.
\textit{Structure-based seeding policies} are agnostic to a potential target's behavioral characteristics, and they select a set of seeds according to the nodes' centrality in the relevant social network~\cite{valente1999accelerating,banerjee2013diffusion} or to structural properties of the set~\cite{borgatti2006identifying,morone2015influence,lu2016vital,ott2018strategic,muller2019effect}. On the other hand, \textit{behavior-based policies} (i.e., policies that take into account individual behavioral characteristics 
such as the individual thresholds) assume that some individuals are more amenable to change and, therefore, willing to initiate the adoption process for a new product or behavior~\cite{valente1996social,valente1998mass,catalini2017early,efferson2020promise}. 
The underlying hypothesis is that as low-threshold individuals are willing to adopt a new idea earlier than their peers, they should be engaged early on to create sufficient early momentum and form a critical mass for the desired behavioral change~\cite{valente1996social,valente2012network}.
A different way to leverage individual thresholds to initiate change is
the recently-introduced complex centrality~\cite{guilbeault2021topological}, which is a hybrid method that factors in both network structure and individual thresholds. But the input thresholds for its calculation are either assumed to be known or extracted from arbitrary distributions, which can lead to suboptimal decisions as shown below.
In sum, the estimated thresholds allow us to address the open questions:
Under which theoretical conditions can behavior-based (or hybrid) policies outperform structure-based ones?

\begin{figure*}
    \centering
    \includegraphics[width=\textwidth]{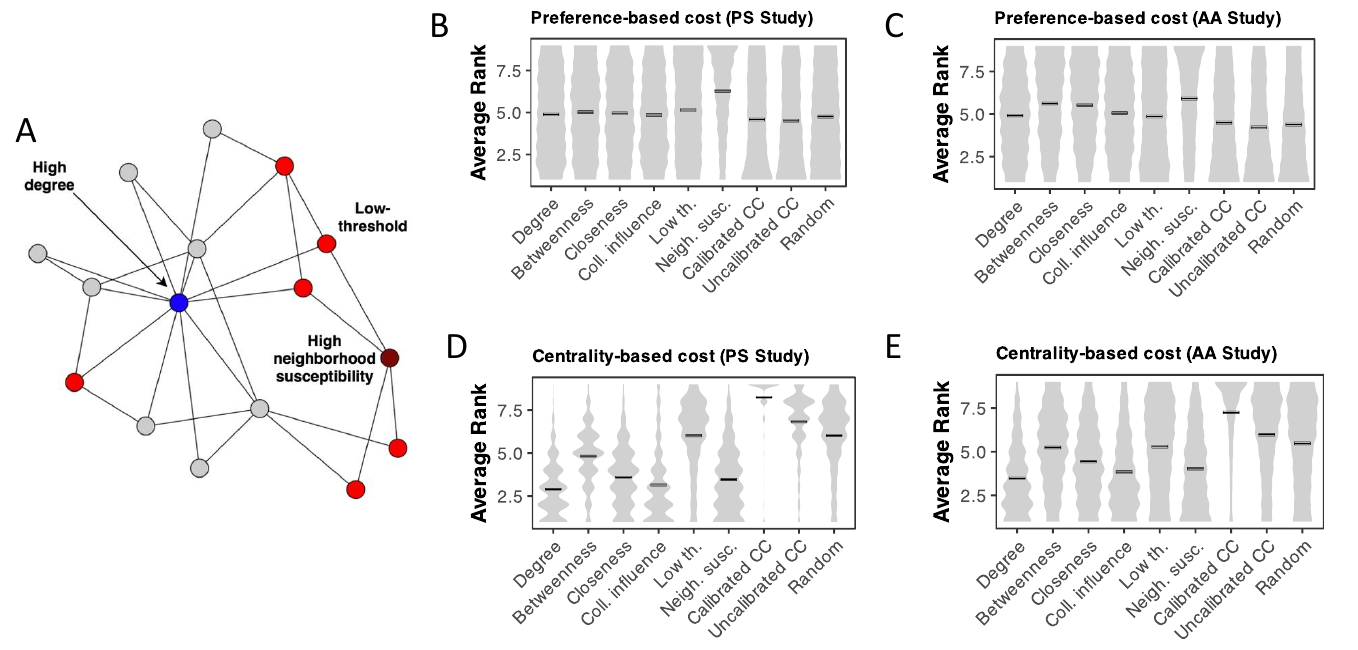}
    \caption{\textbf{Relative performance of seeding policies.} 
    \textbf{(A)} Nodes selected by different policies in an illustrative network structure: The highest-degree node (in blue) has the largest number of connections, but the highest-neighborhood susceptibility node (in dark red) has the largest number of connections to low-threshold nodes (in red).
    \textbf{(B, C)} Relative performance of seeding policies under a preference–based cost structure, measured through the average rank defined in the main text (higher values indicate better performance), for the policy support experiment and the app adoption experiment, respectively. The neighborhood susceptibility policy based on the estimated thresholds significantly outperforms the other policies. \textbf{(D, E)} Relative performance of seeding policies under a centrality–based cost structure for the policy support experiment and the app adoption experiment, respectively. The complex centrality policy based on the estimated thresholds significantly outperforms the other policies. In panels \textbf{B--E}, crossbars represent the mean together with the $95\%$ confidence intervals over $3,240$ runs (see Methods).
    }
    \label{fig:seeding}
\end{figure*}

To address this question, we rely on agent-based spreading simulations based on the threshold model of new product diffusion~\cite{watts2002simple}. 
In the simulations, given a network composed of $N$ nodes, we assign at random $N$ agents to the network's nodes. 
We analyze the largest components of a subset of the Add Health networks considered by Guilbeault and Centola~\cite{guilbeault2021topological}, which include several independent communities that exhibit high variation in network topological properties (see Supplementary Note~\ref{secSI:network_sampling}).
We study the independent diffusion of the 36 products analyzed in the AA experiment and a sample of 36 products from those analyzed in the PS experiment (see Supplementary Note~\ref{secSI:product_sampling} for the sampling details).
For each product, we assume that the agents make adoption choices based on their \textit{ground-truth thresholds}, which we set to be equal to those estimated from the survey's participants (see Methods). 
While the ground-truth thresholds used for agents' adoption decisions are known to the agents, they are not known to the social change practitioner. 
We assume that the practitioner has been able to survey the nodes of the network with a choice-based conjoint survey and use the collected choice data to estimate the individual thresholds.  
For our theoretical analysis, as detailed in Supplementary Note~\ref{secSI:synthetic_choice_data}, we simulate the agents' answers to the survey based on the parameters estimated from the two experiments, and we use the generated choice data to estimate the individual thresholds (\textit{estimated thresholds}, hereafter) with the same method used for the experimental data (see Methods and Supplementary Note~\ref{secSI:synthetic_choice_data}). 
In Supplementary Note~\ref{secSI:partial}, we discuss the robustness of the results in scenarios where only a fraction of the agents can be surveyed. 

Under these assumptions, the optimal seeds of behavior-based policies are determined by the agents' estimated thresholds. 
The threshold estimation enables the design of various behavior-based policies: low-threshold policy (where the focal seed is the lowest threshold node); high neighborhood susceptibility policy (where the focal seed is the node with the largest number of low-threshold neighbors); high calibrated complex centrality policy (where the focal seed is the node with the largest complex centrality~\cite{guilbeault2021topological} calculated using the estimated thresholds as inputs).
We use various structure-based policies as baselines, which we label according to the metric used to select the focal seed: degree~\cite{goldenberg2009role}, collective influence~\cite{morone2015influence}, betweenness~\cite{freeman1977set}, closeness~\cite{bavelas1950communication}, and uncalibrated complex centrality (i.e., the nodes' average complex centrality score over an ensemble of homogeneous threshold values~\cite{guilbeault2021topological} not estimated from actual choice data, see Supplementary Note~\ref{secSI:seeding_policies}).
We also consider a policy where the focal node is selected at random~\cite{efferson2020promise}. We refer to Supplementary Note~\ref{secSI:seeding_policies} for all the policies' implementation. We measure each policy’s performance as number of adopters discounted for the intervention cost, rank them within each configuration (defined by the empirical network being analyzed, the product, and the realization of the simulated conjoint survey and node assignment), and average these ranks across configurations to yield the mean policy rank (see Methods).


We find that while there is no universally best-performing policy, the estimated thresholds are key to identifying the best-performing seeds. 
The relative performance of the policies critically depends on the structure of the cost function, but in all the analyzed scenarios, the best-performing policy is one that takes into account the estimated thresholds. 
Specifically, we first consider a cost function such that the cost to successfully initiate the spreading only depends on the seed nodes' preferences, but it is independent of the nodes' centrality (preference-based cost; see Methods). 
This scenario is especially relevant to contexts where the seeds mostly adopt for personal motivations, and there is no reason why it would require more effort or monetary incentive to persuade central individuals to early adopt~\cite{peres2010innovation}. In these contexts, high-resistance individuals might require stronger interventions to be persuaded to initiate the spreading (e.g., more intervention cycles in a field intervention, or higher monetary incentives in marketing contexts; see Methods). 
We find that in the preference-based cost scenario, the neighborhood susceptibility policy significantly outperforms all the others (see Figs.~\ref{fig:seeding}B--C). 
Differences in performance between the neighborhood susceptibility and all other policies are statistically significant ($p < 0.001$ for all one-tailed Wilcoxon signed-rank tests; the smallest effect size is $r=0.07$ in the AA Study in the comparison against the betweenness centrality). 
Through the estimated thresholds, the neighborhood susceptibility allows the identification of social neighborhoods that are more amenable to change and require a lower cost to successfully initiate the diffusion.
Among the remaining policies, two structure-based policies (betweenness and closeness) perform relatively well yet they underperform with respect to the neighborhood susceptibility. 
In this cost scenario, both the calibrated and the uncalibrated complex centrality generally underperform the other policies.

The relative performance of the policies changes dramatically as we consider a different cost function such that the cost to target a node grows with the node’s centrality, independently of the node's preferences (see Methods).
This scenario is highly relevant to all those applications where there is a strategic advantage of initiating the diffusion from high-centrality nodes, which is often the case in online influencer marketing and other interventions where central nodes are highly sought-after because of their large follower bases and, therefore, are less responsive~\cite{lanz2019climb} and charge higher prices for endorsing new products~\cite{watts2007influentials,bakshy2011everyone,leung2022influencer}. 
We find that in the centrality–dependent cost scenario, the calibrated complex centrality significantly outperforms all the other policies [see Figs.~\ref{fig:seeding}D--E, 
one-tailed Wilcoxon signed-rank tests comparing the performance of calibrated complex centrality to each alternative are statistically significant in all cases ($p < 0.001$), the smallest effect size is $r=0.37$ in the AA Study in the comparison with the uncalibrated complex centrality].
This is due to its capability to identify low-degree nodes with high spreading capacity for threshold-based diffusion~\cite{guilbeault2021topological}. At the same time, the uncalibrated complex centrality -- which neglects the estimated thresholds -- fails to reach a similar performance and, for a smaller seed set size, even to outperform random seeding (Figs.~\ref{fig:smallz1}-\ref{fig:smallz2}).
Therefore, the complex centrality is consistently effective at identifying optimal seeds, but only if estimates of the individual–level thresholds are available to the social change practitioner.
Structure-based policies other than the uncalibrated complex centrality typically exhibit a lower performance than the random seeding policy. 
This confirms the argument that when the higher connectedness of high-degree nodes affects the seeding cost, the social hubs' spreading ability might not be large enough to justify the extra efforts required to detect and target them~\cite{watts2007influentials}.
In this scenario, only the calibrated complex centrality leads to consistently higher seeding performance.

\section*{Discussion}

To effectively steward collective behaviors in social systems, scholars from diverse disciplines~\cite{peres2010innovation,smith2018simulating,schweitzer2018sociophysics,bak2021stewardship,galesic2021human,galesic2021integrating} suggested the need for integrating individual-level behavioral models with simulations of collective behavior, but data-driven efforts in this direction remain rare. Our work contributes to such integration by (1) demonstrating that, building on discrete choice theory, one can estimate individual-level thresholds for behavioral change, which we validate across two different contexts of varying degrees of social influence; (2) leveraging the estimated thresholds to inform seeding policies aimed at initiating social change.

At the individual level, our experimental results indicate that once the thresholds are estimated, individual choices in out-of-sample tasks exhibit significant predictability. 
This finding not only provides a validation of the estimated thresholds, but might also inspire future studies aimed at improving the predictive power.
Besides, the estimated thresholds allowed us to disentangle an experimental choice context where 
most individuals choose independently of social signals from one where most individuals are susceptible to social influence.

Future empirical works could leverage the proposed method for more refined tests of social contagion theories. In our work, all participants were exposed to the percentage of friends who selected each alternative, which necessarily leaves out the variety of social cues encountered in field applications~\cite{borgatti2009network}. This suggests two natural directions for generalizing our findings, related to the characteristics of the decision maker and the received social cues, respectively. At the decision-maker level, future studies could examine how the size of an individual's personal network and their individual role within a community affect their thresholds.
 As for the social cues, future studies could examine whether different types of social cues -- e.g., coming from more similar individuals or individuals with different roles in the community -- affect individual responses. Addressing these factors in the field might require more sophisticated conjoint designs potentially using vignettes~\cite{bicchieri2016norms}. 
 We envision that these efforts could systematically identify which influence mechanisms and contexts cause social signals to play (or not play) a critical role in shaping individual choices, and what the best policy response could be.

In this work, to illustrate how the thresholds could be estimated from empirical data, we collected data using a conjoint design due to its simplicity for eliciting stated choices that mimic real-world decisions. 
For real-world interventions informed by our work, practitioners may face two potential barriers. It is beyond the scope of this work to fully address them, but we offer a few comments on how future research could overcome them. 

The first concern is whether hypothetical decisions in the conjoint survey 
can predict responses to social signals in a real-world intervention. 
Prior evidence shows that, at least in some contexts, insights from hypothetical conjoint studies match remarkably well those from behavioral data~\cite{hainmueller2015validating}.
In general, potential biases induced by the hypothetical nature of conjoint experiments can be alleviated by various means -- e.g., by using vignettes with fictitious characters to reduce social desirability bias~\cite{bicchieri2016norms}, or through incentive-aligned designs~\cite{miller2011should} where responses are consequential and, therefore, participants have a direct incentive to reveal their true preferences. 
Broader concerns about the external validity of laboratory experiments can be mitigated by using observational data -- if available -- instead of or as a complement to experimental data. 
In this respect, the individual-level choice framework adopted here is not limited to data collected from conjoint experiments, as it can in principle be adapted to any setting where data on stated choices is available.

The second concern is feasibility, as conjoint experiments might be too costly to conduct at scale. 
Of course, the feasibility of collecting the necessary data depends on the size of the target population of the social change intervention. 
In many cases (e.g., rural communities~\cite{alexander2022algorithms,banerjee2013diffusion} and school classes~\cite{valente2007peer,ehlert2020human,paluck2016changing}), the target population is small enough that it is feasible to survey it entirely (or to survey a sizeable portion of it). 
When the population is too large to be surveyed, one can survey small samples to estimate the thresholds at the group level, or rely on observational data which are already available in some contexts, e.g., for promoting sustainable behaviors~\cite{wef2020antforest} or informing public-health interventions~\cite{oliver2020mobile}. 
In scenarios where rich observational data are available at the individual level, one could adapt techniques to estimate peer effects from observational data~\cite{ma2015latent,tran2022heterogeneous} to the threshold measurement.
Adapting the utility-based framework to the specifics of different data sources is a promising avenue for further research.

Moving from the individual to the collective level, our findings reveal the circumstances under which seeding policies that incorporate behavioral models can outperform traditional, structure-based policies. 
For example, the recently-introduced complex centrality~\cite{guilbeault2021topological} identifies effective seeds only when the change practitioner has been able to collect individual threshold data, and the cost of successfully targeting a node grows with her network centrality.
The simpler neighborhood susceptibility policy -- relying on the estimated number of susceptible neighbors a given node has -- only outperforms the other methods when the cost of successfully targeting a node depends on their intrinsic preferences. At the collective level, estimating the threshold distribution through the proposed methodology and comparing the resulting predictions against real-world network diffusion processes is an important direction for future research.

The boundaries to the effectiveness of the utility-based framework adopted here deserve as well to be further investigated in future research.
While we assumed a threshold-based diffusion model, aligned with the main paradigm in the social contagion literature~\cite{valente1996social,centola2007complex,guilbeault2018complex}, we show in Supplementary Note~\ref{secSI:linking} that a utility-based framework under different assumptions that those adopted above still recovers interpretable social spreading models -- e.g., the traditional Bass diffusion~\cite{bass1969new} and logistic diffusion models~\cite{maccoun2012burden} -- which could be investigated in future empirical research. Further, we assumed that the timescale that governs the diffusion of the new product or behavior is significantly faster than the timescale that governs changes in individual-level preferences (i.e., thresholds) and network structure. 
This assumption is in line with state-of-the-art social influence maximization works~\cite{centola2007complex,aral2018social,guilbeault2018complex,efferson2020promise}.
In real complex social systems, however, stronger feedback loops may exist:
collective diffusion outcomes, top-down interventions, and interactions between multiple diffusion processes can alter individual-level preferences. In some cases, preference changes could occur at similar timescales as diffusion processes~\cite{moore2022determinants}. 
Adapting the utility-based framework to these more complex scenarios and overall, determining the theoretical limits to the proposed approach remain fascinating challenges for future studies.

\clearpage

\section*{Methods}

\subsubsection*{Threshold estimation}

We describe here the threshold estimation method, based on Eq.~\eqref{eq:th}.
Consider an individual $n$ who needs to select a product from $I$ alternatives. 
Each alternative $i \in\{1,\dots,I\}$ brings individual $n$ a certain utility $U_{ni}$. In line with discrete choice theory~\cite{train2009discrete}, we assume that individual $n$ chooses alternative $i$ if and only if $U_{ni} > U_{nj}$, $\forall j \neq i$. 
The utility can be decomposed into an observable, systematic component ($U^{*}_{ni}$) and a random, unobserved component ($\epsilon_{ni}$) such that $U_{ni} = U^{*}_{ni} + \epsilon_{ni}$. 
The probability that individual $n$ chooses alternative $i$ over $j$ is $\text{Prob}(U_{ni} > U_{nj}) = \text{Prob}(\epsilon_{nj} - \epsilon_{ni} < U^{*}_{ni} - U^{*}_{nj})$. 
As in main text, we assume -- as typical in discrete choice modeling~\cite{train2009discrete} -- that the attribute utility is linear and separable in the attributes. Following the notation in main text,
the systematic utility of individual $n$ from adopting product $i$ is expressed as: 
\begin{equation}
U^{*}_{ni} = \sum_{k=1}^{K} \beta_{nk} x_{ki} + \gamma_n s_{ni}
\label{eq:utility_representative_composition}
\end{equation}
Under the assumption that $\epsilon_{ni}$ is i.i.d. Gumbel distributed, the model parameters ($\beta_{nk}$, $\gamma_{n}$) can be estimated from observed choices (e.g., in a choice-based conjoint experiment~\cite{rao2014applied}) using a Hierarchical Bayes (HB) algorithm~\cite{allenby2006hierarchical}. 
In this model, the probability of individual \textit{n} to select an item \textit{i} from a choice set is expressed as a multinomial logit whose parameters are estimated for each individual using Markov Chain Monte Carlo (MCMC) methods.
In this work, we used the R package \url{ChoiceModelR} \cite{ChoiceModelR} for the HB estimation. 
See Supplementary Note~\ref{secSI:conjoint_exp_description} for more details on the estimation method used in the two experiments.  
The estimation yields individual-level estimates and thus allows to calculate the adoption threshold for every individual and every product.

Inevitably, the estimation results are to some extent model-dependent, as they reflect the assumptions embedded in the choice model specification and the HB estimation. 
An additive and separable model rules out interaction effects, which can be important in some real-world applications.
Although interaction models can still yield a simple analytic threshold (see Supplementary Note~\ref{secSI:complex_contagion_theory}), estimating models with a large number of interaction terms can pose substantial challenges due to model complexity~\cite{green1990conjoint}.
In applications where interactions are essential, this concern may be partly alleviated by restricting interactions to a small set of theory-motivated effects~\cite{rao2014applied}, and by adopting conjoint designs and estimation techniques tailored to identify them~\cite{chrzan2000overview,rao2014applied}.
More generally, utility-based contagion models can be extended in several directions, e.g., by including negative externalities~\cite{lopez2008social} or anticonformist agents~\cite{mittal2024anticonformists} to allow for downward thresholds (an agent abandons a given alternative when a threshold fraction of her social contacts already adopted it) and even multi-threshold responses.

For the HB estimation, the modeling assumptions are embedded in the Bayesian priors, whose specification can influence the resulting estimates.
This dependence diminishes as the number of choice tasks per participant increases, since a broad range of priors will result in similar inferences in large datasets~\cite{rossi2024bayesian}. 
However, when the number of choice tasks per individual is small relative to the number of parameters to be estimated, the choice of priors might have an impact.
Supplementary Notes \ref{secSI:sensitivity_prior_individual}--\ref{secSI:sensitivity_prior_collective} show that our main results are robust across four model specifications that differ in the assumptions on the prior distributions made in the HB model.
With the exception of the extreme case of an uninformative prior, the choice of the prior distribution has a marginal impact on the accuracy of the partworth utility estimates (Fig.~\ref{figSI:hb_priors_ind}A,D).
Further, except for the uninformative prior, the accuracy of adoption threshold estimates and their correlation with the true thresholds remain high and consistent across prior specifications (Fig.~\ref{figSI:hb_priors_ind}B--C;E--F). 
As a result, our conclusions on the relative performance of behavior-based seeding strategies is unaffected by prior choice (Figs.~\ref{figSI:hb_priors_coll_ps}--\ref{figSI:hb_priors_coll_aa}). 
Therefore, while the estimated adoption thresholds should not be interpreted as model-free estimates, our sensitivity analysis shows they remain robust across specifications and can reliably inform the design of social-change interventions.

\subsubsection*{Validation of estimated thresholds}

We aim to assess the predictive power of the individual thresholds estimated from choice-based conjoint data. The core idea is to estimate the thresholds based on a training set, and quantify how many times, in the validation set, the participants make predictable choices, i.e., choices that do not contradict their threshold values. 
For each task in the survey, we fit the HB model (see previous paragraph) on a training set consisting of all tasks except one hold-out task, and use the fitted model to predict the choices in the validation set consisting of the hold-out task. 
We then calculate the average number of predictable choices over all the training-validation splits. 
Specifically, for a given choice task, we quantify the estimated appeal of product alternative $i$ to participant $n$ as $\hat{\Delta}_{ni}:=s_{i}-\hat{\tau}_{ni}$, where $s_{i}$ denotes the social signal displayed for alternative $i$ and $\hat{\tau}_{ni}$ denotes the estimated threshold (based on the training set) of participant $n$ for alternative $i$. 
The appeal of the default alternative (nonadoption) is zero, as by definition, any product alternative would be more appealing than the default if the respective social signal would exceed the corresponding estimated threshold.

Formally, given that participant $n$ selected alternative $i^*$ in task $\mathcal{Z}_{z}$ ($z\in \{1,\dots, Z\}$ with $Z$ being the number of choice tasks) we say that compared to a different alternative $i\in \mathcal{Z}_{z}$, $i^*$ is a predictable choice based on $n$'s estimated thresholds if $\hat{\Delta}_{ni^*}>\hat{\Delta}_{ni}$, where $\hat{\Delta}_{ni}$ denotes the estimated appeal of alternative $i$ to participant $n$.
As each task $\mathcal{Z}_{z}$ has $I=4$ alternatives (three products and one default option), we can measure the degree of predictability of $n$'s choice for task $\mathcal{Z}_{z}$ as
\begin{equation}
a_{nz}=\frac{1}{I-1}\sum_{i \in \mathcal{Z}_{{z}} \setminus \{i^*\}}\Theta(\hat{\Delta}_{ni^*}-\hat{\Delta}_{ni})
\end{equation}
where the sum runs over the $I-1$ alternatives in task $\mathcal{Z}_{z}$ that were not selected by $n$; $\Theta(x)=1$ if $x>0$, $\Theta(x)=0$ otherwise. 
Basically, an observed choice is perfectly predicted by the $n$'s estimated thresholds if none of its alternatives was more appealing according to the thresholds. 
Pooling the predictability levels over all participants and over all training-validation splits (and considering that there is only one hold-out task in each validation set), we define the accuracy of the estimated thresholds as 
\begin{equation}
   A=\frac{1}{NZ}\sum_{n=1}^{N}\sum_{z=1}^{Z}a_{nz}
   \label{accuracy}
\end{equation}
where $N$ denotes the number of participants and $Z$ denotes the number of choice tasks in the experiment (here equal to the number of training-validation splits). 
Note that the accuracy function can be connected with the choice probabilities of a logit choice model in the limit of strong marginal utility of social signal: under the logit model (see Supplementary Note~\ref{secSI:complex_contagion_theory}), in the limit $|\gamma|\gg 1$, one has indeed $P_{ni^*}/P_{ni}\simeq \Theta(\hat{\Delta}_{ni^*}-\hat{\Delta}_{ni})$.

\subsubsection*{Data-driven spreading simulations}
We consider threshold-based spreading simulations where the agents' thresholds are set based on the experimental results.
For the simulation, we set the the ground-truth threshold of a given agent for a given product to the corresponding value estimated in the choice experiment for a randomly-selected participant (sampled with or without replacement conditional on whether the number of nodes occupied by the agents, $N$, is larger or smaller, respectively, than the number of participants, $N_R$).
We additionally assume that the hypothetical decision maker has been able to survey the nodes of the network with a choice-based conjoint experiment. 
We simulate the answers of the agents in the conjoint experiment based on the individual-level utilities estimated in the choice experiments. 
We then proceed with the proposed threshold estimation procedure to estimate the thresholds (see Supplementary Note~\ref{secSI:synthetic_choice_data}).
We denote by $\tau_{ni}$ and $\hat{\tau}_{ni}$ the ground-truth and estimated threshold, respectively, of node $n$ for product $i$.
We simulate diffusion processes that unfold according to the same fractional threshold model studied by Ref.~\cite{guilbeault2021topological} using the ground-truth thresholds $\{\tau_{i\alpha}\}$.

We run the diffusion processes throughout the largest connected components of 18 empirical Add Health networks sampled from those analyzed by Guilbeault and Centola~\cite{guilbeault2021topological} (see Supplementary Note~\ref{secSI:network_sampling} for the sampling procedure).
For each network structure, we measure the nodes' centralities according to the degree, betweenness, closeness, and collective influence (see Supplementary Note~\ref{secSI:seeding_policies_network}). 
Besides, for each network structure, product, realization of the simulated conjoint survey and agent assignment, we measure the behavior-based metrics defined in Supplementary Note~\ref{secSI:seeding_policies}, using the estimated thresholds $\{\hat{\tau}_{i\alpha}\}$ as input.
Motivated by the effectiveness of clustered seeding for effective threshold-based diffusion~\cite{guilbeault2021topological}, for each policy, we adopt a clustered seeding protocol where we target one focal seed selected by the metric and a subset of their closest neighbors such that $\lfloor z\,N \rfloor$ of the nodes are initially active (if the focal seed node has not enough neighbors to reach this proportion of seeds, a recursive search on the higher-order neighbors is performed~\cite{guilbeault2021topological}; we also set a minimum seed set size of four nodes). In main text, we set $z=0.025$; we show in Supplementary Note~\ref{secSI:different_seed_size} that our conclusions are mostly robust with respect to other small values of $z$.

In the preference-based cost scenario, the cost of successfully seeding 
product $i$ with the selected seed set, $\mathcal{S}$, is defined as 
\begin{equation}
C_i(\mathcal{S})=c_0+\sum_{n\in\mathcal{S}:R_{ni}>0}R_{ni},
\label{eq:preference_cost}
\end{equation}
where $R_{ni}=U_{n}^{(0)}-U^{(A)}_{ni}$ and $c_0=1$ is a fixed implementation cost per intervention. This cost function embodies the idea that beyond a fixed implementation cost, $c_0$, it is more costly to successfully target individuals with higher resistance to the product, which can be motivated through a simple choice-theory argument (see Supplementary Note~\ref{secSI:complex_contagion_theory}).
In the centrality-based cost scenario, the seeding cost is defined as 
\begin{equation}
C_i(\mathcal{S})=c_0+\sum_{n\in\mathcal{S}}d_{n},
\label{eq:centrality_cost}
\end{equation}
i.e., the cost of successfully seeding a node $n$ grows with her degree $d_n$~\cite{bakshy2011everyone,lanz2019climb,leung2022influencer}.
The performance of the seeding policy that targets seed set $\mathcal{S}$, $P_i(\mathcal{S})$, is assumed to be the final number of adopters, $A_i(\mathcal{S})$, discounted for the intervention's cost: $P_i(\mathcal{S})=A_i(\mathcal{S})/C_i(\mathcal{S})$.

To assess the relative performance of the examined seeding policies, for each analyzed configuration $\mathcal{C}$, we rank the nine policies by their simulated performance. 
Here a configuration $\mathcal{C}$ is identified by the network being analyzed, the product that spreads through the network, the particular realization of the simulated conjoint survey performed by the change practitioner to estimate the thresholds, and the agents' random assignment to the network's nodes. For a given seeding policy, considering $18$ empirical networks, $36$ products, and $5$ realizations of the simulated conjoint survey and random node assignment for the simulated participants (see Supplementary Notes~\ref{secSI:data}--\ref{secSI:survey_simulation}), we run $3,240$ independent simulations.
Therefore, for each configuration $\mathcal{C}$, 
each method $m$ is assigned to a rank variable $r_{m}(\mathcal{C})$, with higher values indicating better performance (in case of ties, we assign the average of the ranks that would have been assigned without ties). We compare the methods' average rank across all configurations,
\begin{equation}
    \overline{r}_m=\frac{1}{|\mathcal{C}|}\sum_{\mathcal{C}} r_{m}(\mathcal{C}),
\end{equation}
as done in Friedman's statistical test~\cite{friedman1937use}. We also consider an alternative metric where for each configuration $\mathcal{C}$, each method $m$ is assigned to a performance variable that quantifies the method's performance compared to the best-performing method~\cite{zhou2019fast}, $s_{m}(\mathcal{C})=P_m(\mathcal{C})/\max_{m'}\{P_{m'}(\mathcal{C})\}$. Again, we average this variable over all configurations.
For the best-performing policies in the preference-based cost scenario (i.e., the neighborhood susceptibility) and the centrality-based cost scenario (i.e., the calibrated complex centrality), we perform pairwise one-tailed Wilcoxon tests~\cite{guilbeault2021topological} between the ranks of the best-performing policy and those all the other policies, to ensure that the differences are statistically significant ($p< 0.05$).

\subsubsection*{IRB authorization}
The research complies with all relevant ethical regulations and was approved by the University of Zurich review board (OEC IRB 2020-081 from 03.11.2020). 
Before the experiments, all participants provided electronic informed consent. 
Participants in the PS and AA studies (respectively: $N = 296$ and $N=300$; $49\%$ and $54\%$ female; modal age groups 55–64 and 35–44 years) were compensated £1.80 and £1.65, respectively.

\clearpage

\onecolumngrid 

\bigskip
\noindent
{\bf \large Data availability}

The data collected in this study 
are publicly available at \url{https://doi.org/10.5281/zenodo.17841193}. 

\vspace{0.5cm}
\noindent
{\bf \large Code availability}

The code used to obtain the findings of this study 
is publicly available at \url{https://doi.org/10.5281/zenodo.17841193}. 

\vspace{0.5cm}

\noindent
{\bf \large Acknowledgements}

RT and MSM acknowledges financial support from the URPP Social Networks of the University of Zurich; MSM acknowledges financial support from the Swiss National Science Foundation (Grants No. 100013--207888, 100013-236802).
The funders had no role in study design, data collection and analysis, decision to publish or preparation of the manuscript.
We wish to thank Mingmin Feng for her detailed feedback on the code and the manuscript, Fei Wang for his feedback on the choice modeling approach, Luca Lazzaro and Tulasi Agnihotram for their help in conducting the conjoint studies and for their feedback provided in the process. 

\vspace{0.5cm}

\noindent
{\bf \large Author contributions statement}

All authors conceived and designed the research; RA and RT conceived the use of conjoint analysis for the threshold estimation; RT implemented the choice experiments and performed the analysis of experimental data; MSM and RT performed the simulation analysis;
all the authors analyzed data; MSM and RT wrote the first draft of the paper; all authors revised the paper.

\vspace{0.5cm}
\noindent
{\bf \large Competing interests statement} 

The authors declare no competing interests. 

\vspace{0.5cm}
\noindent


\onecolumngrid 
\vspace{0.5cm}
\noindent
{\bf \large Figure legends}
\vspace{0.5cm}

{\bf Fig. 1.} \textbf{Estimating individual–level thresholds.} \textbf{(A)} Threshold-based diffusion models assume that individual–level adoption choices are determined by the decision-makers' threshold. On the other hand, individual-level perspectives focus on estimating individuals' utilities of adopting from choice data. Utility-based approaches to behavioral change reconcile the two perspectives by reinterpreting the individual-level thresholds in terms of individuals' attribute and social utilities~\cite{goldenberg2010chilling,yang2022sociocultural}. \textbf{(B)} For a susceptible adopter, the status-quo utility is initially larger than the utility from adopting. As the number of adopters increases, so does her social utility. The threshold is defined as the minimal level of social signal at which the utility from adopting exceeds the utility from not adopting. \textbf{(C)} For both experiments, the individual-level thresholds estimated from experimental data hold out-of-sample predictive power, as illustrated by their superior accuracy compared to a random-threshold baseline. In these box plots, the central line indicates the median; box limits represent the first and third quartiles (interquartile range, IQR); whiskers go from the edges of the box to the farthest data points that are within $1.5\times\text{IQR}$. Any points beyond that limit are shown as individual outliers. The number of observations used to calculate the accuracy of each method is 4,363 and 4,183 for the PS and AA Study, respectively. \textbf{(D)} In general, different products exhibit different threshold distributions, as illustrated by the two examples provided here (instant messaging app in blue; energy policy in orange). \textbf{(E)} The distribution of the proportion of independent adopters (as opposed to susceptible adopters) is significantly lower for the app adoption experiment (AA, in blue) than for the policy support experiment (PS, in orange), which highlights the importance of context for the distribution of individual thresholds. \textbf{(F)} An individual is susceptible for adopting a given product when her resistance is positive and lower than the marginal utility of  social signal, which corresponds to the gray stripe in the $\gamma-R$ diagram. There are significantly more observations that fall within the gray stripe in the AA experiment than in the PS experiment, which explains the higher percentage of susceptible adopters in the AA experiment. Data in this panel is based on a sample of products, as described in Supplementary Note~\ref{secSI:product_sampling}.

{\bf Fig. 2.} \textbf{Relative performance of seeding policies.} 
    \textbf{(A)} Nodes selected by different policies in an illustrative network structure: The highest-degree node (in blue) has the largest number of connections, but the highest-neighborhood susceptibility node (in dark red) has the largest number of connections to low-threshold nodes (in red).
    \textbf{(B, C)} Relative performance of seeding policies under a preference–based cost structure, measured through the average rank defined in the main text (higher values indicate better performance), for the policy support experiment and the app adoption experiment, respectively. The neighborhood susceptibility policy based on the estimated thresholds significantly outperforms the other policies. \textbf{(D, E)} Relative performance of seeding policies under a centrality–based cost structure for the policy support experiment and the app adoption experiment, respectively. The complex centrality policy based on the estimated thresholds significantly outperforms the other policies. In panels \textbf{B--E}, crossbars represent the mean together with the $95\%$ confidence intervals over $3,240$ runs (see Methods).

\clearpage

\noindent
{\bf \large References}
\bibliography{tp}

\clearpage

\onecolumngrid

\centerline{ \bf \large Supplementary Material for:}

\centerline{\bf Integrating behavioral experimental findings into dynamical models to inform social change interventions}

\medskip

\centerline{Radu Tanase, René Algesheimer, Manuel Sebastian Mariani}

\tableofcontents

\setcounter{section}{0}
\renewcommand{\thesection}{S\arabic{section}}

\setcounter{figure}{0}
\renewcommand{\thefigure}{S\arabic{figure}}

\setcounter{table}{0}
\renewcommand{\thetable}{S\arabic{table}}

\clearpage

\addtocontents{toc}{\protect\setcounter{tocdepth}{2}}

\section{Choice-based conjoint experiments}
\label{secSI:conjoint_exp_description}

\subsection{Brief introduction to choice-based conjoint experiments}
\label{secSI:intro_conjoint}

Conjoint experiments are a widely-used technique \cite{rao2014applied} to measure how the total evaluation of a product or behaviour (further referred as product) is related to the aspects describing the product (further referred as product attributes).
Choice-based conjoint experiments (CBC) are a specific type of conjoint experiments where the product evaluation is done in terms of a discrete choice. 
Specifically, participants in CBC experiments are faced with a number of choice tasks. 
In each choice task, participants are presented a number of products (i.e, choice alternatives) which differ along several dimensions (i.e., product attributes) and are asked which one they would choose (if any) in a given (often hypothetical) situation. 
From observing such choices, it is then possible to infer which product attributes individuals value most in making the decision. 
This builds on the premise that when making the choice, participants trade-off different aspects of the choice alternatives. 

The theoretical foundation for CBC lies in the random utility theory \cite{mcfadden1974conditional}. 
In such models, an individual is faced with a number of choice alternatives in a choice set, each bringing her a certain utility. 
The individual will choose the alternative with the highest utility. 
While the utility of each alternative is known to the individual, it is not known to the researcher. 
Therefore, from the researcher's point of view, an individual's (random) utility can be decomposed into a deterministic part and a random component \cite{train2009discrete}. 
The deterministic part is a function (often a linear combination) of the utility corresponding to each product attribute (i.e., partworth utilities or simply attribute utilities).
The random component is unknown and requires some assumptions about its distribution. 
These assumptions determine which model will be used to describe the probability that a product is chosen. 
The most used model is the multinomial logit, which assumes the random component follows an extreme value distribution \cite{rao2014applied}. 

The Hierarchical Bayes (HB) model~\cite{allenby2006hierarchical} (used in this article) is a variation of the multinomial logit that yields individual-level estimates of the attribute utilities (i.e., one estimate for each individual and each level of any product attribute).
Based on the individual level estimates, it is then possible to calculate the total utility of a product for an individual. 
The HB model assumes that the individual partworth utilities 
follow the same distribution for all the individuals in the sample, 
with the distribution's parameters drawn from a population distribution.
To estimate the partworth utilities, it first assumes an initial value of the parameters describing the distribution (i.e., prior belief) and then updates it based on data in a Bayesian manner. As the resulting equations cannot be solved analytically, the estimation is done using Markov Chain Monte Carlo (MCMC) methods \cite{rao2014applied, allenby2006hierarchical}. 
We refer the interested reader to ref.~\cite{rao2014applied} for an applied introduction into conjoint analysis and to ref.~\cite{train2009discrete} for a comprehensive read which includes extensive details into how various models are formulated and estimated. 

\subsection{The policy support study (PS)}
\label{secSI:conjoint_exp_description_PS}

\paragraph*{Study design and data.}
We recruited 296 US participants (300 participants in total, but 4 participants have been excluded for not completing the survey) from a demographically (i.e., gender and ethnicity) representative US sample on Prolific to take part in a choice-based conjoint survey on support for carbon capturing policies. 
No statistical methods were used to pre-determine the sample size but our sample size is similar to those reported in previous publications~(e.g., ref.~\cite{miller2011should}). 
Participants were told to imagine a carbon capture and storage policy might be implemented in their state and they need to select which alternative from a choice set they would endorse. 
Each choice set contained three policies and a none option. 
The policies presented were described by five attributes (policy type, policy costs, beginning of the policy implementation, required distance to residential areas and organisation endorsing the policy) with four levels each, closely following \citet{pianta2021carbon}.
Additionally, we included a sixth attribute representing the (hypothetical) percentage of one's friends who endorse the policy. 
Participants were presented with a total of 15 choice tasks, each participant receiving a different set of choice tasks (generated by the Sawtooth software using the balanced overlap method \cite{SawtoothBalancedOverlap}). 
An overview of the attributes and levels used in the conjoint design can be seen in Table \ref{tableSI:energy_study_design}.
A sample of the conjoint survey administered to the participants is attached to this Supplementary Material.

\begin{table}[]
\centering
\resizebox{\textwidth}{!}{%
\begin{tabular}{lccccc}
\hline
Attributes                                                              & \multicolumn{5}{c}{Levels}                                                                                                                                             \\ \hline
\multicolumn{1}{l|}{Policy type}                                        & \multicolumn{1}{c|}{Ban}                      & \multicolumn{1}{c|}{Subsidies}  & \multicolumn{1}{c|}{Tax}              & \multicolumn{1}{c|}{}                 &      \\
\multicolumn{1}{l|}{Policy costs}                                       & \multicolumn{1}{c|}{\$4}                      & \multicolumn{1}{c|}{\$9}        & \multicolumn{1}{c|}{\$14}             & \multicolumn{1}{c|}{\$19}             &      \\
\multicolumn{1}{l|}{Begining of policy implementation}                  & \multicolumn{1}{c|}{2025}                     & \multicolumn{1}{c|}{2035}       & \multicolumn{1}{c|}{2045}             & \multicolumn{1}{c|}{2055}             &      \\
\multicolumn{1}{l|}{Required distance to residential areas}             & \multicolumn{1}{c|}{2 miles}                  & \multicolumn{1}{c|}{5 miles}    & \multicolumn{1}{c|}{10 miles}         & \multicolumn{1}{c|}{50 miles}         &      \\
\multicolumn{1}{l|}{Policy endorsement by}                              & \multicolumn{1}{c|}{Carbon Capture Coalition} & \multicolumn{1}{c|}{Greenpeace} & \multicolumn{1}{c|}{Democratic Party} & \multicolumn{1}{c|}{Republican Party} &      \\
\multicolumn{1}{l|}{\% of your friends who endorse the policy scenario} & \multicolumn{1}{c|}{1\%}                      & \multicolumn{1}{c|}{23\%}       & \multicolumn{1}{c|}{45\%}             & \multicolumn{1}{c|}{76\%}             & 98\% \\ \hline
\end{tabular}%
}
\caption{\textbf{Attributes and levels used in the Carbon Capturing Policy Study.}}
\label{tableSI:energy_study_design}
\end{table}

\paragraph*{Estimation.}
We estimated the individual partworth coefficients using the Hierarchical Bayes \cite{allenby2006hierarchical} algorithm. 
We used the function \url{choicemodelr} from the R package \url{ChoiceModelR} \cite{ChoiceModelR} with 30'000 MCMC iterations (10'000 for burn in and the remaining 20'000 for parameter estimation, from each every tenth draw was retained). 
We considered the attribute \textit{Percentage of friends who endorse the policy} as numeric and the remaining attributes (including the none option) as categorical, for which we used effects coding. 
Note that in effects coding all levels of a variable represent deviations from the overall mean (as opposed to a specific reference level when using dummy coding) with the constraint that all levels sum to zero.
For the none option, the effects coding implies we have one level for choosing the status quo (\url{none = True}) and one level for choosing any of the alternatives in the choice set (\url{none = False}). 
We note that in computing the utility of an alternative in the choice set, the value of \url{none = False} needs to be added to (the sum of) the utilities of the attributes describing the alternative. Thus individual $n$’s utility from adopting product (or behavior) $i$ is $U_{ni} = U^{(A)}_{ni} + U^{(S)}_{ni} + U_n^{(\Bar{0})}$, where $U_n^{(\Bar{0})}$ is the utility corresponding to \url{none = False}. 
As this is just an artifact of the coding system used in estimation, we referred everywhere in the text to $U_{ni}$ as $U_{ni} = U^{(A)}_{ni} + U^{(S)}_{ni}$, which correspond to a dummy coding of the none option.  
The HB algorithm does not directly result in point estimates (single values of a parameter) but rather in a distribution of values for each parameter (one value being generated in each MCMC draw used for parameter estimation). 
Therefore, we calculated the parameter estimates as the sample mean of the distribution of the MCMC draws. 
We input the resulting estimated parameters ($\hat{\beta_{n}}, \hat{\gamma_n}, \hat{U}^{(0)}_n$) into Eq~1 in main text to obtain the estimated threshold for product $i$.

\paragraph*{Model quality.}
First, we measured the reliability of the conjoint data by including two fixed choice tasks  (on positions 5 and 10) that were identical within respondents, and measured how many respondents gave the exact same answer to both tasks~\cite{miller2011should}. 
We found the answers to the two questions were the same for $77 \%$ of the respondents, which is substantially larger than the expected fraction ($25\%$) in a baseline model where individuals make decisions at random.

Second, we evaluated how well the model predicts choices out of sample. 
To this end, 
we split the data into a held-out test set consisting of one choice task and a training set consisting of the remaining choice tasks\footnote{To avoid data leakage (i.e., information from the test set leaking into the training set), when the held-out test set included one of the two fixed choice tasks, we removed the other from the training data. 
As there are two identical fixed tasks, removing one of them is essential to avoid that the same task is present in both the training and test sets.}. 
We fitted the model only on the training set, predicted the choices in the test set, and compared the predictions against the observed choices. 
We repeated this process for all choice tasks. 
We found an average out of sample accuracy rate of $68\%$, which is significantly larger than the accuracy rate of a baseline model making choices at random ($25\%$, as each tasks consists of 3 policy alternatives + a none option).

\begin{figure}[h]
\centering
\includegraphics[scale=0.9]{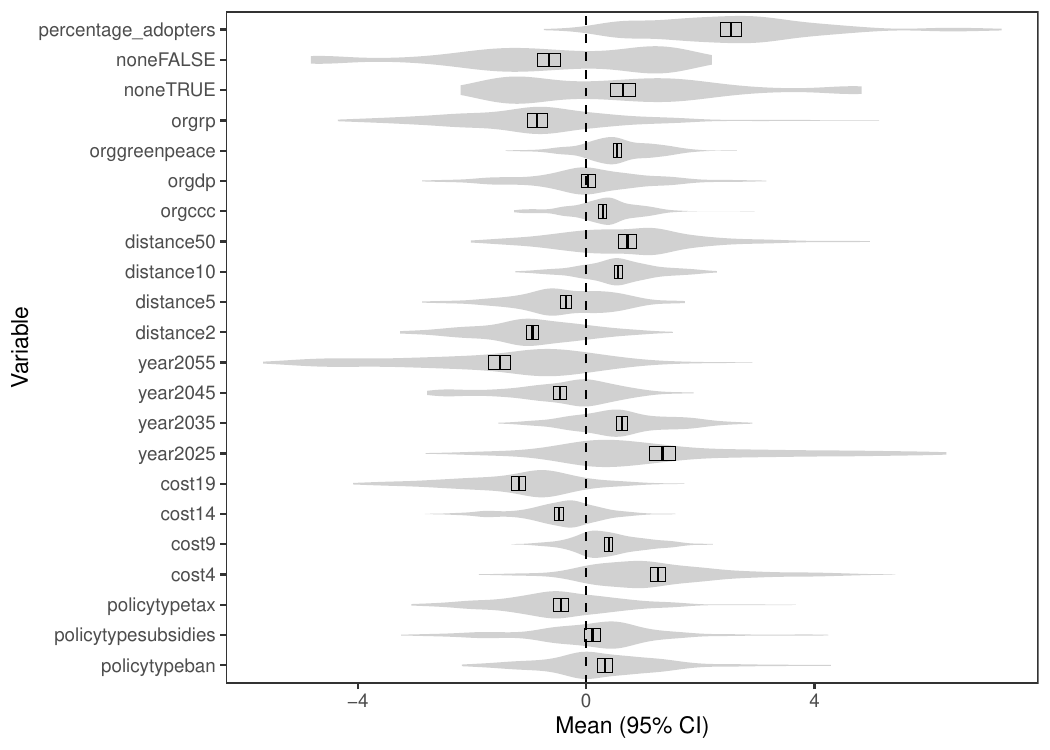}
\caption{\textbf{Average utilities across respondents in the PS Study.} The y axis contains all attributes and levels. The x axis shows the distribution of the marginal utility of each attribute level together with the mean and 95 $\%$ confidence interval (crossbars), based on $N=296$ observations. On average, the percentage of friends endorsing the policy (labeled as \textit{percentage\_adopters}) has a positive partworth utility.}
\label{figSI:energy_study_utilities}
\end{figure}

\begin{figure}[h]
\centering
\includegraphics[width=\textwidth]{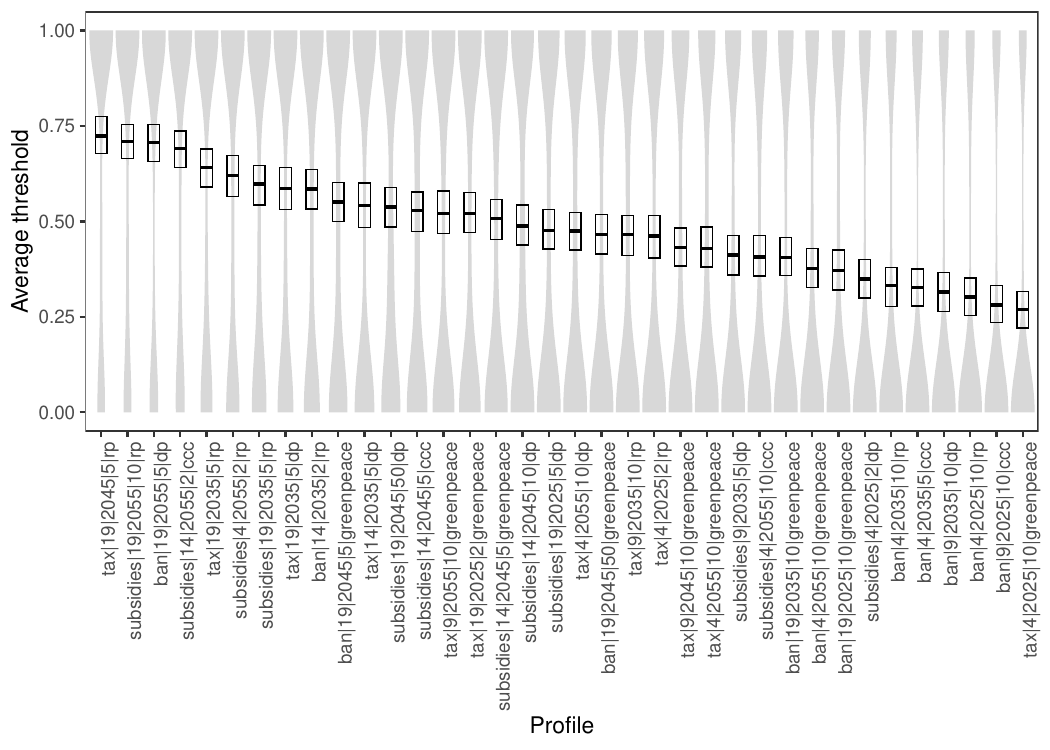}
\caption{\textbf{Threshold distribution across individuals (y axis) for each policy (x axis) in the PS Study}. Crossbars represent the mean together with the 95$\%$ confidence interval based on $N=291$ observations. The policies are sorted by the mean threshold. We observe that some policies have a lower mean threshold, which implies that they are more attractive to the respondents.}
\label{figSI:energy_study_avg_threshold}
\end{figure}

\paragraph*{Results.}
We asses the impact of each attribute on the respondents' choices by measuring the attributes' relative importance. 
To measure the attribute importance for an individual, we measure for each attribute $a$, the difference between the maximum and minimum marginal utility across all levels of $a$. 
We normalize the obtained attribute importance values so that they sum to one for each individual. 
To obtain the results in Table \ref{tableSI:energy_study:avg_variable_importance}, we subsequently average the obtained individual attribute importance values over all individuals \cite{SawtoothImportanceScore}. 
We found the attribute with the highest relative importance is the beginning of policy implementation, followed by policy costs, required distance to residential areas, organisation endorsing the policy, percentage of friends endorsing the policy, and policy type (see Table \ref{tableSI:energy_study:avg_variable_importance}). 
On average, cheaper, stringent (ban) policies that start soon and are further away from the residential area are more appealing than expensive, lax policies that start late and close to the residential area (see Fig. \ref{figSI:energy_study_utilities}). 
The ranking of the attribute level utilities within the same attribute are largely consistent with the study we draw from \cite{pianta2021carbon}, the only difference being in the ranking of organisations endorsing the policy. 
We observe that 291 out of the 296 valid respondents have a positive partworth utility of the social signal (with the remaining 5 having a negative value). 
We note that an interpretation of the individual-level threshold is only compatible with the social contagion literature if the partworth utility of the social signal is not negative, which is true for almost all respondents. 
We remove from our analysis the 5 respondents with a negative partworth utility of the social signal, and leave the study of these special cases to further research. 
We measure the individual-level thresholds using the methodology described in the Methods section of the article for the 36 sampled products (see Supplementary Note~\ref{secSI:product_sampling}) and calculate the average threshold per product. 
The results can be seen in Figure \ref{figSI:energy_study_avg_threshold}.

\paragraph*{Linearity of the social signal.}

\begin{figure}[h]
\centering
\includegraphics[scale=0.9]{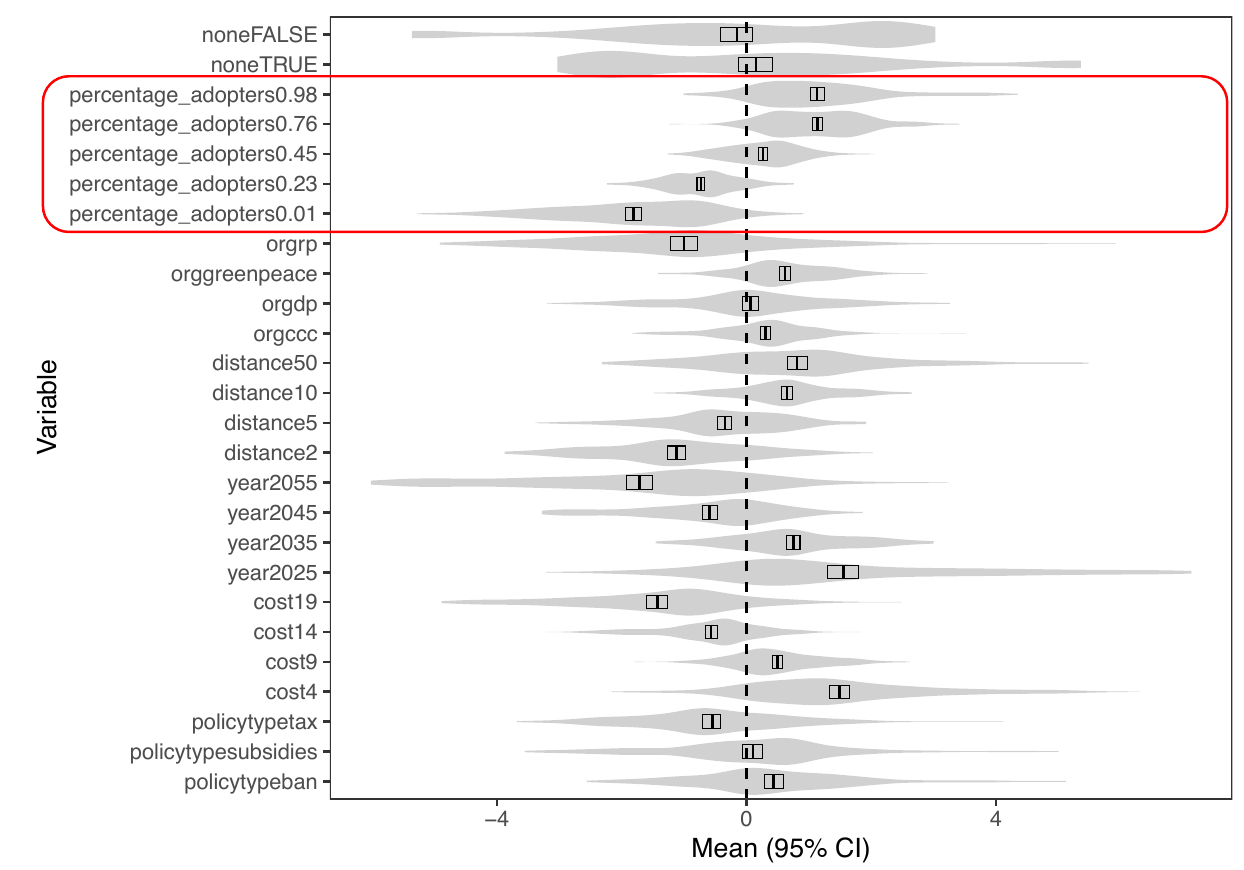}
\caption{\textbf{Average utilities across respondents treating the percentage of adopters as categorical variable (PS Study).} The y axis contains all attributes and levels. The x axis shows the distribution of the marginal utility of each attribute level together with the mean and 95 $\%$ confidence interval (crossbars) based on $N=296$ observations. On average, the utility of the percentage of adopters increases across the first three levels and plateaus at the highest level, where utilities for the last two levels are comparable (red rectangle).}
\label{figSI:energy_study_avg_utilities_si_factor}
\end{figure}

To test the plausibility of assuming a linear coefficient of the social signal, we re-estimate the model treating the \textit{percentage of friends who endorse the policy scenario} as categorical. Fig.~\ref{figSI:energy_study_avg_utilities_si_factor} shows that the average utility of the social signal increases with the social signal for all levels but the last one ($98\%$), which exhibits a similar utility to the previous value ($76\%$) (Pearson’s $r(1478) = 0.77$, $p < 0.001$, $95\%$ CI=$[0.75, 0.79]$).
Give the high positive correlation and that the effect is largely monotonic, for simplicity, we decided to use a linear coefficient in main text. 
Note that in principle, as long as the effect of social signal on the adoption probability is monotonously increasing, the threshold estimation is still feasible even in scenarios where the linearity assumption is not supported by the data -- e.g., by adapting Eq.~(1) in main text to a general nonlinear functional form for the social utility\footnote{For example, if $U^{(S)}_{ni}=\phi_{\gamma_n}(s_{ni})$, where $\phi_{\gamma_n}(x)$ is a monotonically increasing function in $[0,1]$ characterized by parameter $\gamma_n$, one readily obtains $\tau_{ni}=\phi_{\gamma_n}^{-1}(R_{ni})$.} or by treating the social signal as categorical and adapting the interpolation procedure in ref.~\citep{miller2011should} to accommodate a nonparametric representation of the social utility.


\begin{table}[]
\begin{tabular}{lll}
\hline
{\textbf{Variable}}                  & {\textbf{Average Importances}} & {\textbf{Standard Deviation}} \\ \hline
{Beginning of policy implementation}& {0.2084293}& {0.14176090}\\
{Policy costs}                       & {0.1743773}& {0.11072475}\\
{Required distance to residential areas}& {0.1620187}& {0.09583969}\\
Policy endorsement by& {0.1606038}& {0.08421775}\\
\% of your friends who endorse the policy scenario& {0.1495264}& {0.08918514}\\
Policy type& {0.1450446}& {0.09645266}\end{tabular}
\caption{\textbf{Average variable importance in the Policy Support Study}. The most important is beginning of policy implementation, followed by policy costs, required distance to residential areas, organisations who endorses the policy scenario, percentage of friends who endorse the policy scenario and policy type.}
\label{tableSI:energy_study:avg_variable_importance}
\end{table}

\clearpage

\subsection{The app adoption study (AA)}
\label{secSI:conjoint_exp_description_AA}
\paragraph*{Study design and data.}
In this study, we surveyed respondents about their use of a new instant messaging applications. 
The setup was identical to the PS study. 
We recruited 300 US participants from Prolific (54$\%$ female). 
The (fictive) messaging apps were described by four attributes (accessibility, authentication, customisation level and video call support) with two to three levels each. 
Similarly to the PS Study, we included a fifth attribute representing the (hypothetical) percentage of one's friends who are already using the app. 
Participants were presented with a total of 14 choice tasks, each participant receiving a different set of choice tasks (generated by the Sawtooth software using the balanced overlap method \cite{SawtoothBalancedOverlap}). 
 An overview of the attributes and levels used in the conjoint design can be seen in Table \ref{tableSI:energy_study_design}. 
A sample of the conjoint survey administered to the participants is attached to this Supplementary Material.

\begin{table}[]
\begin{tabular}{lccccc}
\hline
Attribute                                                     & \multicolumn{5}{c}{Level}                                                                                                                  \\ \hline
\multicolumn{1}{l|}{Accessibility}                            & \multicolumn{1}{c|}{Mobile}       & \multicolumn{1}{c|}{Web}        & \multicolumn{1}{c|}{}             & \multicolumn{1}{c|}{}     &      \\
\multicolumn{1}{l|}{Authentication}                           & \multicolumn{1}{c|}{Simple}       & \multicolumn{1}{c|}{Two-factor} & \multicolumn{1}{c|}{Multi-factor} & \multicolumn{1}{c|}{}     &      \\
\multicolumn{1}{l|}{Customisation level}                      & \multicolumn{1}{c|}{Low}          & \multicolumn{1}{c|}{Medium}     & \multicolumn{1}{c|}{High}         & \multicolumn{1}{c|}{}     &      \\
\multicolumn{1}{l|}{Video calls}                              & \multicolumn{1}{c|}{Multi-person} & \multicolumn{1}{c|}{One-to-one} & \multicolumn{1}{c|}{}             & \multicolumn{1}{c|}{}     &      \\
\multicolumn{1}{l|}{\% of your friends already using the app} & \multicolumn{1}{c|}{1\%}          & \multicolumn{1}{c|}{23\%}       & \multicolumn{1}{c|}{45\%}         & \multicolumn{1}{c|}{76\%} & 98\% \\ \hline
\end{tabular}
\caption{\textbf{Attributes and levels used in the App Adoption Study.}}
\label{tableSI:app_study_design}
\end{table}

\paragraph*{Estimation.}
We used again Hierarchical Bayes with 30'000 MCMC iterations, the first 10'000 for burn in and the remaining 20'000 for parameter estimation (from each every tenth draw was retained). 
We considered the attribute \textit{Percentage of friends already using the app} as numeric and the remaining attributes as categorical, for which we used effects coding. 
We calculated the parameter estimates as the sample mean of the distribution generated by the MCMC draws. 

\paragraph*{Model quality.}
To evaluate the model quality we performed the out of sample prediction validity test described in the PS Study. 
We found an average prediction accuracy of $77\%$ which is substantially larger than that of a baseline model making choices at random ($25\%$, as before). 

\begin{figure}[h]
\centering
\includegraphics[scale=0.9]{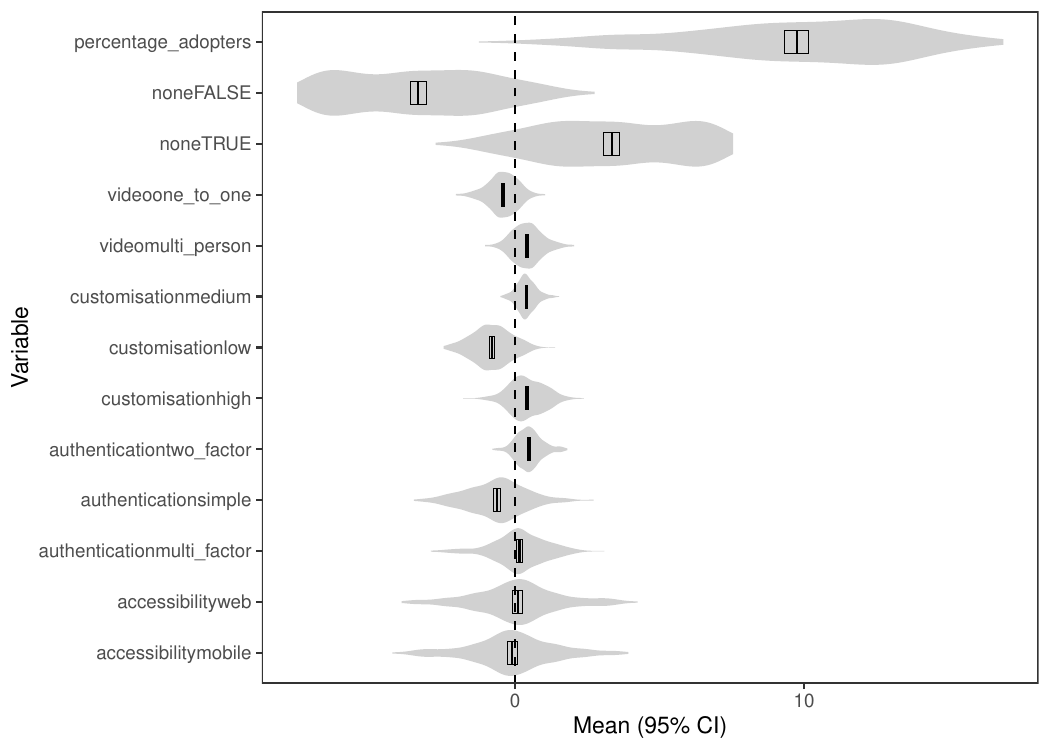}
\caption{\textbf{Average utilities across respondents in the AA Study.} The y axis contains all attributes and levels. The x axis shows the distribution of the marginal utility of each attribute level together with the mean and 95 $\%$ confidence interval (crossbars) based on $N=300$ observations. On average, the percentage of adopters has a large, positive utility.}
\label{figSI:app_study_average_utilities}
\end{figure}

\begin{figure}[h]
\centering
\includegraphics[width=\textwidth]{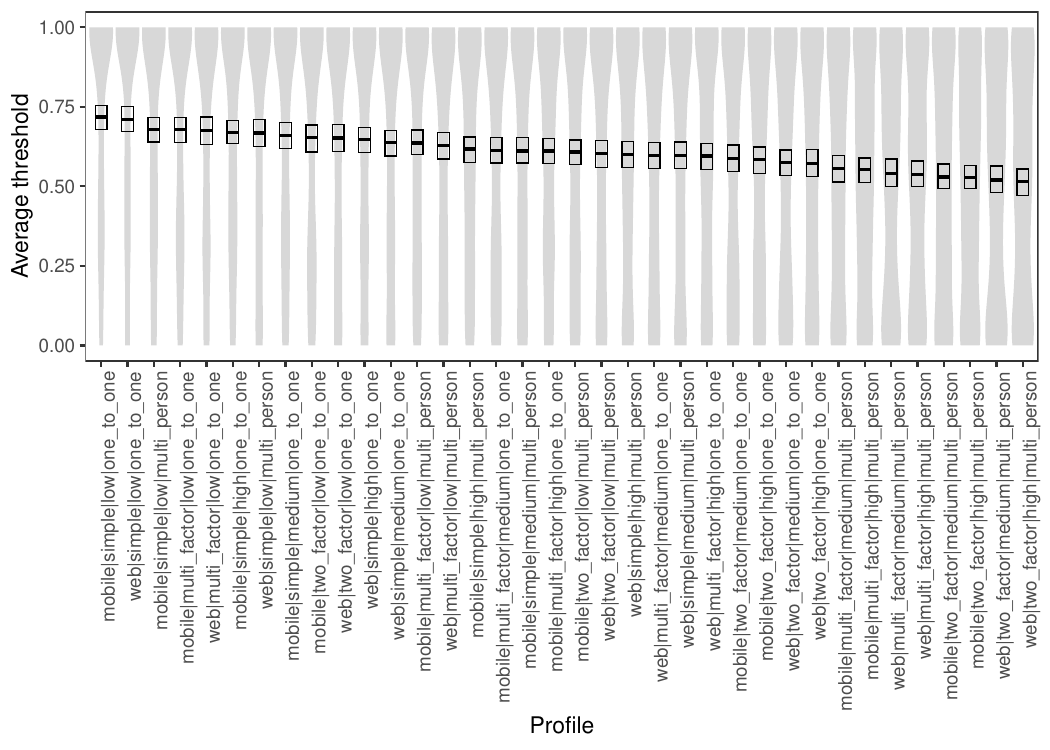}
\caption{\textbf{Threshold distribution across individuals (y axis) for each product (x axis) in the AA Study}. Crossbars represent the mean together with the 95$\%$ confidence interval based on $N=299$ observations. The products are sorted by the mean threshold. We can observe some products have a lower mean threshold, implying they are more attractive. However, compared to the PS Study, the difference between the most and the least attractive product is considerably smaller.}
\label{figSI:app_study_avg_threshold}
\end{figure}

\paragraph*{Results.}
Overall, compared to the PS study, we find higher variation in average importance between the attributes. 
The highest ranked attribute is the percentage of friends already using the app, followed by accessibility, authentication, customisation level, and video calls (see Table \ref{tableSI:app_study:avg_variable_importance}). 
On average, messaging apps that are also web accessible, have two or multi-factor authentication and have a medium-high level of customisation are more appealing than mobile only apps with simple authentication and low customisation level (see Fig. \ref{figSI:app_study_average_utilities}). 
We observe that 299 out of the 300 respondents have a positive partworth utility of the social signal, with the remaining one having a negative utility. 
For the reasons outlined in describing the PS study, we did not consider further in the analysis the participant with a negative partworth utility of the social signal.
We measure the individual-level threshold using the methodology described in the Methods section of main text, and calculate the average threshold per product.
The results can be seen in Figure \ref{figSI:app_study_avg_threshold}.  

\begin{table}[htp]
\begin{tabular}{lcc}
\hline
Variable                            & \multicolumn{1}{l}{Average Importance} & \multicolumn{1}{l}{Standard Deviation} \\ \hline
\% of friends already using the app & 0.56692052                             & 0.17081138                             \\
Accessibility                       & 0.13926414                             & 0.12564616                             \\
Authentication                      & 0.12267676                             & 0.08752978                             \\
Customisation level                 & 0.10365652                             & 0.06592403                             \\
Video calls                         & 0.06748205                             & 0.05570236                            
\end{tabular}
\caption{\textbf{Average variable importance in the App Adoption Study}. The variables are ranked by importance.}
\label{tableSI:app_study:avg_variable_importance}
\end{table}

\begin{figure}[h]
\centering
\includegraphics[scale=0.9]{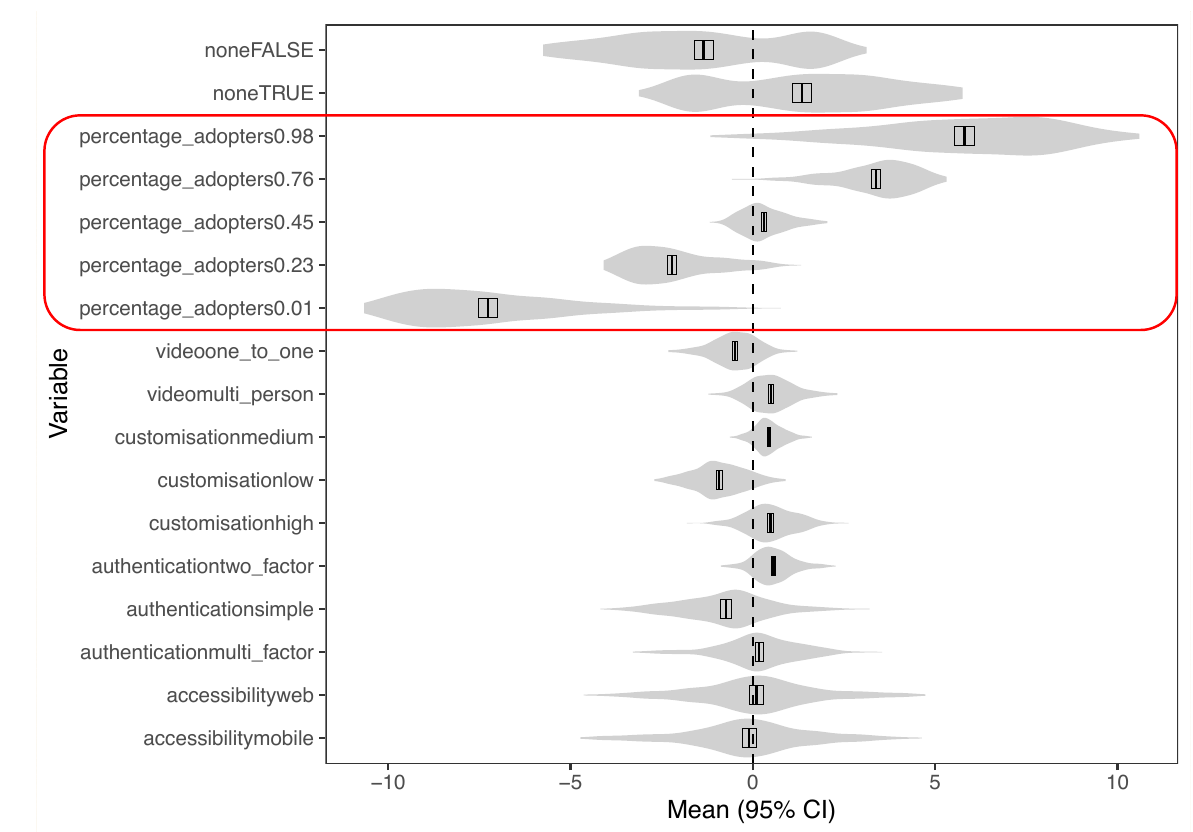}
\caption{\textbf{Average utilities across respondents treating the percentage of adopters as categorical variable (AA Study).} The y axis contains all the attributes and their levels. The x axis shows the distribution of the utility of an attribute level together with the mean and 95 $\%$ confidence interval (crossbars) based on $N=300$ observations. On average, the utility of the percentage of adopters increases with the level (red rectangle).}
\label{figSI:app_study_avg_utilities_si_factor}
\end{figure}

\paragraph*{Linearity of the social signal.}
To test the plausibility of assuming a linear coefficient of the social signal, we re-estimate the model treating the \textit{percentage of friends already using the app} as a categorical variable. Figure~\ref{figSI:app_study_avg_utilities_si_factor} shows a close to linear increase of the utility of the social signal as a function of the number of adopters (Pearson’s $r(1498) = 0.92$, $p < 0.001$, $95\%$ CI=$[0.92, 0.93]$), supporting the use of a linear coefficient.

\clearpage

\section{Data for the data-driven simulations}
\label{secSI:data}

\subsection{Social network data}
\label{secSI:network_sampling}

The original Add Health data \cite{guilbeault2021topological} contains 85 networks. To save computational time, we restrict the analysis to a sample of 18 networks
spanning diverse network properties. 
We constructed the sample as follows. 
For every network, we computed  the number of nodes and the global transitivity index of the graph (i.e., the number of observed transitive triads divided by the number of potential transitive triads~\cite{barabasi2016network}). 
For each of the two metrics, we computed the $33\%$ and 67$\%$ quantiles and labelled each network with respect to the metric as: low (metric $<33\%$ quantile), medium (metric lies between the $33\%$ and the $67\%$ quantiles) or high (metric $>67\%$ quantile). 
In this way, each network received two labels, one for the number of nodes and the other for transitivity. 
We created all possible combinations of the two labels (9 in total) and sampled two networks from each combination. 

\subsection{Product sampling}
\label{secSI:product_sampling}
The total number of products that can be generated from a conjoint design is equal to all distinct combinations of attribute levels (not considering the social signal). 
Thus, there are 768 and 36 possible products in the PS and AA study, respectively. 
To save computational time, we decided to restrict our analysis to: (1) all 36 products in the AA study and (2) a subset of 36 products from the PS study. 
We selected the 36 products from the PS study such that the average threshold over respondents per product spans a broad range of observed values.
Specifically, we calculated the threshold of every respondent for every product as described in the Methods section in the main text. 
Then we calculated the average threshold per product over respondents.
We divided the range of the average threshold values in 6 equally-sized intervals (based on the $16.7\%$, $33\%$, $50\%$, $67\%$ and $83\%$ quantiles) and sampled 6 products from each interval. 

\section{Simulated conjoint surveys}
\label{secSI:synthetic_choice_data}

The main objective of this Supplementary Note is to describe the procedure to generate simulated conjoint surveys (Section~\ref{secSI:survey_simulation}) and the obtained results, which concern both the accuracy and robustness of the threshold estimation procedure (Sections~\ref{secSI:recovery}--\ref{secSI:sensitivity_prior_individual_bimodal}). The simulated surveys are generated as described in Section~\ref{secSI:survey_simulation} are also used to inform the seeding simulations (see main text).

\subsection{Survey simulation}
\label{secSI:survey_simulation}

We consider a scenario where a change practictioner needs to estimate individual thresholds to inform behavior-based seeding policies. 
As explained in the main text, we initially assume that the change practitioner has been able to survey the nodes of the network. 
For our simulation study, we simulate this survey.
To simulate the survey, we follow a three step procedure, similar to the one described in \citet{hein2020analyzing}: 
(1) We calibrate the agents' partworth utilities on the empirical data from the two conjoint experiments (PS and AA);
(2) We generate choice tasks; (3) We simulate the agents' choices in the choice tasks based on their calibrated partworth utilities. 
To ensure that our results do not depend on a specific realization of the data, for each of the two empirical conjoint studies (PS and AA), we generate 5 independent realizations of the choice simulation following the procedure detailed below.

\paragraph*{1. Calibration of partworth utilities.}

\begin{figure*}[t]
    \centering\includegraphics[width=\textwidth]{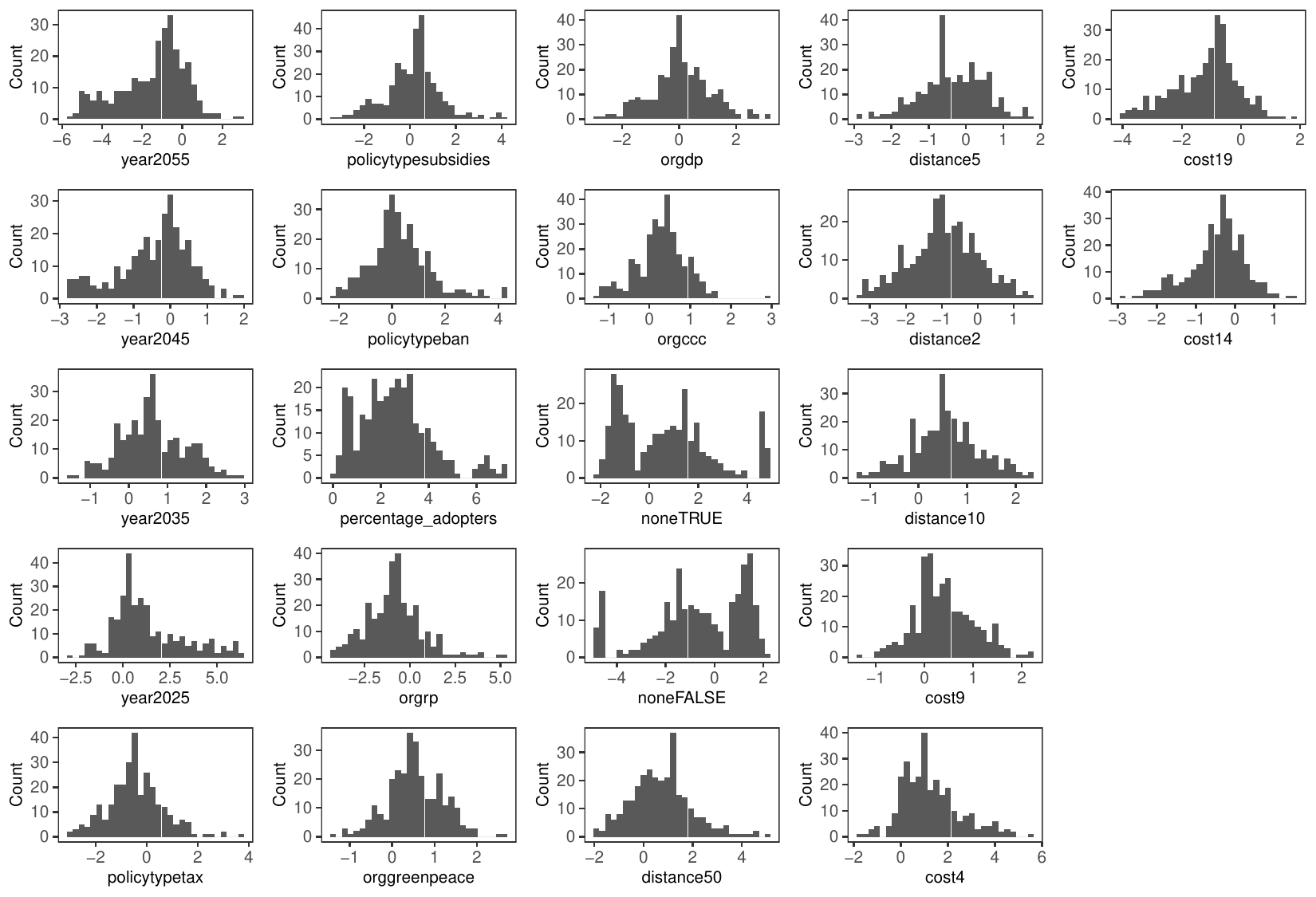}
    \caption{\textbf{Distribution of ground-truth partworth utilities for the simulation calibrated with the results of the PS Study ($N=291$ observations).}}
    \label{figSI:distribution_coefs_ps}
\end{figure*}

\begin{figure*}[t]
    \centering
    \includegraphics[width=\textwidth]{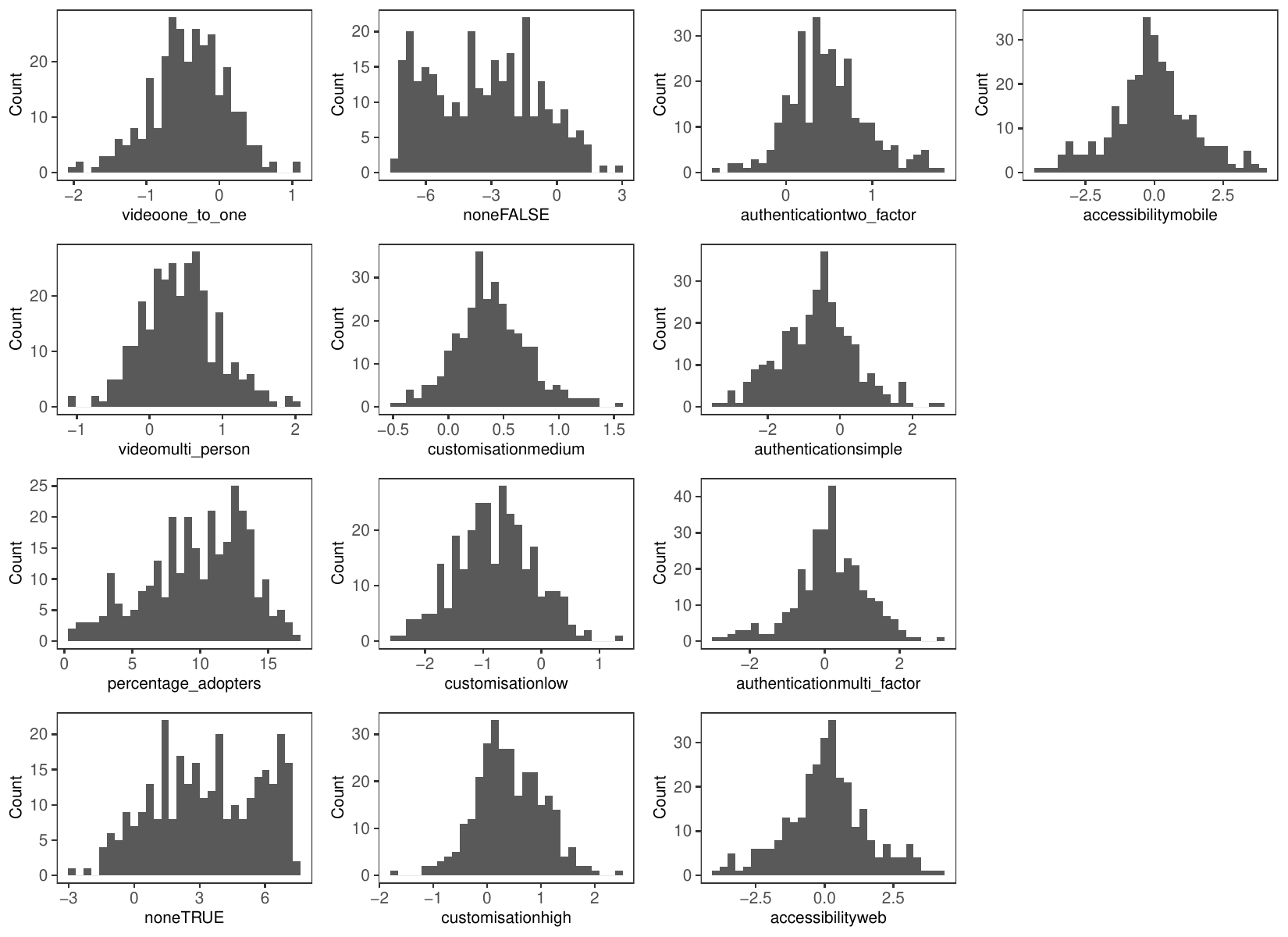}
    \caption{\textbf{Distribution of ground-truth partworth utilities for the simulation calibrated with the results of the AA Study ($N=299$ observations).}}
    \label{figSI:distribution_coefs_aa}
\end{figure*}

Each agent is characterized by an $n_{par}$-dimensional vector whose elements constitute the partworth utilities. 
Thus $n_{par} = 24$ in the PS study and $n_{par} = 13$ in the AA study (one parameter for each level of each attribute, one parameter for the social signal, and two parameters for the none option). 
To calibrate the choice simulations with the empirical data from the conjoint experiments, we endow each agent with a ground-truth partworth utility vector equal to the estimated partworth utilities of a participant selected at random from the pool of real-world participants\footnote{Excluding the few participants with $\hat{\gamma}<0$, see Section~\ref{secSI:conjoint_exp_description}.} without replacement.
The resulting distributions of partworth utilities across agents are shown in Figs.~\ref{figSI:distribution_coefs_ps}$-$\ref{figSI:distribution_coefs_aa}.

\paragraph*{2. Generation of choice tasks.}

We generate $Z = 15$ choice tasks per agent, each with a choice set consisting of two product alternatives and a none option. 
Each product alternative in the choice set is described as a unique combination of the product attributes from the focal conjoint study, and one out of three levels of the social signal: $0.1$, $0.5$ or $0.9$.  
For every agent and every choice set, we randomly sample the two alternatives from the set of all possible product configurations. 

\paragraph*{3. Choice simulation.}

We simulate the conjoint survey by simulating the choices made by the agents.
To simulate which alternative was selected in each choice set, we compute the utilities corresponding to each alternative, and the utility of the none option. 
For each agent, we compute the utility of a given alternative by summing up the attribute utilities for that alternative and the utility of the social signal corresponding to the given social signal level. 
We assume the utility of the none option is constant within each agent across choice tasks. 
We further assume there is some noise in the process\footnote{In this context, the presence of noise can reflect scenarios where the attributes included in the conjoint are not sufficient to fully capture the total utility of a product.}, and add to the utility of each alternative (either product or none) a random component drawn from a Gumbel distribution with scale parameter $\theta = 1$ \cite{hein2020analyzing}.
Lastly, we assume that the agent selects the alternative with the highest total utility. 

\subsection{Parameter recovery}
\label{secSI:recovery}

For each simulated choice dataset, we recover the individual partworth utilities using the Hierarchical Bayes algorithm (similarly to what we did in the two real conjoint studies, see Supplementary Note \ref{secSI:conjoint_exp_description}). 
We used 30'000 MCMC iterations, the first 10'000 for burn in and the remaining 20'000 for parameter estimation, from which every tenth draw was retained. 
We considered the social signal attribute as numeric and the remaining attributes as categorical, for which we used effects coding. 
We calculated the parameter estimates as the sample mean of the distribution generated by the MCMC draws. 
We used the estimated parameters to calculate the estimated thresholds through Eq.~(1) in the main text. 

We tested the quality of the parameter recovery in two ways, similarly to \citet{hein2020analyzing}.
First, we calculated the mean Pearson correlation between the ground-truth and the estimated partworth utilities over all parameters. 
As the average of Pearson correlations does not converge to the true correlation, following Hein et al.~\cite{hein2020analyzing}, we first rescaled the correlation coefficients using Fisher’s $z$-transformation [$f(x) = 0.5\, \log((1 + x)/(1 - x))$], averaged the rescaled correlations, and converted the resulting average back to the original scale using the inverse of Fisher's $z$-transform. 
We found the mean correlation ranges over simulated datasets from $0.74$ to $0.76$ in the PS study and from $0.72$ to $0.74$ in the AA study. 

Second, we measured how often the true coefficients lie within the 95$\%$ credible intervals of the estimated coefficients. 
We found the average value over coefficients and agents ranges across simulated datasets
from $95.02\%$ to $96.34\%$ in the PS study and from $94.78\%$ to $96.32\%$ in the AA study.
As the average correlation coefficients are large and the true coefficients lie almost always within the $95\%$ credible intervals of the estimated coefficients, we conclude the parameter recovery is accurate.

\subsection{Sensitivity to prior specification}
\label{secSI:sensitivity_prior_individual}
We also use the simulated conjoint to test the robustness of individual-level estimates with respect to the specification of the prior distributions. 

\paragraph*{Motivation.}
Because of the hierarchical structure, each individual’s partworth utilities are influenced not only by their own choices but also by information from the rest of the sample. 
Specifically, each individual's vector of partworth utilities is assumed to come from a population-level distribution (first-stage prior) whose parameters are themselves estimated from the data conditional on assumptions about their distribution (second-stage priors). 
As a result, each posterior estimate reflects a combination of the respondent’s own data and the group-level information. 
This effect shrinks individual estimates toward the estimated group mean. 
The extent of shrinkage depends both on the amount of data available for each individual and on the prior distributions specified for the group-level parameters. 
Formally, following the notation in main text, the utility of individual $n$ for alternative $i$ is given by $U_{ni} = \sum_{k=1}^{K} \beta_{nk}\,x_{ki} + \gamma_{n}s_{ni} + \epsilon_{ni}$. 
Let $\beta'_{n}$ denote the vector of all parameters ($\beta_{n}$ and $\gamma_{n}$) for individual $n$. 
Under the frequently-used normal prior model, the vector of parameters follows a multivariate normal distribution $\beta'_{n} \sim \mathcal{N}(b, \Sigma)$ (first-stage prior) with $b \sim \mathcal{N}(\bar{b}, A)$ as (second-stage) prior on the population mean and $\Sigma \sim \Psi(\nu, V)$ as (second-stage) prior on the population covariance, where $\Psi(\nu, V)$ denotes the inverted Wishart (IW) distribution with $\nu$ degrees of freedom and scale matrix $V$ \cite{rossi2024bayesian}. 
It is typical to reflect the researcher's uncertainty by controlling the prior on the covariance matrix and assume a very diffuse prior on the population mean (e.g., mean of $b$ at $0$ and very large variance)~\cite{train2009discrete}. 
Tighter priors on $\Sigma$ result in a stronger shrinkage toward the group mean, whereas looser priors allow for more individual-specific variation~\cite{hein2019effect}.
As the amount of data at the individual level increases, the effect of the prior specification diminishes~\cite{rossi2024bayesian}.

A potential concern is that different specifications of the priors on $\Sigma$ (different degrees of freedom and/or scale matrices in $\Psi(\nu, V)$) might lead to different estimated partworth utilities and consequently to different adoption thresholds.  
The effect of prior specification of the covariance matrix on the accuracy of recovering individual-level parameters has been addressed in an extensive simulation study by Ref.~\cite{hein2019effect}. 
Covering a wide range of conditions, the study finds that 
while the performance is influenced by the specifications of the priors, it generally remains high -- except in specific cases, such as when there is a substantial mismatch between the prior variance and the true heterogeneity in the data \cite{hein2019effect}. 
Using a similar but simplified setup, our objective here is to determine whether the main results in main text are robust with respect to the tightness of the prior and consequently to the degree of shrinkage it induces.

\paragraph*{Method.}
To address this concern, we repeat the simulation setup, generating datasets with varying numbers of choice tasks per individual $Z \in [15, 30, 100]$ consisting of the original dataset ($Z=15$ tasks) and two additional simulated datasets (corresponding to $Z=30$ and $Z=100$ tasks, respectively).
From each dataset, we estimate the partworth utilities using HB with four prior specifications.
Following a standard approach, for all HB models, we use a diffuse prior on the population mean~\cite{rossi2024bayesian, train2009discrete}, as implemented in the default option in the software package: $\bar{b} = \vec{0}$ (vector of zeroes) and $A = 100 \cdot \Sigma$ \cite{ChoiceModelR}. 
We vary the prior on the covariance matrix $\Sigma$ as follows. 
Following the notation in the documentation of the software package~\cite{ChoiceModelR}, $\nu = df + K'$ and $V = V_0 \cdot \nu \cdot v$, where $K'$ is the number of parameters to be estimated per individual (16 in the PS Study and 8 in the AA Study), $V_0$ is an $K' \times K'$ block diagonal matrix (see \cite{ChoiceModelR} for its specification), $v>0$ controls the prior variance and $df\geq2$ controls the degrees of freedom of the IW distribution \cite{ChoiceModelR}. 
Therefore, in the current software implementation, to control the tightness of the prior on the covariance matrix, the researcher can adjust the degrees of freedom (through $df$) and the prior variance (through $v$ and indirectly through $df$).
Lower values of $df$ and higher values of $v$ lead to looser priors, reducing the influence of the group average~\cite{hein2019effect}.  
We consider four specifications: default ($df=5$, $v=2$), tight ($df=10$, $v=2$), loose ($df=2$, $v=2$) and uninformative ($df=2$, $v=100$).
For illustration, in Table \ref{tableSI:meanIWAA} we show the mean of the IW distribution with the above specifications, from which the covariance matrices are drawn in the AA Study. 
The tightness of the prior can be understood by comparing the variance (diagonal) entries to the range of the partworth coefficients (see Fig. \ref{figSI:app_study_average_utilities} for the AA Study). 
Standard deviations (i.e., the square roots of variances in the diagonal entries) that are large relative to the typical magnitude of the partworths correspond to loose priors, while smaller standard deviations correspond to tight priors.

\begin{figure*}[t]
    \centering
    \includegraphics[width=\textwidth]{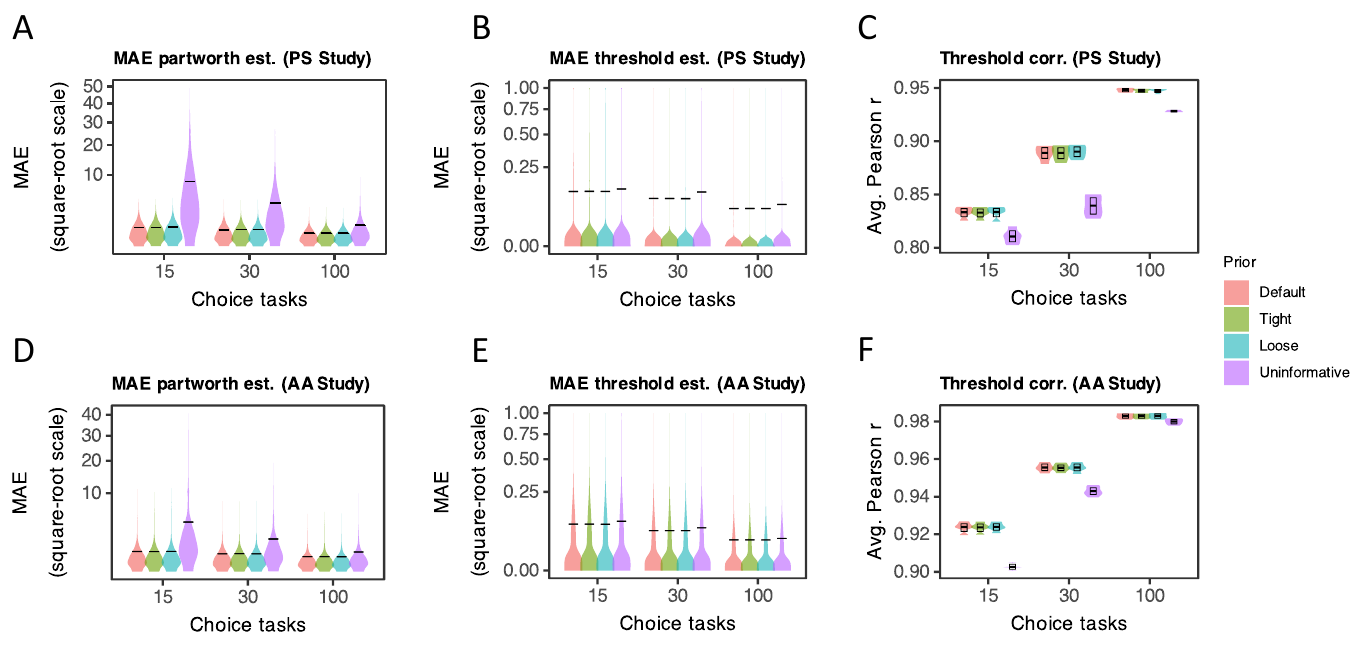}
    \caption{\textbf{Estimation accuracy under different prior specification for the PS Study (top row) and the AA Study (bottom row).}
    All but the uninformative prior result in similar accuracy of recovering the partworth utilities \textbf{(A, D)} and adoption thresholds \textbf{(B, E)} and similar correlations between the true and the estimated thresholds \textbf{(C, F)}. Crossbars represent the mean together with the $95\%$ confidence intervals based on the following number of observations: $32,010$ (A), $46,548 - 52,236$ (B); $19,435$ (D); $53,424 - 53,712$ (E); 5 (C, F). 
    }
    \label{figSI:hb_priors_ind}
\end{figure*}

For the MCMC estimation we use $30'000$ iterations ($10'000$ for burn in and $20'000$ for estimation) for all except the uninformative prior, where we use $50'000$ ($30'000$ for burn in and $20'000$ for estimation) due to slower convergence. 
In all cases, we retained every 10th draw for parameter estimation. 
    We use three measures to calculate the estimation accuracy: 
    \begin{itemize}
        \item Mean absolute error (MAE) in recovering the ground-truth partworth utilities,  defined as the average absolute value of the difference between the true and the estimated partworth utilities: $\text{MAE}=1/(HN K')\sum_{h=1}^H\sum_{n=1}^N\sum_{k=1}^{K'} |\hat{\beta'}_{hnk} - \beta'_{hnk}|$, where $H=5$ is the number of realizations of the simulated choice data, $N$ is the number of simulated individuals, $K'$ is the number of parameters (attribute utilities and utility of the social signal), $\beta'_{hnk}$ is the true value in simulated choice data $h$ of parameter $k$ for simulated individual $n$, $\hat{\beta'}_{hnk}$ is the corresponding estimated value. 
    \item Mean absolute error of recovering the individual thresholds. 
    Similarly as above, it is defined as the average absolute value of the difference between the true and the estimated thresholds over all simulated individuals, products and realizations of the simulated choice data\footnote{For all priors, the estimated partworth utility of the social signal is negative for a small subset of simulated individuals. As previously noted, in these situations, thresholds are not defined under the adopted choice framework. These cases are therefore removed from the analysis. As the number of removals differs slightly across priors, the sample size used for computing the threshold-estimation MAE also differs. This leads to the small variation in observation counts reported in Figs.~\ref{figSI:hb_priors_ind}B,E and \ref{figSI:hb_priors_ind_bimodal}B,E.}.
    \item Average Pearson's correlation between the true and the estimated thresholds.
    The correlation is computed based on data pooled from all individuals and all products. The correlations are then rescaled using Fisher's $z$-transformation, averaged over the five realizations of the simulated choice data and scaled back using the inverse Fisher's $z$-transformation, as described in Supplementary Note~\ref{secSI:recovery}.
    \end{itemize}
\paragraph*{Results.}
We find that all but the uninformative prior produce similarly accurate partworth estimation (consistent with~\cite{hein2019effect}), and all methods' mean absolute error decreases as the data size increases (Figs.~\ref{figSI:hb_priors_ind}A,D). 
The differences between the models' accuracy is even smaller for the threshold estimation task (Figs.~\ref{figSI:hb_priors_ind}B,E).
Importantly, the thresholds estimated through the different models are highly correlated with the ground-truth thresholds (Figs.~\ref{figSI:hb_priors_ind}C,F). 
For all priors and number of choice tasks, the average Pearson correlations are larger than $0.81$ and $0.90$ in the PS and AA Study, respectively.
Taken together, these findings indicate that as long as the choice of priors is sensible, the estimated thresholds are robust with respect to prior specification.

\begin{table}
\centering
\caption{\textbf{Mean of IW distributions under different priors in the AA study.} Larger variances (diagonal entries) correspond to looser priors. The reason why there are less variables compared to Fig.~\ref{figSI:app_study_average_utilities} is that for categorical variables the utility of the reference level is uniquely determined by the estimated utilities of the other levels.}
\label{tableSI:meanIWAA} 

\subfloat[Tight prior (v=2, df=10)]{
\begin{tabular}{lrrrrrrrr}
 & acc. mobile & auth. simple & auth. two\_factor & cust. low & cust. medium & video multi\_person & $\%$ adopters & none \\
acc. mobile & 2.00 & 0.00 & 0.00 & 0.00 & 0.00 & 0.00 & 0.00 & 0.00 \\
auth. simple & 0.00 & 2.67 & -1.33 & 0.00 & 0.00 & 0.00 & 0.00 & 0.00 \\
auth. two\_factor & 0.00 & -1.33 & 2.67 & 0.00 & 0.00 & 0.00 & 0.00 & 0.00 \\
cust. low & 0.00 & 0.00 & 0.00 & 2.67 & -1.33 & 0.00 & 0.00 & 0.00 \\
cust. medium & 0.00 & 0.00 & 0.00 & -1.33 & 2.67 & 0.00 & 0.00 & 0.00 \\
video multi\_person & 0.00 & 0.00 & 0.00 & 0.00 & 0.00 & 2.00 & 0.00 & 0.00 \\
$\%$ adopters & 0.00 & 0.00 & 0.00 & 0.00 & 0.00 & 0.00 & 4.00 & 0.00 \\
none & 0.00 & 0.00 & 0.00 & 0.00 & 0.00 & 0.00 & 0.00 & 2.00 \\
\end{tabular}
\label{tab:iw_2}
}

\vspace{0.5cm}

\subfloat[Default prior (v=2, df=5)]{
\begin{tabular}{lrrrrrrrr}
 & acc. mobile & auth. simple & auth. two\_factor & cust. low & cust. medium & video multi\_person & $\%$ adopters & none \\
acc. mobile & 3.25 & 0.00 & 0.00 & 0.00 & 0.00 & 0.00 & 0.00 & 0.00 \\
auth. simple & 0.00 & 4.33 & -2.17 & 0.00 & 0.00 & 0.00 & 0.00 & 0.00 \\
auth. two\_factor & 0.00 & -2.17 & 4.33 & 0.00 & 0.00 & 0.00 & 0.00 & 0.00 \\
cust. low & 0.00 & 0.00 & 0.00 & 4.33 & -2.17 & 0.00 & 0.00 & 0.00 \\
cust. medium & 0.00 & 0.00 & 0.00 & -2.17 & 4.33 & 0.00 & 0.00 & 0.00 \\
video multi\_person & 0.00 & 0.00 & 0.00 & 0.00 & 0.00 & 3.25 & 0.00 & 0.00 \\
$\%$ adopters & 0.00 & 0.00 & 0.00 & 0.00 & 0.00 & 0.00 & 6.50 & 0.00 \\
none & 0.00 & 0.00 & 0.00 & 0.00 & 0.00 & 0.00 & 0.00 & 3.25 \\
\end{tabular}
\label{tab:iw_1}
}

\vspace{0.5cm}

\subfloat[Loose prior (v=2, df=2)]{
\begin{tabular}{lrrrrrrrr}
 & acc. mobile & auth. simple & auth. two\_factor & cust. low & cust. medium & video multi\_person & $\%$ adopters & none \\
acc. mobile & 10.00 & 0.00 & 0.00 & 0.00 & 0.00 & 0.00 & 0.00 & 0.00 \\
auth. simple & 0.00 & 13.33 & -6.67 & 0.00 & 0.00 & 0.00 & 0.00 & 0.00 \\
auth. two\_factor & 0.00 & -6.67 & 13.33 & 0.00 & 0.00 & 0.00 & 0.00 & 0.00 \\
cust. low & 0.00 & 0.00 & 0.00 & 13.33 & -6.67 & 0.00 & 0.00 & 0.00 \\
cust. medium & 0.00 & 0.00 & 0.00 & -6.67 & 13.33 & 0.00 & 0.00 & 0.00 \\
video multi\_person & 0.00 & 0.00 & 0.00 & 0.00 & 0.00 & 10.00 & 0.00 & 0.00 \\
$\%$ adopters & 0.00 & 0.00 & 0.00 & 0.00 & 0.00 & 0.00 & 20.00 & 0.00 \\
none & 0.00 & 0.00 & 0.00 & 0.00 & 0.00 & 0.00 & 0.00 & 10.00 \\
\end{tabular}
\label{tab:iw_3}
}

\vspace{0.5cm}

\subfloat[Uninormative prior (v=100, df=2)]{
\begin{tabular}{lrrrrrrrr}
 & acc. mobile & auth. simple & auth. two\_factor & cust. low & cust. medium & video multi\_person & $\%$ adopters & none \\
acc. mobile & 500.00 & 0.00 & 0.00 & 0.00 & 0.00 & 0.00 & 0.00 & 0.00 \\
auth. simple & 0.00 & 666.67 & -333.33 & 0.00 & 0.00 & 0.00 & 0.00 & 0.00 \\
auth. two\_factor & 0.00 & -333.33 & 666.67 & 0.00 & 0.00 & 0.00 & 0.00 & 0.00 \\
cust. low & 0.00 & 0.00 & 0.00 & 666.67 & -333.33 & 0.00 & 0.00 & 0.00 \\
cust. medium & 0.00 & 0.00 & 0.00 & -333.33 & 666.67 & 0.00 & 0.00 & 0.00 \\
video multi\_person & 0.00 & 0.00 & 0.00 & 0.00 & 0.00 & 500.00 & 0.00 & 0.00 \\
$\%$ adopters & 0.00 & 0.00 & 0.00 & 0.00 & 0.00 & 0.00 & 1000.00 & 0.00 \\
none & 0.00 & 0.00 & 0.00 & 0.00 & 0.00 & 0.00 & 0.00 & 500.00 \\
\end{tabular}
\label{tab:iw_4}
}
\end{table}



\subsection{Understanding the effect of partworth modality on estimation accuracy}
\label{secSI:sensitivity_prior_individual_bimodal}
\paragraph*{Motivation.}

To derive the results of the sensitivity analysis in Section~\ref{secSI:sensitivity_prior_individual}, we calibrated the choice simulations by setting the ground-truth partworth utilities equal to the individual-level estimates from the two conjoint studies. 
From the so-calibrated ground-truth utilities, we generated the simulated choice datasets, which we subsequently used to test the recovery of the partworth utilities via HB estimation under different specifications of the normal prior.
One concern is that in those simulations, the input parameters (see Figs.~\ref{figSI:distribution_coefs_ps}$-$\ref{figSI:distribution_coefs_aa} for their distribution) were themselves estimated through the same HB method whose sensitivity we evaluate. 
As a result, the analysis may not reflect the method's sensitivity in scenarios where the underlying preference distribution is more complex or irregular.
To address this concern, we extend the sensitivity analysis to settings in which the underlying distribution of ground-truth partworth utilities is not entirely derived from the estimation method. 
Specifically, we consider the more complex case of multimodal distributions, which would occur in scenarios where the population consists of multiple latent segments with distinct preferences.

\begin{figure*}[t]
    \centering
    \includegraphics[scale=0.7]{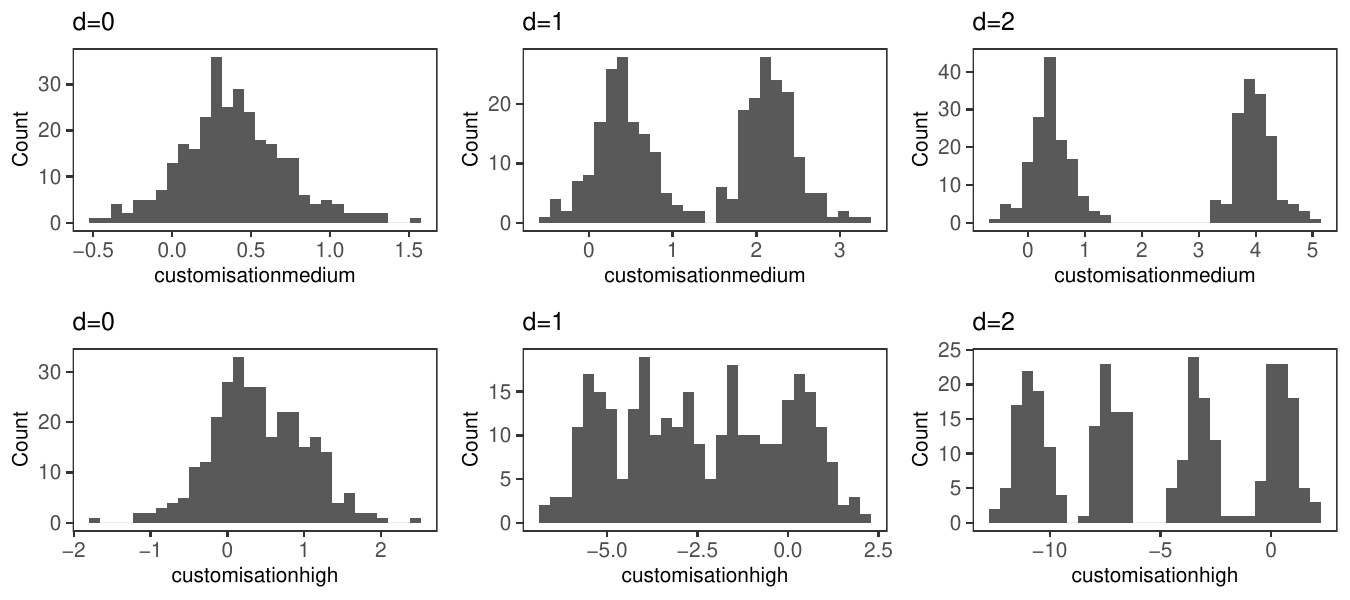}
    \caption{\textbf{Distribution of two ground-truth partworth utilities for different values of the shifting factor ($d\in\{0,1,2\}$).}
    The top row shows the distribution of the partworth utility for the \textit{medium} level of the \textit{customisation} attribute in the AA study. Here $d=0$ corresponds to the calibrated distribution used as ground-truth in Section~\ref{secSI:sensitivity_prior_individual}. As $d$ increases (from right to left), the distributions becomes increasingly more bimodal; when $d=2$, the two modes are more widely separated.
    The bottom row shows the distribution of the reference level (\textit{high}) of the same attribute.
    Because reference-level utilities under effect coding are computed from the other levels, the resulting distribution can have more than two modes (here four modes, that are clearly separated for $d=2$). Each panel is based on $N=299$ observations.}
    \label{figSI:hist_coefs_diff_d}
\end{figure*}

\paragraph*{Method.}
We repeat the analysis in Section~\ref{secSI:sensitivity_prior_individual}, considering $15$ choice tasks per agent. 
In Section~\ref{secSI:sensitivity_prior_individual}, the ground-truth partworth utilities corresponded to the individual-level HB estimates. 
We further refer to them as the \textit{calibrated distributions}.
Here, we generate multimodal ground-truth distributions by perturbing the calibrated distributions as follows. 
For each partworth utility, we randomly select $50\%$ of the agents and shift their partworth utilities by an amount equal to the range of the calibrated distribution (maximum minus minimum) multiplied by a shift factor $d \in \{1,2\}$.
Fig.~\ref{figSI:hist_coefs_diff_d} illustrates the resulting distributions of selected ground-truth partworth utilities for different values of $d$.
Based on the so-obtained ground-truth partworth utility distributions, we conduct five independent choice simulations for each $d$ value, using the same procedure described in Section~\ref{secSI:sensitivity_prior_individual}. 
For the HB estimation of the generated choice data, we use the same prior specifications and the same evaluation metrics described in Section~\ref{secSI:sensitivity_prior_individual}.
When evaluating the accuracy of recovering the ground-truth partworth utilities, we normalize the MAE by $d+1$ to account for the differences in coefficient scales across scenarios.

\begin{figure*}[t]
    \centering
    \includegraphics[width=\textwidth]{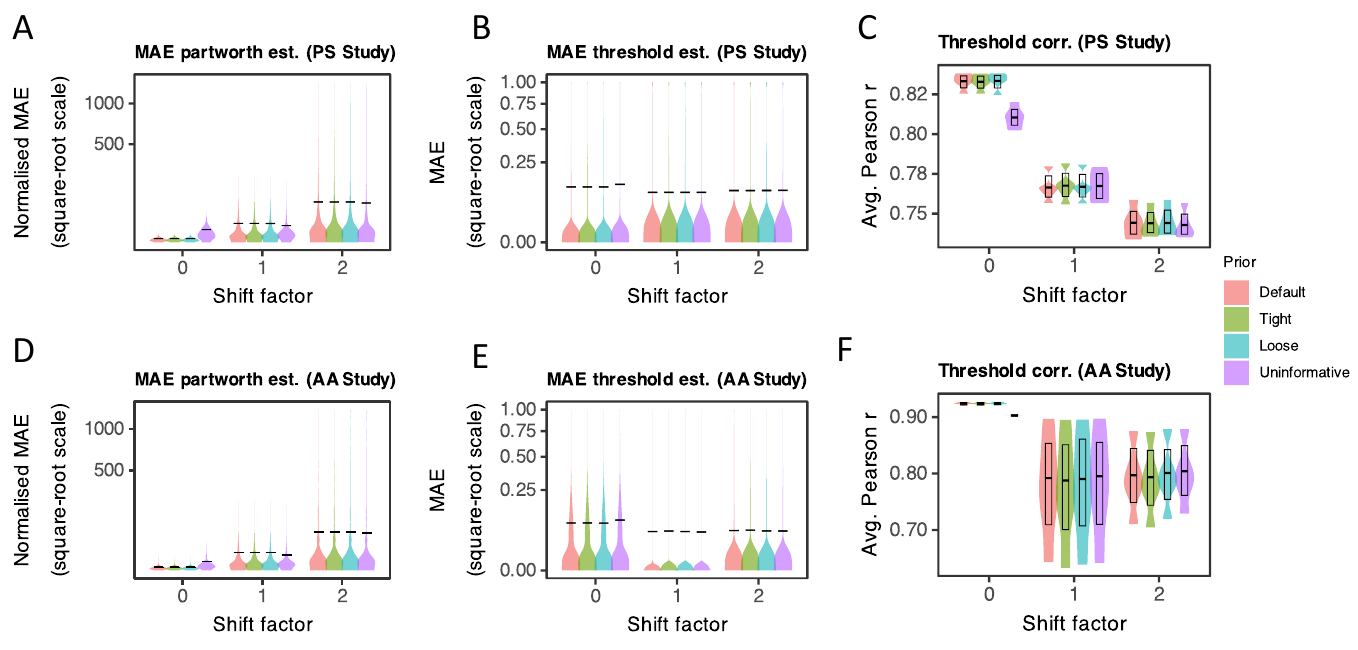}
    \caption{\textbf{Estimation accuracy under different prior specification for the PS Study (top row) and the AA Study (bottom row)}. The x-axis represents the value of the shift factor $d$ used to generate the ground-truth partworth distribution. 
    All but the uninformative prior result in similar accuracy of recovering the partworth utilities \textbf{(A, D)} and adoption thresholds \textbf{(B, E)} and similar correlations between the true and the estimated thresholds \textbf{(C, F)}.
     represent the mean together with the $95\%$ confidence intervals based on the following number of observations: $32,010$ (A), $50,508 - 52,344$ (B); $19,435$ (D); $52,992 - 53,748$ (E); $5$(C, F).}
    \label{figSI:hb_priors_ind_bimodal}
\end{figure*}

\paragraph*{Results.}
The results for the informative priors are in qualitative agreement with those in Section~\ref{secSI:sensitivity_prior_individual}, which are shown for reference in the charts for $d=0$. 
All priors except the uninformative one produce similarly accurate estimates of the partworth utilities (Figs.~\ref{figSI:hb_priors_ind_bimodal}A,D), while all priors produce similarly accurate estimates of the adoption thresholds (Figs.~\ref{figSI:hb_priors_ind_bimodal}B,E) and comparable correlations between the true and estimated thresholds (Figs.~\ref{figSI:hb_priors_ind_bimodal}C,F).

In the multimodal scenarios ($d\in\{1,2\}$), the uninformative prior outperforms the rest in partworth estimation. 
 This arguably occurs because, when the true distribution of partworths is multimodal, a tighter normal prior induces stronger shrinkage of individual estimates toward the global mean, thereby masking the underlying multimodality and increasing bias~\cite{rossi2024bayesian}. 
Similarly, the disadvantage of the uninformative prior for threshold estimation observed in the calibrated scenarios (Figs.~\ref{figSI:hb_priors_ind_bimodal}B--C; E--F; bar charts at $d=0$) vanishes in the multimodal scenarios ($d\in\{1,2\}$).

Overall, these results show that under more challenging, multimodal preference structures, informative normal priors do not necessarily outperform an uninformative prior.
As expected, the correlation between estimated and ground-truth thresholds decreases relative to the calibrated simulations\footnote{The findings for the MAE of the threshold estimation (Figs.~\ref{figSI:hb_priors_ind_bimodal}B,E) seem in apparent contradiction with this result, as they show a decreasing MAE as the shifting factor increases. This apparent contradiction can be resolved by noting that for larger $d$, the threshold distribution becomes more degenerate (i.e., it consists of more $0$s and $1$s).
In such cases, minimizing the MAE is inherently simpler than in the original case ($d=0$), as a naive model that always ``guesses" the dominant value can achieve artificially high accuracy.
To illustrate this point, we additionally evaluate the performance of a naive baseline model where we estimate the HB model, predict the thresholds of all the individuals for a given product, and then assign to all the individuals the most frequently predicted value among $0$ or $1$. 
The performance of this baseline model improves substantially as the shift factor increases: by $42\% (30\%)$ for $d=1$ compared to $d=0$, and by $44\% (32\%)$ for $d=2$ compared to $d=0$ in the PS (AA) Study. Hence, while the MAE decreases as $d$ increases, the MAE reduction compared to the baseline's MAE actually shrinks, in qualitative agreement with Figs.~\ref{figSI:hb_priors_ind_bimodal}C,F.} (Figs.~\ref{figSI:hb_priors_ind_bimodal}C,F). 
Nevertheless, the correlations remain high across all scenarios [larger than $r = 0.74$ ($r = 0.79$) in the PS (AA) Study], pointing to a reasonable robustness of the estimated thresholds against different preference scenarios.
We note that in scenarios where underlying multimodal distributions are to be expected, extensions of the HB model (such as those using mixtures of normals \cite{rossi2024bayesian}) might improve estimation accuracy.

\clearpage

\section{Seeding policies}
\label{secSI:seeding_policies}

\subsection{Structure-based seeding policies}
\label{secSI:seeding_policies_network}

\paragraph*{Degree policy.} The degree assumes that a node is central if she has many connections. A node's degree is simply defined as its number of connections~\cite{newman2018networks}.
Despite the simplicity of this metric, some works have found high-degree nodes to be effective at predicting and triggering large-scale cascades of adoptions~\cite{goldenberg2009role,hinz2011seeding,libai2013decomposing}, although some works found counterexamples in model-based and empirical data, e.g.,~\cite{galeotti2009influencing,watts2007influentials,barbera2015critical,zhou2024beyond}.
We compute it via the \url{igraph} package \cite{igraphPackage}.

\paragraph*{Betweenness policy.} The betweenness centrality assumes that a node is central if she is traversed by many of the network's shortest paths. For a focal node $i$, it averages over all pairs of nodes $(s,t)$ the fraction of shortest paths between $s$ and $t$ that pass through $i$~\cite{newman2018networks}. 
Similarly to high-degree nodes, some works have found high-betweenness nodes to be effective at triggering large-scale cascades of adoptions~\cite{hinz2011seeding}, although some works found counterexamples in model-based and empirical data, e.g.,~\cite{yoganarasimhan2012impact,rossman2021network}.
We compute it via the \url{igraph} package \cite{igraphPackage}.

\paragraph*{Closeness policy.} 
The closeness centrality assumes that a node is central if it is close to many other nodes in the network. It is defined as the inverse of the sum of shortest-path distances from the focal node to all the other nodes in the network~\cite{bavelas1950communication}.
Some works have found high-closeness nodes to be effective at triggering large-scale cascades of adoptions, especially for simple contagions above their critical point~\cite{zhou2019fast}.
We compute it via the \url{igraph} package \cite{igraphPackage}.

\paragraph*{Collective Influence policy.} The collective influence metric is derived analytically as an approximate solution to a network dismantling problem. In this problem, one seeks to find the minimal set of nodes whose removal from the network causes the collapse of the network's largest component~\cite{morone2015influence}. We compute it via our own function.

\subsection{Behavior-based seeding policies}

For each configuration $\mathcal{C}$ (identified by 
the network being analyzed, the product that spreads through the network, the agents' random assignment to the network's nodes, and the particular realization of the simulated conjoint survey performed by the change practitioner to estimate the thresholds), we measure various node-level behavior-based metrics that take into account the estimated thresholds (see main text and Supplementary Note~\ref{secSI:synthetic_choice_data}). The underlying assumption is that the social change practitioner has been able to survey the nodes of the network, estimate their thresholds for different products, and use the estimated thresholds to implement the seeding policies defined below. 

\paragraph*{Low threshold policy.} The low-threshold policy simply selects as the focal seed the node with the lowest estimated threshold in a given configuration; In case of ties, the focal node is selected at random among those with the lowest estimated threshold.

\paragraph*{Neighborhood susceptibility policy.} The neighborhood susceptibility (NS) of a given node $i$ is defined as her number of low-threshold (highly-susceptible) social contacts. Following Watts~\cite{watts2002simple}, we define a given node $j$ is a low-threshold node if $\hat{\tau}_j\leq d_j^{-1}$, where $\hat{\tau}_j$ is the threshold estimated by the social change practitioner and $d_j$ is the degree of node $j$; if this is the case, a single adoption event in $j$'s neighborhood is sufficient to trigger their adoption. Therefore, high-NS nodes are not necessarily the most central ones in the network according to the degree, but they are surrounded by a local ``mass" of highly-susceptible individuals.
Differently than the degree and other standard centrality metrics, the NS metrics takes into account individual-level behavioral information, namely, individuals' thresholds.

\paragraph*{Calibrated and uncalibrated complex centrality (CC) policies.} 
The complex centrality metric has been introduced by Guilbeault and Centola~\cite{guilbeault2021topological} to identify the optimal network locations where to initiate a complex contagion process under a clustered seeding strategy.
In Guilbeault and Centola~\cite{guilbeault2021topological}'s code, the complex centrality metric is measured by running the simulation of the complex contagion process under the ground-truth thresholds, and then measuring the average length of the shortest paths that connect a given node to the reachable nodes in the ``activated network" composed of the nodes who adopted.
We adapt their code to our study by measuring the complex centrality based not on the ground-truth thresholds (which are typically unknown to the decision maker), but on different assumptions on which individual-level behavioral data is available to the social change practitioner. For simplicity, we consider here two scenarios.
\begin{itemize}
    \item \textit{Calibrated complex centrality (CCC)}.
    We assume that the change social practitioner has been able to survey the network's nodes, and uses the estimated thresholds to run the simulations behind the complex centrality calculation.
    \item \textit{Uncalibrated complex centrality (UCC)}. We assume that the change practitioner has no behavioral information about the network's nodes. Following the same procedure used by Guilbeault and Centola~\cite{guilbeault2021topological} in their analysis of Indian villages, we average each node's complex centrality over an ensemble composed of five homogeneous absolute threshold scenarios ($\tau_i \, d_i=\theta$ where $\theta\in\{2,3,4,5,6\}$). As the uncalibrated CC assumes no knowledge of the estimated thresholds, and it averages the CC over a range of arbitrarily-selected threshold values, we classify it as a structure-based policy in the main text.
\end{itemize}
We implemented our own function to generate synthetic choice data and estimate the thresholds from the simulated data (see Supplementary Note \ref{secSI:synthetic_choice_data}); we implemented the two types of CC-based policies by simply passing the corresponding threshold vectors into a function adapted from the code by Guilbeault and Centola~\cite{guilbeault2021topological}.

\clearpage

\section{Additional results for seeding policies}
\label{secSI:addition_results_seeding}

\subsection{Results for the ratio-to-best metric}

\begin{figure*}[h]
    \centering
    \includegraphics[width=\textwidth]{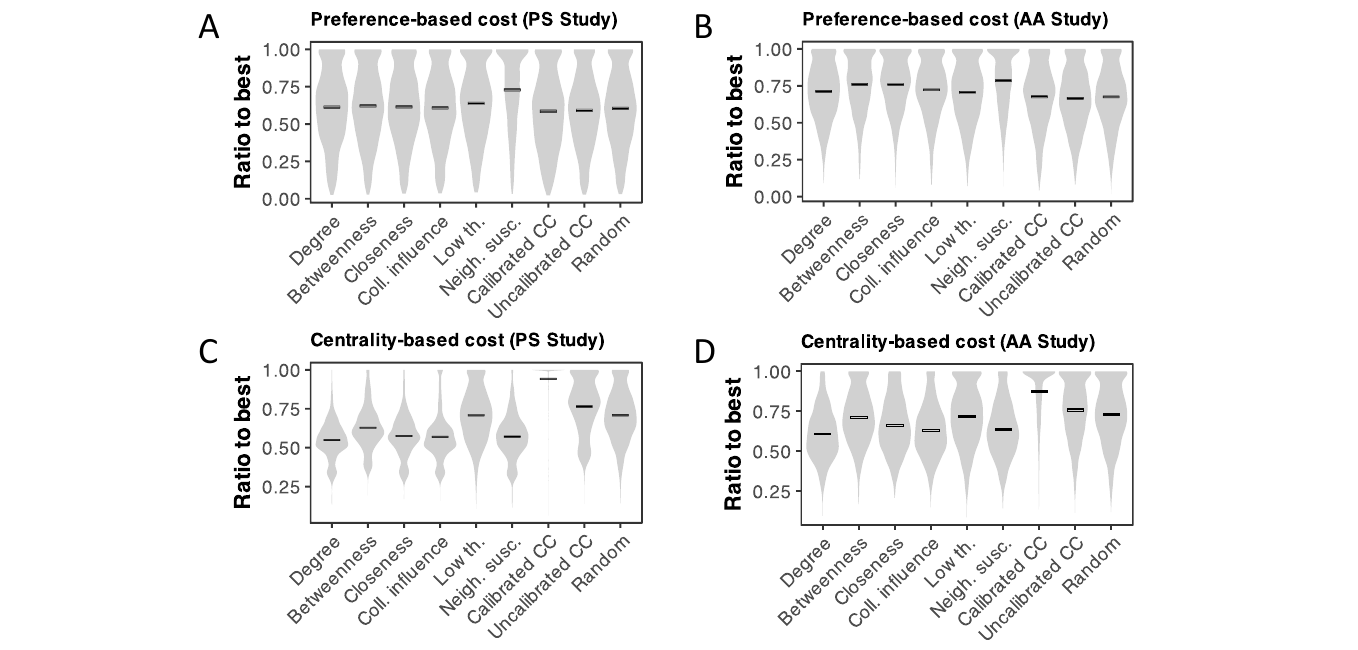}
    \caption{\textbf{Relative performance of seeding policies according to the ratio-to-best metric.} 
    \textbf{(A, B)} Relative performance of seeding policies under a preference–based cost structure, measured through the ratio-to-best metric defined above, for the policy support experiment and the app adoption experiment, respectively. The neighborhood susceptibility policy based on the estimated thresholds significantly outperforms the other policies (one-tailed Wilcoxon signed-rank test, all p-values are smaller than $p=0.001$, the smallest effect size is $r=0.09$ in the AA Study in the comparison against the betweenness centrality). \textbf{(C, D)} Relative performance of seeding policies under a centrality–based cost structure for the policy support experiment and the app adoption experiment, respectively. The complex centrality policy based on the estimated thresholds significantly outperforms the other policies (one-tailed Wilcoxon signed-rank test, all p-values are smaller than $p=0.001$, the smallest effect size is $r=0.42$ in the AA Study in the comparison against the uncalibrated complex centrality). 
    Overall the results are in qualitative agreement with those obtained for the mean rank metric (Fig.~2 in the main text).
    In all the panels, crossbars represent the mean together with the $95\%$ confidence intervals over $3,240$ runs.}
    \label{figSI:rtb}
\end{figure*}

We show here the results for the ratio-to-best metric. Specifically, to assess the relative performance of the examined seeding policies, for each analyzed configuration $\mathcal{C}$, we measure the simulated performance $P$ of the nine policies. 
For each configuration $\mathcal{C}$, each method $m$ is assigned to a performance variable that quantifies the method's performance compared to the best-performing method~\cite{zhou2019fast}, $s_{m}(\mathcal{C})=P_m(\mathcal{C})/\max_{m'}\{P_{m'}(\mathcal{C})\}$. We average this variable across all configurations.
The results, shown in Fig.~\ref{figSI:rtb}, are largely consistent with those shown in Fig.~2 of main text.

\clearpage

\subsection{Results for different seed set sizes}
\label{secSI:different_seed_size}

\begin{figure*}[h]
    \centering   
    \includegraphics[width=\textwidth]{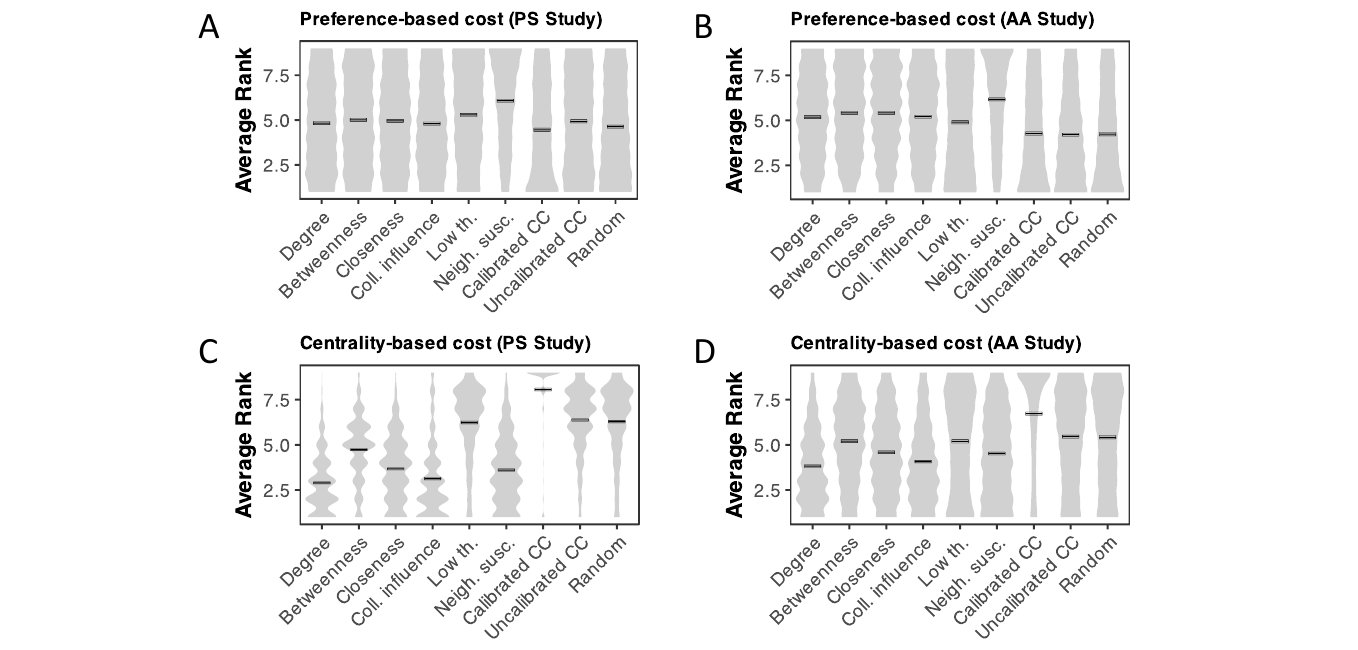}
    \caption{\textbf{Relative performance of seeding policies for a smaller seed set size ($z=0.0125$), according to the mean rank metric.} 
    \textbf{(A, B)} Relative performance of seeding policies under a preference–based cost structure, measured through the mean rank metric used in main text, for the policy support experiment and the app adoption experiment, respectively. The neighborhood susceptibility policy based on the estimated thresholds significantly outperforms the other policies (one-tailed Wilcoxon signed-rank test, all p-values smaller than $p = 0.001$, the smallest effect size is $r=0.19$ in the AA Study in the comparison against the closeness centrality). 
    \textbf{(C, D)} Relative performance of seeding policies under a centrality–based cost structure for the policy support experiment and the app adoption experiment, respectively. The complex centrality policy based on the estimated thresholds significantly outperforms the other policies (one-tailed Wilcoxon signed-rank test, all p-values smaller than $p=0.001$, the smallest effect size is $r=0.32$ in the AA Study in the comparison against the uncalibrated complex centrality). 
    Overall the results are in qualitative agreement with those obtained with $z=0.025$ (Fig.~2 in the main text).
    In all the panels, crossbars represent the mean together with the $95\%$ confidence intervals over $3,240$ runs.
    }
    \label{fig:smallz1}
\end{figure*}

\begin{figure*}[h]
    \centering    
    \includegraphics[width=\textwidth]{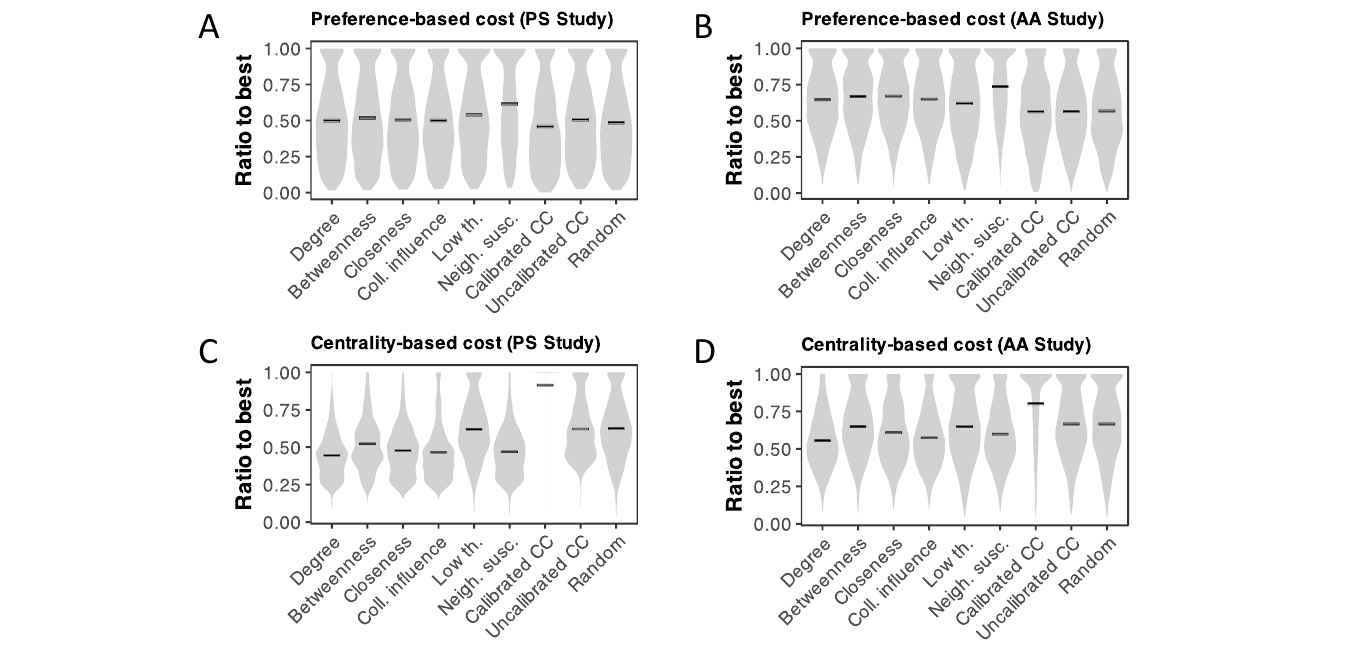}
    \caption{\textbf{Relative performance of seeding policies for a smaller seed set size ($z=0.0125$), according to the ratio-to-best metric.} 
    \textbf{(A, B)} Relative performance of seeding policies under a preference–based cost structure, measured through the ratio-to-best metric defined above, for the policy support experiment and the app adoption experiment, respectively. The neighborhood susceptibility policy based on the estimated thresholds significantly outperforms the other policies
    (one-tailed Wilcoxon signed-rank test, all p-values smaller than $p = 0.001$, the smallest effect size is $r=0.20$ in the comparison against the low-threshold policy in the PS Study).
    \textbf{(C, D)} Relative performance of seeding policies under a centrality–based cost structure for the policy support experiment and the app adoption experiment, respectively. The complex centrality policy based on the estimated thresholds significantly outperforms the other policies (one-tailed Wilcoxon signed-rank test, all p-values are smaller than $p = 0.001$, the smallest effect size is $r=0.37$ in the AA Study in the comparison against the random policy). Overall the results are in qualitative agreement with those obtained with $z=0.025$ (Fig.~2 in main the text). In all the panels, crossbars represent the mean together with the $95\%$ confidence intervals over $3,240$ runs.}
    \label{fig:smallz2}
\end{figure*}

We show here the results for different seed set sizes. Compared to the seed set size used in main text ($z=0.025$), we consider here both a smaller seed set size ($z=0.0125$) and a larger one ($z=0.05$). 
The results for the smaller seed set size are in qualitative agreement with those in the main text (see Figs.~\ref{fig:smallz1}--\ref{fig:smallz2}). The only exception is the lack of significant differences in the performance of the uncalibrated complex centrality and random seeding in all setups except the preference-based cost in the PS Study (one-tailed Wilcoxon signed-rank test, all p-values are larger than $p=0.23$; the largest effect size is $r=0.04$ for the PS Study, ratio-to-best metric, centrality-based cost), which supports the conclusion that the complex centrality can only realize its full potential when calibrated with the estimated thresholds. 

\begin{figure*}[h]
    \centering    
    \includegraphics[width=\textwidth]{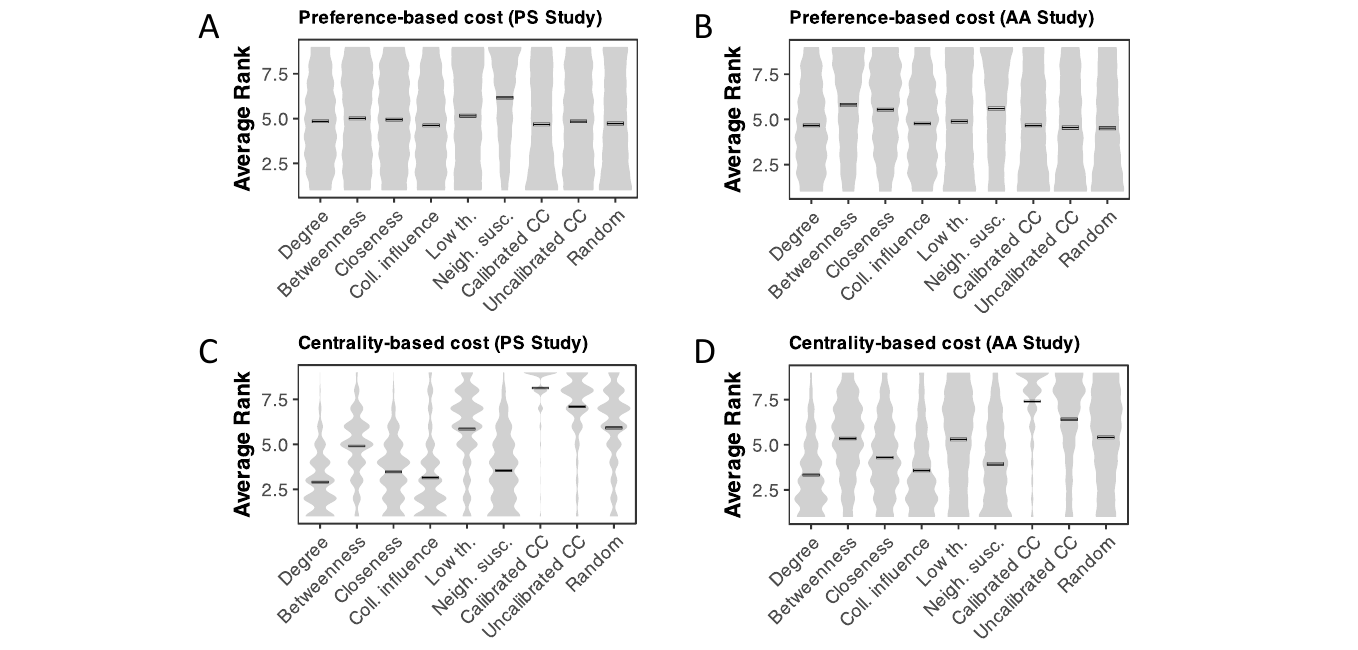}
    \caption{\textbf{Relative performance of seeding policies for a larger seed set size ($z=0.05$), according to the mean rank metric.} 
    \textbf{(A, B)} Relative performance of seeding policies under a preference–based cost structure, measured through the mean rank metric used in main text, for the PS Study and the AA Study, respectively. 
    In the simulations calibrated with the AA Study results, the neighborhood susceptibility policy outperforms all the other metrics except the betweenness and closeness centralities (one-tailed Wilcoxon signed-rank test, all p-values are smaller than $p = 0.001$, the smallest effect size is $r=0.18$ in the comparison against the low threshold): For larger seed sets, the estimated thresholds remain informative but no longer fully sufficient to identify more effective focal seeds than the betweenness and closeness centrality.
    \textbf{(C, D)} Relative performance of seeding policies under a centrality–based cost structure for the policy support experiment and the app adoption experiment, respectively. The complex centrality policy based on the estimated thresholds significantly outperforms the other policies (one-tailed Wilcoxon signed-rank test, all p-values are smaller than $p = 0.001$, the smallest effect size is $r=0.32$ in the AA Study in the comparison against the uncalibrated complex centrality); this result is in qualitative agreement with those obtained with $z=0.025$ (Fig.~2 in the main text).
    In all the panels, crossbars represent the mean together with the $95\%$ confidence intervals over $3,240$ runs.}
    \label{fig:largez1}
\end{figure*}

\begin{figure*}[h]
    \centering
    \includegraphics[width=\textwidth]{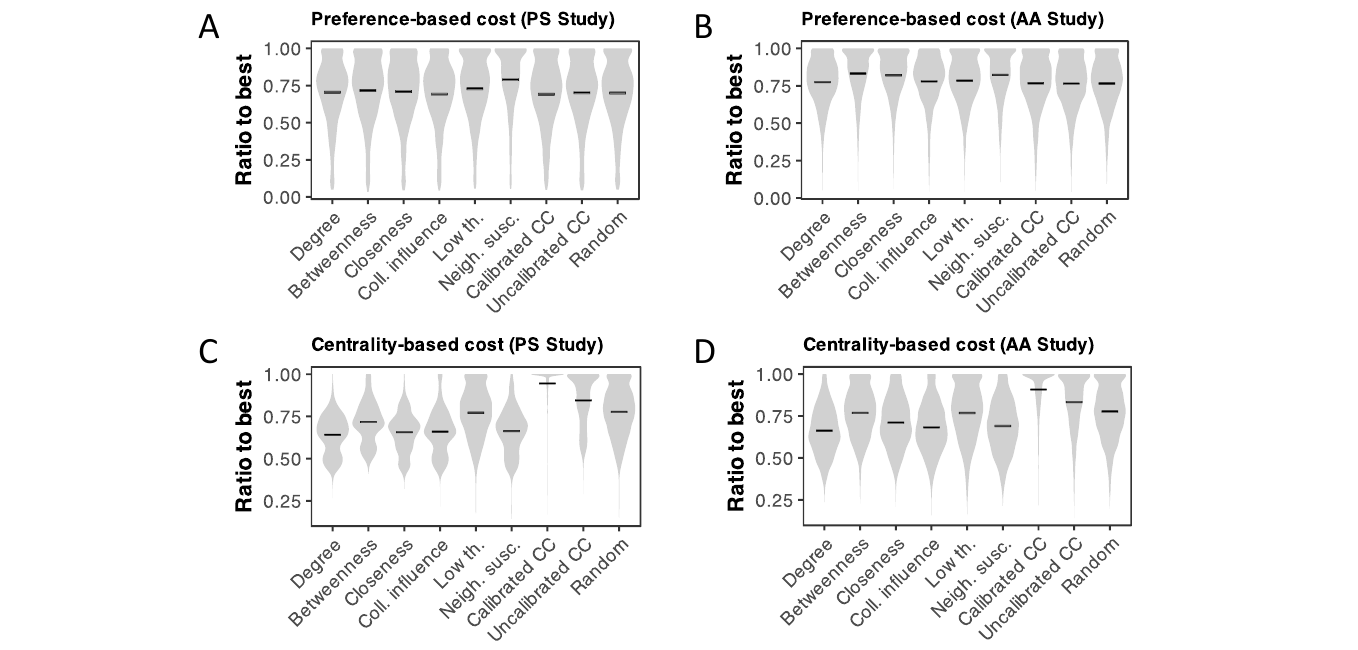}
    \caption{\textbf{Relative performance of seeding policies for a larger seed set size ($z=0.05$), according to the ratio-to-best metric.} 
    \textbf{(A, B)} Relative performance of seeding policies under a preference–based cost structure, measured through the ratio-to-best metric defined above, for the policy support experiment and the app adoption experiment, respectively. Again, in the simulations calibrated with the AA study results, the neighborhood susceptibility policy outperforms all the other metrics except the betweenness and closeness centralities (one-tailed Wilcoxon signed-rank test, all p-values are smaller than $p = 0.001$, the smallest effect size is $r=0.19$ in the comparison against the low threshold): For larger seed sets, the estimated thresholds remain informative but no longer fully sufficient to identify more effective focal seeds than the betweenness and closeness.
    \textbf{(C, D)} Relative performance of seeding policies under a centrality–based cost structure for the policy support experiment and the app adoption experiment, respectively. The complex centrality policy based on the estimated thresholds significantly outperforms the other policies (one-tailed Wilcoxon signed-rank test, all p-values are smaller than $p = 0.001$, the smallest effect size is $r=0.36$ in the AA Study in the comparison against the uncalibrated complex centrality). In all the panels, crossbars represent the mean together with the $95\%$ confidence intervals over $3,240$ runs.}
    \label{fig:largez2}
\end{figure*}

The results for the larger seed set size are in qualitative agreement with those in the main text (see Figs.~\ref{fig:largez1}--\ref{fig:largez2}), except for the relative performance of the methods in the preference-based cost scenario for the AA Study.
In this scenario, we find that while the neighborhood susceptibility still outperforms the degree and collective influence (one-tailed Wilcoxon signed-rank test, all p-values smaller than $p = 0.001$; the smallest effect size is $r=0.21$ in the comparison against the collective influence for the average rank metric), it no longer significantly outperforms the betweenness and closeness centralities (one-tailed Wilcoxon signed-rank test, all p-values are larger than $p=0.12$; the largest effect size is $r=0.05$ in the comparison against the betweeness centrality for the average rank metric). The reason is arguably that with only one focal seed node selected by the neighborhood susceptibility metric, there is no guarantee that the additional seeds around the focal node require low cost for being successfully targeted. 
This limitation might be attenuated by network homophily, which we will investigate in a follow-up article. On the other hand, the calibrated complex centrality benefits from maximizing spreading size by taking into account clustered seeding in its calculation.

\clearpage

\subsection{Results in scenarios with partial threshold data}
\label{secSI:partial}

We examine the relative performance of the seeding policies in scenarios where only a portion of the nodes have been surveyed, and a fraction of the thresholds is missing. 
For the behavior-based policies, we narrow our focus to the low-threshold and neighborhood susceptibility policies (and consequently to the preference-based cost where they outperform the other metrics), as the calibrated complex centrality requires all the thresholds as input.
For simplicity, we assume that when selecting the seeds according to these two policies, the social change practitioner neglects completely information from the nodes that she has not been able to survey. 
Therefore in this scenario, the focal seed according to the low-threshold policy is the one with the lowest threshold among the surveyed nodes; the focal seed according to the neighborhood susceptibility policy is the one with the largest number of low-threshold neighbors among the surveyed nodes.

We find that for both experiments, the neighborhood susceptibility policy maintains an advantage over the other policies even where a sizeable fraction of the nodes (e.g., $10\%$ or $25\%$) have not been surveyed (see Supplementary Figs.~\ref{figSI:missing_10}-\ref{figSI:missing_25}).
When data is highly incomplete ($50\%$ or more non-surveyed nodes), the significant advantage of the neighborhood susceptibility policy is still robust in the PS experiment, but not in the AA experiment (see Supplementary Figs.~\ref{figSI:missing_50}-\ref{figSI:missing_90}).
This is arguably due to the larger number of low-threshold nodes in the PS experiment, which makes the identification of low-threshold nodes simpler even with highly incomplete data.
Note that, for simplicity, we restricted the set of potential seeds to the surveyed nodes, even in scenarios where they only constitute a small portion of the population. 
An alternative approach to deal with data incompleteness would rely on  threshold measurements at group level.
The Hierarchical Bayes algorithm adopted here~\cite{allenby2006hierarchical} can indeed deliver group-level parameter distributions (where the groups are determined, e.g., based on gender, ethnicity or any relevant discrete covariates), from which  representative individuals can be drawn and their thresholds used as input for the behavior-based seeding algorithms. This possibility, which we leave to future research, might further improve the detection of nodes with high neighborhood susceptibility and potentially enable an estimation of the complex centrality in scenarios with incomplete data.



\begin{figure*}[h]
    \centering
    \includegraphics[width=\textwidth]{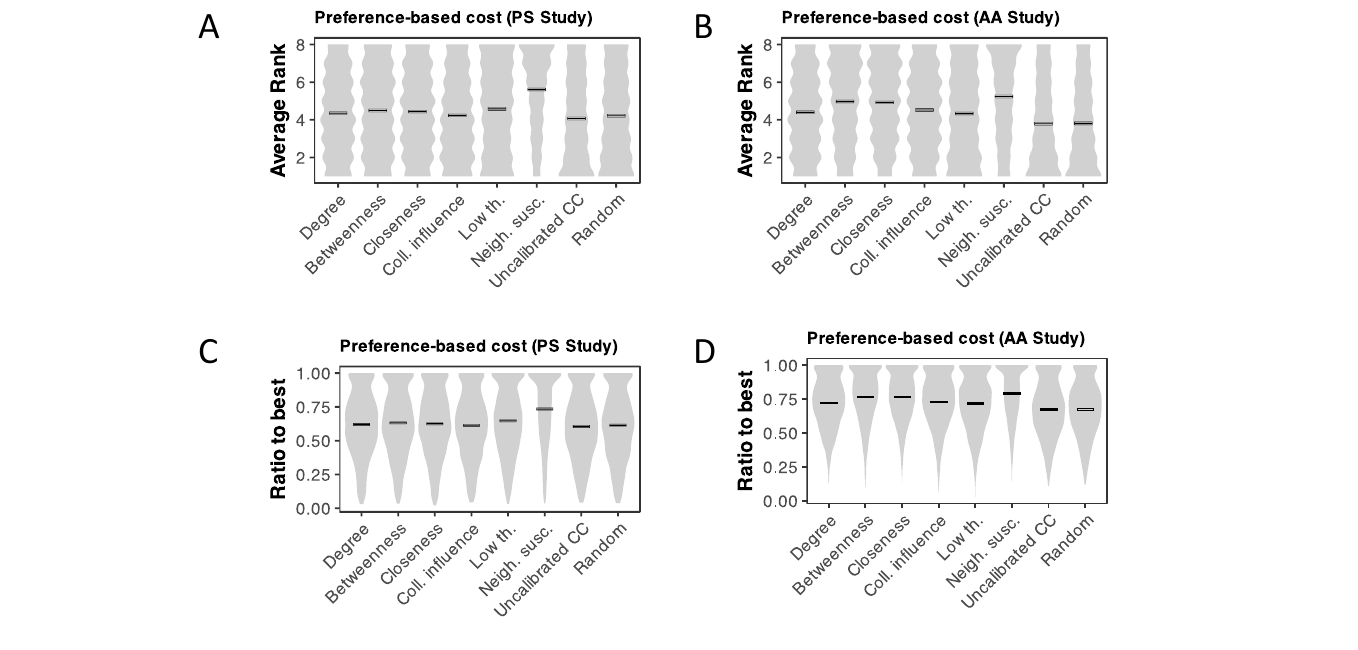}
    \caption{\textbf{Relative performance of seeding policies when $90\%$ of the nodes are surveyed.} 
    Relative performance of seeding policies under a preference–based cost structure, measured through the mean rank metric used in main text \textbf{(A, B)} and through the ratio-to-best metric defined above \textbf{(C, D)} for the policy support experiment and the app adoption experiment, respectively. The neighborhood susceptibility policy based on the estimated thresholds outperforms the other policies (one-tailed Wilcoxon signed-rank test, all p-values are smaller than $p = 0.001$, the smallest effect size is $r=0.07$ in the AA Study in the comparison against the betweenness centrality for the average rank metric).
    In all the panels, crossbars represent the mean together with the $95\%$ confidence intervals over $3,240$ runs.
}
\label{figSI:missing_10}
\end{figure*}

\begin{figure*}[h]
    \centering
    \includegraphics[width=\textwidth]{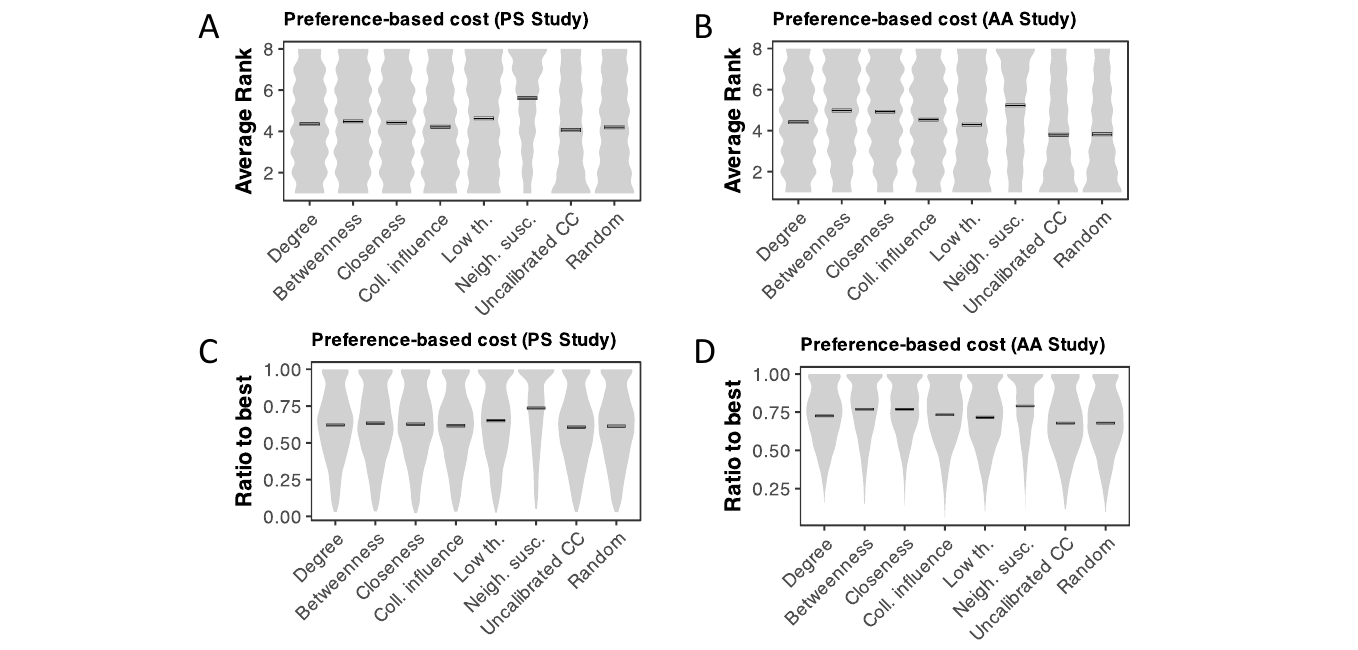}
    \caption{\textbf{Relative performance of seeding policies when $75\%$ of the nodes are surveyed.} 
    Relative performance of seeding policies under a preference–based cost structure, measured through the mean rank metric used in main text \textbf{(A, B)} and through the ratio-to-best metric defined above \textbf{(C, D)} for the policy support experiment and the app adoption experiment, respectively. The neighborhood susceptibility policy based on the estimated thresholds outperforms the other policies (one-tailed Wilcoxon signed-rank test, all p-values are smaller than $p = 0.001$, the smallest effect size is $r=0.07$ in the AA Study in the comparison against the betweenness centrality for the average rank metric).
    In all the panels, crossbars represent the mean together with the $95\%$ confidence intervals over $3,240$ runs.
}
    \label{figSI:missing_25}
\end{figure*}

\begin{figure*}[h]
    \centering
    \includegraphics[width = \textwidth]{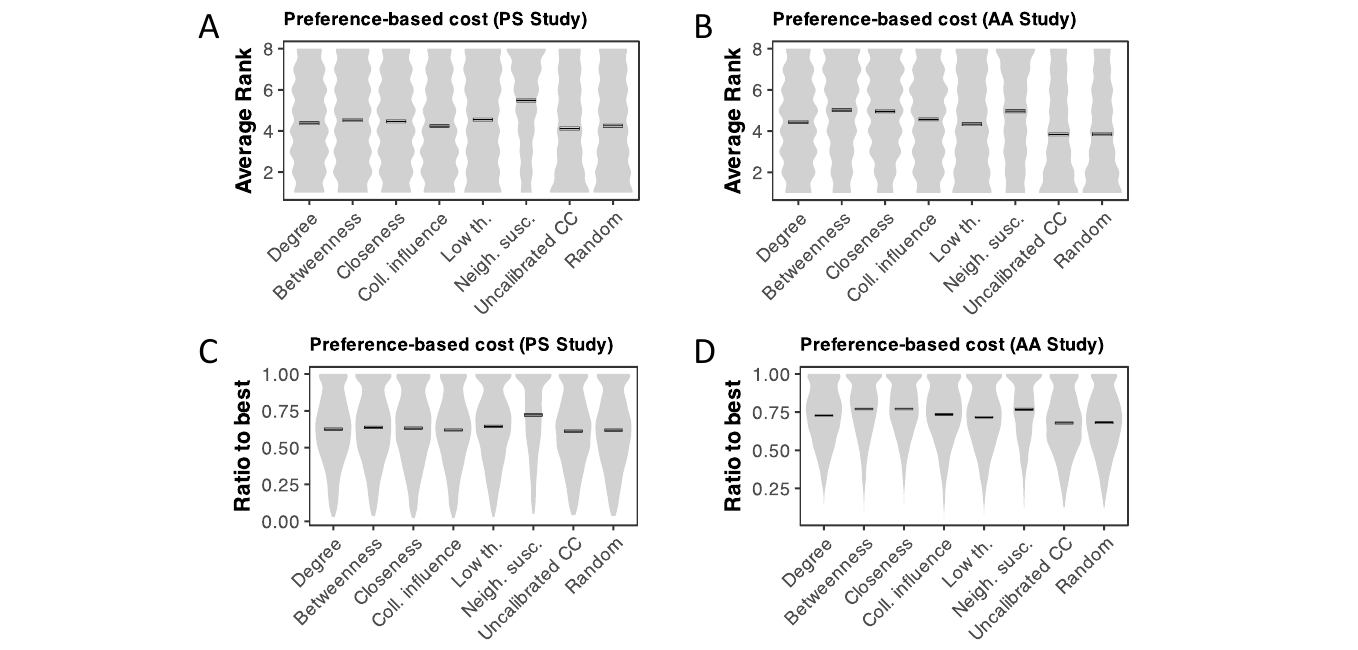}
    \caption{\textbf{Relative performance of seeding policies when $50\%$ of the nodes are surveyed.} 
    Relative performance of seeding policies under a preference–based cost structure, measured through the mean rank metric used in main text \textbf{(A, B)} and through the ratio-to-best metric defined above \textbf{(C, D)} for the policy support experiment and the app adoption experiment, respectively. 
    In the simulations calibrated with the AA experiment, the neighborhood susceptibility policy based on the estimated thresholds no longer outpeforms the betweenness and closeness centrality (one-tailed Wilcoxon signed-rank test, all p-values are larger than $p = 0.48$, the largest effect size is $r=0.0007$ in the comparison against the closeness centrality for the average rank metric).
    In all the panels, crossbars represent the mean together with the $95\%$ confidence intervals over $3,240$ runs.
    }
\label{figSI:missing_50}
\end{figure*}

\begin{figure*}[h]
    \centering    
    \includegraphics[width = \textwidth]{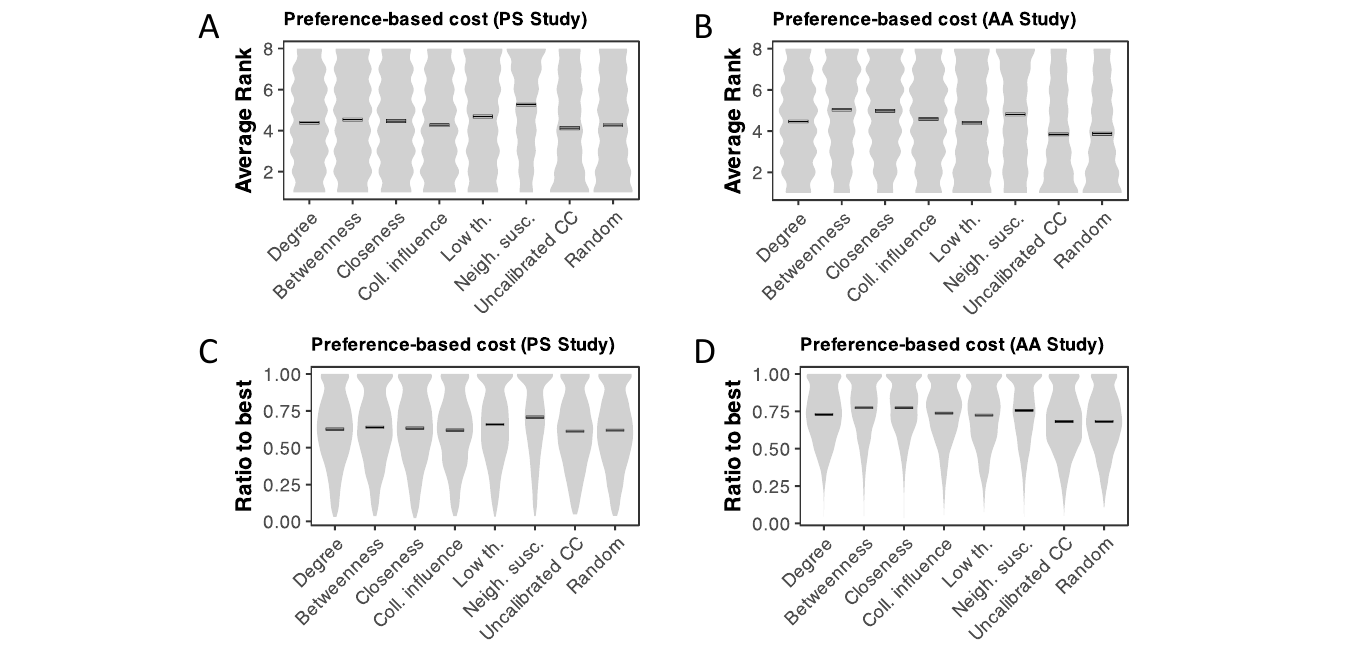}
    \caption{\textbf{Relative performance of seeding policies when $25\%$ of the nodes are surveyed.} 
    Relative performance of seeding policies under a preference–based cost structure, measured through the mean rank metric used in main text \textbf{(A, B)} and through the ratio-to-best metric defined above \textbf{(C, D)} for the policy support experiment and the app adoption experiment, respectively. 
    In the simulations calibrated with the AA experiment, the neighborhood susceptibility policy based on the estimated thresholds no longer outperforms betweenness and closeness centrality (one-tailed Wilcoxon signed-rank test, all p-values are larger than $p = 0.997$, the largest effect size is $r=0.07$ in the comparison against the betwenness centrality for the ratio-to-best metric).
    In all the panels, crossbars represent the mean together with the $95\%$ confidence intervals over $3,240$ runs.
    }    
    \label{figSI:missing_75}
\end{figure*}

\begin{figure*}[h]
    \centering
    \includegraphics[width = \textwidth]{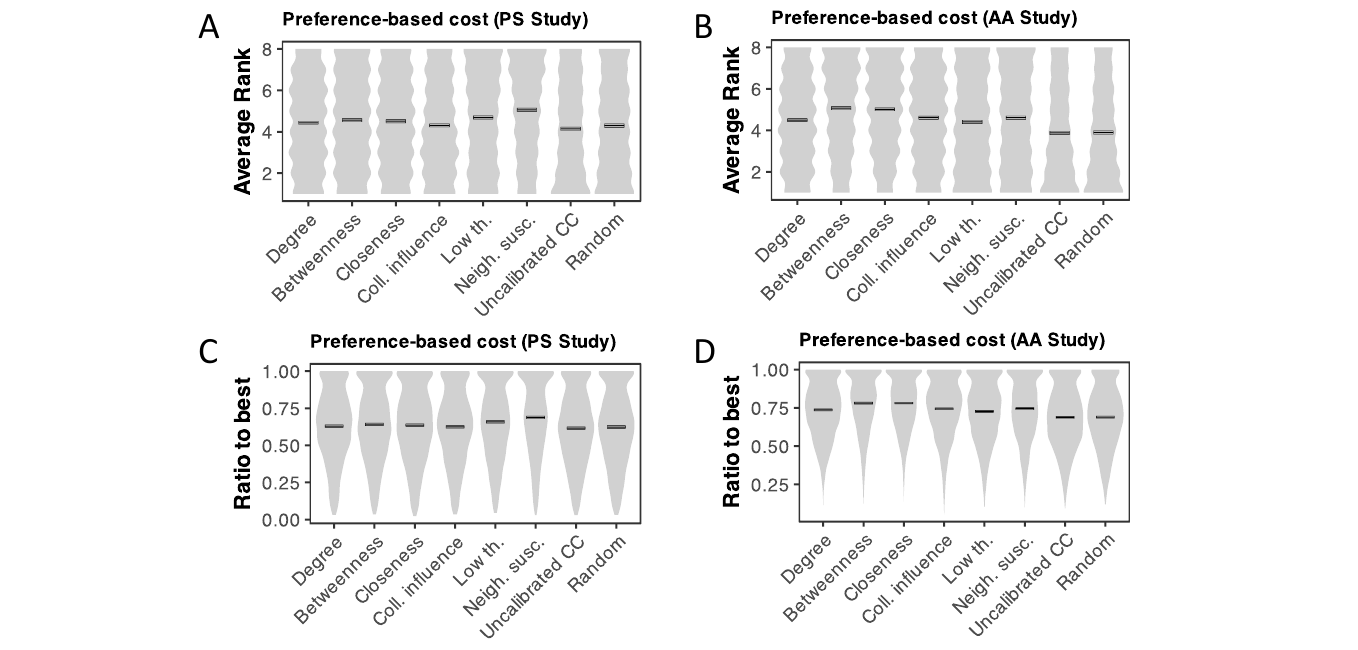}
    \caption{\textbf{Relative performance of seeding policies when $10\%$ of the nodes are surveyed.} 
    Relative performance of seeding policies under a preference–based cost structure, measured through the mean rank metric used in main text \textbf{(A, B)} and through the ratio-to-best metric defined above \textbf{(C, D)} for the policy support experiment and the app adoption experiment, respectively. 
    In the simulations calibrated with the AA experiment, the neighborhood susceptibility policy based on the estimated thresholds no longer outperforms betweeness centrality, closeness centrality and collective influence (one-tailed Wilcoxon signed-rank test, all p-values are larger than $p = 0.47$, the largest effect size is $r=0.14$ in the comparison against the betweenness centrality for the ratio-to-best metric).
    In all the panels, crossbars represent the mean together with the $95\%$ confidence intervals over $3,240$ runs.
    }
\label{figSI:missing_90}
\end{figure*}

\clearpage

\subsection{Effect of prior specification on the identification of effective seeds}
\label{secSI:sensitivity_prior_collective}

\begin{figure*}[h]
    \centering
    \includegraphics[width=\textwidth]{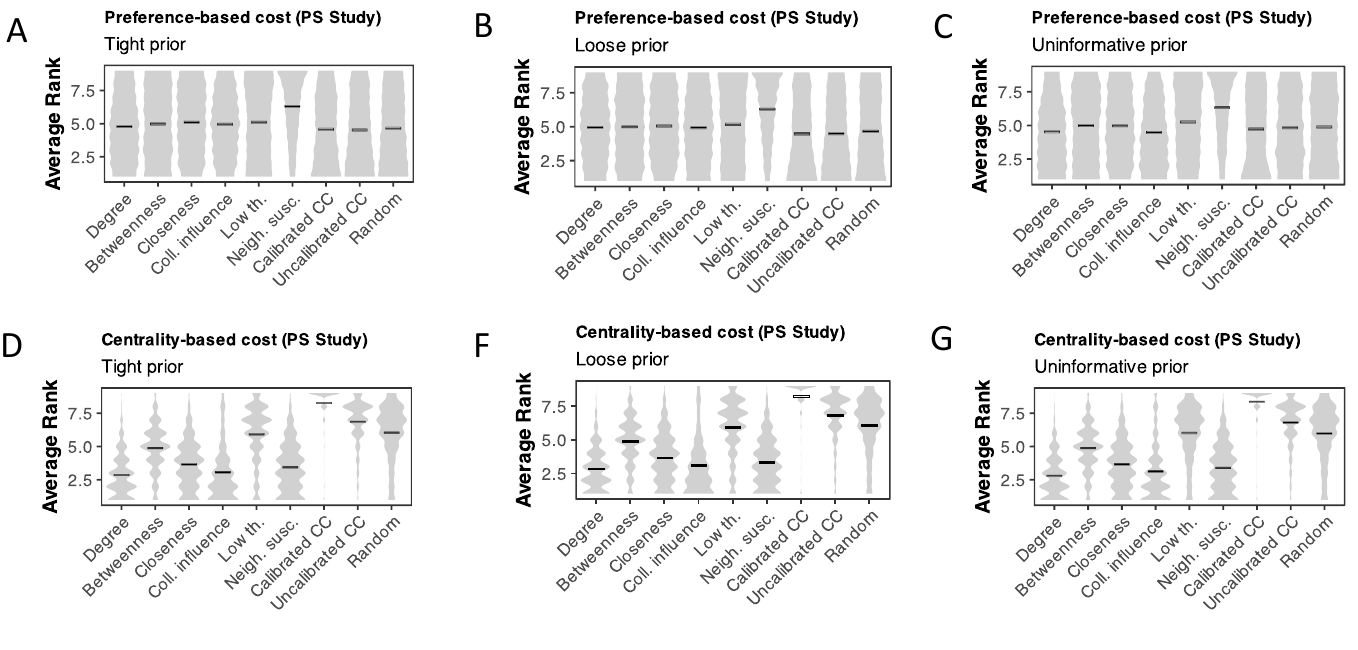}
    \caption{\textbf{Relative performance of seeding policies in the PS Study under different priors and cost structures according to the mean rank metric.} 
    Overall the results are in qualitative agreement with those obtained in Fig.~2 in the main text.
    }
    \label{figSI:hb_priors_coll_ps}
\end{figure*}

\begin{figure*}[h]
    \centering
    \includegraphics[width=\textwidth]{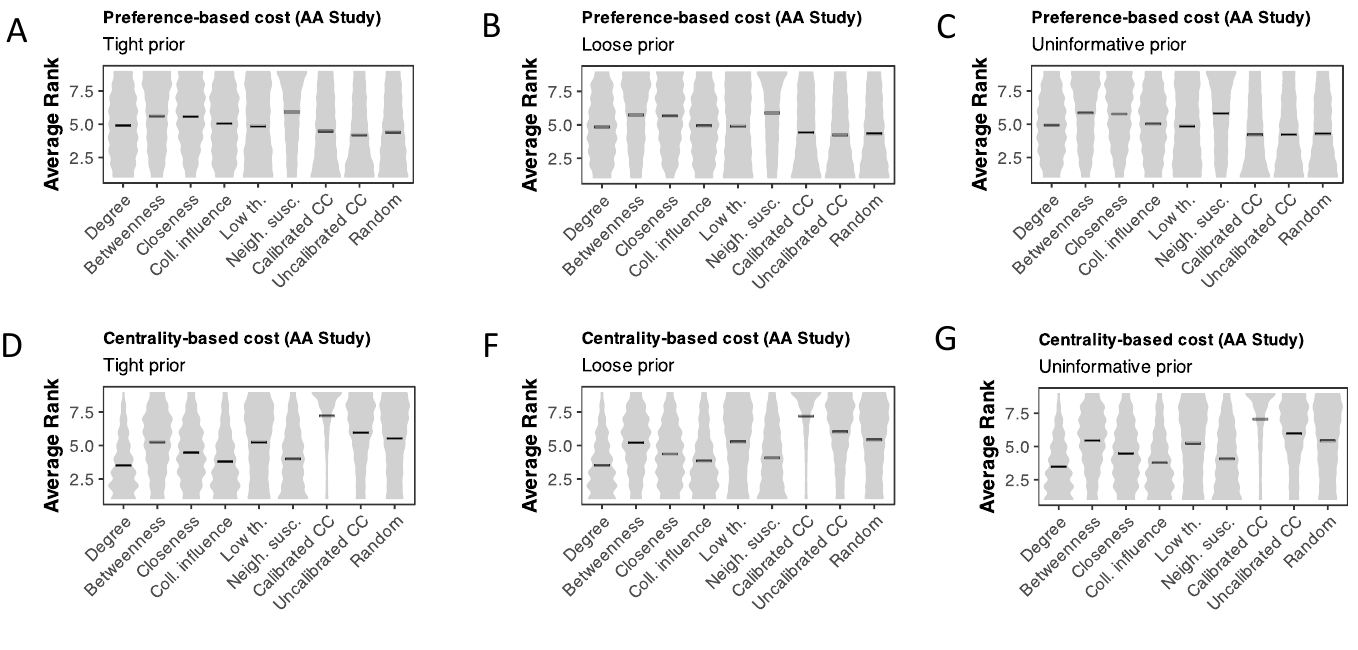}
    \caption{\textbf{Relative performance of seeding policies in the AA Study under different priors and cost structures according to the mean rank metric.} 
    Overall the results are in qualitative agreement with those obtained in Fig.~2 in the main text.
    }
    \label{figSI:hb_priors_coll_aa}
\end{figure*}

In Supplementary Note~\ref{secSI:sensitivity_prior_individual} we showed the estimated thresholds are robust to the priors chosen in the HB estimation (as long as priors are sensible). 
We now examine if the small detected differences in the estimated thresholds across different priors might still lead to substantial differences in the performance of seeding strategies informed by the different estimated thresholds. 
To this end, we replicate the simulation procedure described in the Methods section of the main paper, using the adoption thresholds estimated under the four prior specifications, based on the dataset with 15 choice tasks per respondent (see Supplementary Note~\ref{secSI:sensitivity_prior_individual}). 
The results are shown in Figs.~\ref{figSI:hb_priors_coll_ps} (PS Study) and~\ref{figSI:hb_priors_coll_aa} (AA Study).

For both studies, across all prior specifications, the results are in qualitative agreement with those in Fig.~2 in the main text.  
In the preference-based cost scenario, the neighborhood susceptibility remains the most effective seeding strategy in all cases except for the uninformative prior in the AA Study (one-tailed Wilcoxon signed-rank test, all p-values are smaller than $p=0.02$, the smallest effect size is $r=0.03$ in the comparison with betweenness centrality in the AA Study for the loose prior).
In the centrality-based cost, the calibrated complex centrality performs best (one-tailed Wilcoxon signed-rank test, all p-values are smaller than $p=0.001$, the smallest effect size is $r=0.33$ in the comparison with the uncalibrated complex centrality in the AA Study for the uninformative prior). 
This consistency can be attributed to the relatively small differences in threshold estimation accuracy (Fig.~\ref{figSI:hb_priors_ind}B,E) and the high correlation between true and estimated thresholds across all priors (Fig.~\ref{figSI:hb_priors_ind}C,F), since the adoption threshold is the only estimated parameter used for the seed identification. 

\clearpage

\clearpage

\section{Recovering social spreading models from choice theory}
\label{secSI:linking}

In main text, we assumed a threshold-based diffusion model, aligned with the social contagion literature~\cite{valente1996social,centola2007complex,guilbeault2018complex}.
Here we show that a utility-based framework under different assumptions that those adopted in main text still recovers interpretable social spreading models -- e.g., the traditional Bass diffusion~\cite{bass1969new} and logistic diffusion models~\cite{maccoun2012burden} -- which could be investigated in future empirical research.

In this section, we move from a deterministic choice model to a stochastic one for binary choices (here the binary choice of interest is adopting vs. not adopting the new product or behavior). Following standard random utility theory~\citep{train2009discrete,bouchaud2013crises}, a stochastic choice model is achieved by introducing a noise term into the utility functions. The relative importance of the noise compared to the utility's deterministic components is ruled by a clarity parameter: the lower the clarity, the higher the importance of the noise term~\cite{maccoun2012burden}. 
In Section~\ref{secSI:complex_contagion_theory}, we show that by making the same assumptions about the distribution of noise as those used for estimation (see Methods), we obtain a logistic diffusion model, which can be interpreted as a noisy threshold-based diffusion model~\cite{eckles2024long}. The threshold model emerges as the high-clarity limit (or zero-temperature limit, in the statistical physics terminology~\cite{marsili1999multinomial}) of the resulting model. Notably, in the low-clarity limit, where noise dominates, the logistic model approximates the widely-used Bass model of new product diffusion~\cite{bass1969new} (see Supplementary Note~\ref{secSI:bass_model}).

\subsection{General framework}

In main text, we considered a simple utility-based choice model where an individual adopts product $i$ as soon as the utility from adopting product $i$, $U_{ni}$, exceeds the utility from not adopting, $U^{(0)}_{n}$, which led to a simple formula linking individual threshold and utility coefficients (Eq~1 in main text). Within discrete choice theory, random utility models~\cite{train2009discrete,bouchaud2013crises} provide a natural framework for estimating individual utilities from empirical data (which we already leveraged in main text, see Methods) and for modeling noisy choices, which is the focus of this Section.

In general terms, discrete choice theory~\cite{train2009discrete} considers behavioral models $y=h(x,\epsilon)$ that describes how an individual-level choice $y$ is determined by some observable properties $x$ and some unobservable ones, $\epsilon$, where $\epsilon$ follows a certain distribution, $\epsilon\sim f(\epsilon)$.
The goal is typically to obtain the probability that an individual makes choice $y$, $\Pi(y|x)$, given the observable properties $x$.
The theory assumes that an individual $n$ chooses among $I$ alternatives $i\in\{1,\dots,I\}$ in such a way to maximize her utility~\cite{train2009discrete}
\begin{equation}
    U_{ni}=\lambda\,U^*_{ni}+\epsilon_{ni},
\end{equation}
where $U^*_{ni}$ depends on observable properties of individual $n$ and alternative $i$; $\epsilon_{ni}$ is an error term drawn from a certain distribution $f$: $\epsilon_{ni}\sim f(\epsilon_{ni})$; $\lambda$ denotes a scale parameter~\cite{train2009discrete} also referred to as clarity parameter~\cite{maccoun2012burden}.
One can show that if $\epsilon_{i\alpha}$s are i.i.d. variables extracted from an extreme value distribution (Gumbel, $f(\epsilon)=\exp{(-\epsilon)}\,\exp{(-\exp{(-\epsilon)})}$, then the probability that individual $n$ chooses alternative $i$ is simply~\cite{train2009discrete}
\begin{equation}
    \Pi_{ni}=\frac{\exp{(\lambda\,U^*_{ni})}}{\sum_{j=1}^{I}\exp{(\lambda\,U^*_{nj})}}.
    \label{multiple}
\end{equation}
Eq.~\eqref{multiple} can be also derived via statistical physics methods, by assuming that an individual maximizes her expected utility plus the variety of her available options, represented by the entropy of the choice probability distribution~\cite{marsili1999multinomial}.
For a \textit{binary choice} between not adopting and adopting a given product $i$, we simplify the notation and denote by $\Pi_{ni}$ and $\bar{\Pi}_{ni}=1-\Pi_{ni}$ the probabilities that $n$ adopts and does not adopt product $i$, respectively. We have:
\begin{equation}
    \Pi_{ni}=\frac{1}{1+\exp{(\lambda\,(\bar{U}^*_{ni}-U^*_{ni}))}},
    \label{binary1}
\end{equation}
where we denoted with $U^*_{ni}$ and $\bar{U}^*_{ni}$ the systematic utilities from adopting and not adopting, respectively.
In the following, we shall use these results to derive collective adoption models. 
Clarity parameter $\lambda$ rules the relative importance of the systematic and stochastic components; in the statistical physics language, it can be referred to as inverse temperature~\cite{marsili1999multinomial}.
We can consider two regimes: high clarity ($\lambda\gg 1$) and low clarity ($\lambda\ll 1$). In the binary case, in the high-clarity regime,
\begin{equation}
    \Pi_{ni}=\frac{1}{1+\exp{(\lambda(\bar{U}^*_{ni}-U^*_{ni}))}}\simeq \Theta(U^*_{ni}-\bar{U}^*_{ni}),
    \label{binary}
\end{equation}
where $\Theta$ is the Heaviside function. This represents the threshold rule that has been used in main text to reinterpret the threshold-based diffusion model.
In the low-clarity regime, we use the property that for $\epsilon\sim 0$, one has $(1+\exp{\epsilon})^{-1}\simeq 1/2-\epsilon/4$. Therefore,
\begin{equation}
    \Pi_{ni}\simeq \frac{1}{2}+\frac{\lambda}{4}(U^*_{ni}-\bar{U}^*_{ni})+\mathcal{O}(\lambda^2).
    \label{low_clarity}
\end{equation}
The threshold-like probability obtained in the high-clarity regime and the linear probability obtained in the low-clarity regime can be linked to the adoption probabilities for a complex and a simple social contagion model, respectively, as shown below.

\subsection{High clarity regime: Threshold model}
\label{secSI:complex_contagion_theory}

We discuss here we show how the threshold model analyzed in the main text emerges naturally as the high-clarity limit of the random utility model introduced above. We further demonstrate how the utility-based framework provides intuitive expressions for the cost associated with successfully targeting a given node. Finally, we examine the influence of two-way inter-attribute interactions on the analytical form of the adoption threshold.

\paragraph*{Linking partworth utilities and the adoption threshold.}
Consider a scenario where an individual makes a binary choice about whether to adopt product $i$. As in the main text, suppose that the systematic component of her utility from adopting product $i$ is:
\begin{equation}
    \bar{U}^*_{ni}=\lambda\,\Biggl(\sum_{a}\beta_{na}\,x_{i a}+\gamma_{n}\,s_{ni}\Biggr)=\lambda\,(U^{(A)}_{ni}+\gamma_{n}\,s_{ni}).
    \label{umodel}
\end{equation}
Suppose that the utility from not adopting is fixed, $\bar{U}^*_{n}=U^{(0)}_{n}$, where $U^{(0)}_{n}$ could represent the price, the risk from adopting, or some product-specific cost of adopting. From Eq.~\eqref{binary}, it follows that
\begin{equation}
    \Pi_{ni}=\frac{1}{1+\exp{(\lambda\,(U^{(0)}_{n}-U^{(A)}_{ni}-\gamma_{n}\,s_{ni}))}}.
\end{equation}
For $U^{(A)}_{ni}=0$, only social signals affect individuals' adoption decisions, and we recover the logistic model referred to as burden of social proof model~\cite{maccoun2012burden}.
In the limit of large $\lambda$ (high clarity), $\Pi_{ni}=1$ if and only if
\begin{equation}
   U^{(0)}_{n}-U^{(A)}_{ni}-\gamma_{n}\,s_{ni}>0,
\end{equation}
which can be rewritten as
\begin{equation}
s_{ni}>\gamma_{n}^{-1}(U^{(0)}_{n}-U^{(A)}_{ni})=\tau_{ni},
\label{th}
\end{equation}
from which Eq~1 in the main text. \\

\paragraph*{Motivating the cost functions.}
The proposed interpretation of the threshold model also enables a natural interpretation of the cost or effort needed to persuade an individual to adopt a new product in the absence of social signal ($s_{ni}=0$), which is the case at the seeding stage, when the diffusion has not started yet. In this case, $U^{(S)}=0$; one can interpret the cost of targeting as the utility increase needed such that $U_{ni}>\bar{U}_{ni}$ is verified. We add the effort term, $U^{(E)}_{ni}$, to the utility as 
\begin{equation}
U^{*}_{ni}(s_{ni}=0)=U^{(A)}_{ni}+U^{(E)}_{ni}.
\label{eq:utility_composition}
\end{equation}
Although efforts to incentivize adoption can take various forms~\cite{blair2019motivating}, to fix ideas, we interpret $U^{(E)}_{ni}$ as the result of a monetary incentive offered by the practitioner to individual $n$ in order to persuade her to adopt $i$.
For simplicity, we assume $U^{(E)}_{ni}=\mu_{n}\,c_{ni}$, where $c_{ni}$ denotes the cost of targeting $n$ individual to promote $i$, and $\mu_{n}$ denotes $n$'s sensitivity to the targeting offer. 
Under these assumption, in the beginning of the diffusion, $n$ adopts only if $U^{(A)}_{ni}+U^{(E)}_{ni}\geq U_{n}^{(0)}$; the minimal incentive for this condition to hold is 
\begin{equation}
    \kappa_{ni}=\mu_n^{-1}(U_n^{(0)}-U^{(A)}_{ni}).
    \label{cost}
\end{equation}
This result is intuitive: the cost to effectively target $n$ for product $i$ is higher the higher the resistance ($R_{ni}=U_n^{(0)}-U^{(A)}_{ni}$) between her status-quo utility and her attribute utility. It aligns with the literature on social norms, which posits that successfully targeting initially-resistant agents requires higher incentive~\cite{vogt2016changing,efferson2020promise,constantino2022scaling}. In main text, we set $\mu_n=1$. Note that Eq.~\eqref{cost} outputs negative values of $\kappa_{ni}$ for individuals with $R_{ni}<0$; we assume that the effective cost of successfully targeting these individuals is zero, which leads to
\begin{equation}
    \kappa_{ni}=\max\Bigl\{U_{n}^{(0)}-U^{(A)}_{ni},0\Bigr\}.
    \label{cost}
\end{equation}
Finally, we consider a fixed implementation cost per intervention, $c_0$, which leads to the final expression for the cost in the preference-based scenario: 
\begin{equation}
C_i(\mathcal{S})=c_0+\sum_{n\in\mathcal{S}}\max\Bigl\{U_{n}^{(0)}-U^{(A)}_{ni},0\Bigr\}
\end{equation}
which is Eq.~5 in main text.
In main text, we also consider a scenario where the cost of targeting depends on the centrality of the prospect target, which better aligns with the literature on influencers in social media~\cite{lanz2019climb,leung2022influencer}.
In the centrality-based cost scenario, the seeding cost is defined as 
\begin{equation}
C_i(\mathcal{S})=c_0+\sum_{n\in\mathcal{S}}d_{n},
\end{equation}
where $d_n$ denotes $n$'s degree, which is Eq.~6 in main text, and $c_0$ denotes again a fixed implementation cost per intervention. In this scenario, the cost is independent of $n$'s utility. \\

\paragraph*{Adoption threshold in presence of interactions.}

In main text, Eq.~\eqref{umodel} for the systematic utility assumes both additivity and separability of the partworths. 
One can formally relax these assumptions in various ways. A general model that includes two-way interactions between all partworth utilities is:
\begin{equation}
\bar{U}^*_{ni}=\lambda\,\Biggl(\sum_{a}\beta_{na}\,x_{i a}+\sum_{a,b}\beta'_{nab}\,x_{i a}\,x_{i b}+\gamma_{n}\,s_{ni}+\sum_a\gamma'_{na}\,x_{i a}\,s_{ni}\Biggr)
    \label{umodel_int}
\end{equation}
where $\beta'_{nab}$ captures the interaction between attributes $a$ and $b$; $\gamma'_{na}$ captures the interaction between attribute $a$ and the social signal. 
Eq.~\ref{umodel_int} can be rewritten as:
\begin{equation}
  \bar{U}^*_{ni}=  \lambda\,(\tilde{U}^{(A)}_{ni}+\tilde{\gamma}_{n}\,s_{ni}),
\end{equation}
where we defined the interacting attribute utility,
\begin{equation}
\tilde{U}^{(A)}_{ni}=\sum_{a}\beta_{na}\,x_{i a}+\sum_{a,b}\beta'_{nab}\,x_{i a}\,x_{i b},
\end{equation}
and the interacting susceptibility to social influence,
\begin{equation}
\tilde{\gamma}_{ni}=\gamma_{n}+\sum_a \gamma'_{na}\,x_{i a}.
\end{equation}
With these definitions, if $\tilde{\gamma}_{n}>0$, one can obtain again an expression for threshold, $\tilde{\tau}_{ni}$ from the condition $U^*_{ni}\geq \overline{U}^*_{ni}$, which leads to the following expression: 
\begin{equation}
\tilde{\tau}_{ni}=\frac{U^{(0)}_{n}-\tilde{U}^{(A)}_{ni}}{\tilde{\gamma}_{ni}},
\label{th_int}
\end{equation}
which is similar to Eq.~1 in main text, with the interaction-free attribute utility and susceptibility being replaced by the interacting ones, $\tilde{U}^{(A)}_{ni}$ and $\tilde{\gamma}_{ni}$. We conclude that as long as $\tilde{\gamma}_{n}>0$, the threshold is well-defined (in a similar sense as for the model without interactions).
We leave for future research explorations of how the thresholds might be estimated in presence of significant interaction effects in specific contexts, which might require tailored designs or estimation methods.

\subsection{Low clarity regime: Bass model}
\label{secSI:bass_model}

We show here how the random utility model defined above, in the low-clarity limit, leads to the well-studied Bass model of new product diffusion, and we link the coefficients of the Bass model to the utility and clarity parameters.
The Bass model assumes that faced with the binary choice of whether to adopt a new product or not, the probability that individual $n$ adopts at a given time $t$ is $\Pi_{n}(t)=p_n+q_n\,s_{n}(t)$, where $s_n(t)$ is the social signal received by $n$ at time $t$ about the product, and $p,q$ denote the innovation and innovation coefficient, respectively; for simplicity in this section, we omit the product subscript ($i$) from the notation. In a fully-connected network, $s_{n}(t)=s(t)$ denotes $n$'s cumulative number of adopters at time $t$, which in a homogeneous preference scenario ($p_n=p, q_n=q$), leads to the original formulation of the Bass model~\cite{bass1969new}.
When the potential adopters are embedded in a non-trivial network topology, $s_{n}(t)$ can represent the fraction of $n$'s social contacts who already adopted the product~\cite{rossman2021network}.
Following Centola and Macy's terminology~\cite{centola2007complex}, the Bass model can be interpreted as a \textit{simple contagion model} as individuals have a non-zero probability of adopting even if only one of their social contacts already adopted (even when $p=0$).

We show that the Bass model can be reinterpreted in terms of individual-level utility functions. 
Similarly as above, we assume that individual $n$'s utility from adopting the new product at a given time $t$ is $U^*_{n}(t)=\lambda\,(U^{(A)}_n+\gamma_n \,s_{n}(t))$, where $U^{(A)}_n$ could be expressed, as in the main text, as a linear combination of the product attributes; again, $\gamma_n$ denotes the marginal utility of social signals for $n$. 
We assume again that the status-quo utility from not adopting is fixed: $\bar{U}^*_{n}(t)=U^{(0)}_n$, where $U^{(0)}_n$ is a parameter that could represent the price or risk associated with the adoption. In the low-clarity regime Eq.~\eqref{low_clarity}, we obtain that $\Pi_{n}$ follows the Bass equation with innovation parameter $p_n=(1-\lambda\,(U^{(0)}_n-U^{(A)}_n))/2=1/2-\lambda\,R_n/4=1/2-R_n/T$ and imitation parameter $q_n=\lambda\,\gamma_n/4=\gamma_n/T$, where we defined the temperature parameter~\cite{marsili1999multinomial,bouchaud2013crises} as $T=4/\lambda$. 
The role of temperature parameter $T$ influences the Bass model's parameter: In the infinite-temperature limit, the choice is only determined by the stochastic component of the utility. In this limit, the innovation parameter reflects a coin flip ($1/2$ probability) and the social signal plays no role (zero imitation parameter). As the temperature is large (necessary to derive the linear adoption probability) but finite, $p_n$ deviates from $1/2$ and $q_n>0$. The finite-temperature result matches our intuition: the higher the resistance of individual $n$ for the product, the lower the innovation coefficient; the higher $n$'s marginal utility from the social signal, the higher the imitation coefficient.

This simple argument shows that not only complex contagion models, but also simple contagion ones can be reinterpreted in terms of the proposed micro-level framework. Under standard assumption in discrete-choice theory, if the utility from adopting grows as a power-law of the social signal, the two considered complex and simple contagion models emerge naturally in the high- and low-clarity regimes.
Taken together, these findings suggest that individual-level models might benefit policies to maximize or prevent social spreading for a broader class of models than the complex contagion one studied in the main text.
 We leave the derivation of alternative models of simple contagions (e.g., rumor spreading models~\cite{borge2012absence} and empirical diffusion models~\cite{aral2018social}) for future research.

\end{document}